\documentclass[reprint,superscriptaddress]{revtex4-2}

\usepackage{amssymb,amsmath}
\usepackage{graphicx}
\usepackage{multirow}
\usepackage{appendix}
\usepackage{mathtools}
\usepackage{amsmath,bm}
\usepackage{siunitx}
\usepackage{lineno}
\usepackage{physics}
\modulolinenumbers[5]
\graphicspath{{./IMG/}}
\begin{document}
\title [mode = title]{A First-Principles Explanation of the Luminescent Line Shape of SrLiAl$_3$N$_4$:Eu$^{2+}$  Phosphor for Light-Emitting Diode Applications}

\author{Julien Bouquiaux}
\email{julien.bouquiaux@uclouvain.be}
\affiliation{Institute of Condensed Matter and Nanosciences, Universit\'{e} catholique de Louvain, Chemin des \'{e}toiles 8, bte L07.03.01, B-1348 Louvain-la-Neuve, Belgium}
\author{Samuel Ponc\'{e}}
\affiliation{Institute of Condensed Matter and Nanosciences, Universit\'{e} catholique de Louvain, Chemin des \'{e}toiles 8, bte L07.03.01, B-1348 Louvain-la-Neuve, Belgium}
\author{Yongchao Jia}
\affiliation{Yanshan University,Hebei Key Laboratory of Applied Chemistry, Yanshan University, Hebei Street 438, 066004, Qinhuangdao, P. R. China}
\author{Anna Miglio}
\affiliation{Institute of Condensed Matter and Nanosciences, Universit\'{e} catholique de Louvain, Chemin des \'{e}toiles 8, bte L07.03.01, B-1348 Louvain-la-Neuve, Belgium}
\author{Masayoshi Mikami}
\affiliation{Materials Design Laboratory, Science $\&$ Innovation Center, Mitsubishi Chemical Corporation, 1000,
Kamoshida-cho Aoba-ku, Yokohama, 227-8502, Japan}
\author{Xavier Gonze}
\affiliation{Institute of Condensed Matter and Nanosciences, Universit\'{e} catholique de Louvain, Chemin des \'{e}toiles 8, bte L07.03.01, B-1348 Louvain-la-Neuve, Belgium}

\date{\today}

\begin{abstract}
White light-emitting diodes are gaining popularity and are set to become the most common light source in the U.S. by 2025. 
However, their performance is still limited by the lack of an efficient red-emitting component with a narrow band emission.
The red phosphor SrLiAl$_3$N$_4$:Eu$^{2+}$ is among the first promising phosphors with a small bandwidth for next-generation lighting, but the microscopic origin of this narrow emission remains elusive.
In the present work, density functional theory, the $\Delta$SCF-constrained occupation method, and a generalized Huang-Rhys theory are used to provide an accurate description of the vibronic processes occurring at the two Sr$^{2+}$ sites that the Eu$^{2+}$ activator can occupy.
The emission band shape of Eu(Sr1), with a zero-phonon line at 1.906 eV and a high luminescence intensity, is shown to be controlled by the coupling between the 5d$_{z^2}$-4f electronic transition and the low-frequency phonon modes associated with the Sr and Eu displacements along the Sr channel.
The good agreement between our computations and experimental results allows us to provide a structural assignment of the observed total spectrum.
By computing explicitly the effect of the thermal expansion on zero-phonon line energies, the agreement is extended to the temperature-dependent spectrum.
These results provide insight into the electron–phonon coupling that accompanies the 5d–4f transition in similar UCr$_4$C$_4$-type phosphors. 
Furthermore, these results highlight the importance of the Sr channel in shaping the narrow emission of SrLiAl$_3$N$_4$:Eu$^{2+}$, and they shed new light on the structure–property relations of such phosphors.
\end{abstract}
\maketitle
\section{Introduction}
\label{sec:Intro}

Eco-efficient white-light emitting diodes (WLEDs) rely on one or more phosphor materials to convert the ultraviolet or blue emission from a LED chip into a desired  wavelength emission spectrum.
Phosphors are made of a host material doped with an activator, the latter emitting light, whose wavelength is tuned by the effect of host crystal structure and chemical environment. 
The numerous combinations between host and activator have led to the creation of thousands of possible phosphor materials with a variety of photoluminescence (PL) properties including emission peak wavelength, thermal stability, quantum efficiency or shape of the PL emission spectrum.
In the past decade, a large focus was put on the discovery of phosphors with narrow emission bandwidth in order to improve the color-purity of backlighting LED devices or, in the case of the red phosphors in WLED, to avoid wasting energy in the near-infrared region where human eye is not sensitive \cite{lin2016critical,pust2015revolution,lee20222022}. 

In the search for new-generation phosphors~\cite{fang2022evolutionary}, materials with UCr$_4$C$_4$-type structure, doped with Eu$^{2+}$ have recently attracted  attention~\cite{fang2020cuboid,fang2018control}. 
The first instance of such materials, SrLiAl$_3$N$_4$:Eu$^{2+}$ (SLA) phosphor, discovered by Schnick et al.~\cite{pust2014narrow}, was quickly followed by other nitride-based phosphors such as Sr[Mg$_2$Al$_2$N$_4$]:Eu$^{2+}$ or Sr[Mg$_3$SiN$_4$]:Eu$^{2+}$~\cite{pust2014group,schmiechen2014toward}.
Given the strong nephelauxetic effect of N$_8$ cuboid environment around the Eu activator, the emission color is limited to the red region. 
To lower the emission wavelength, adding O$^{2-}$ in the cuboid environment like in the oxy-nitride Sr[Li$_2$Al$_2$O$_2$N$_2$]:Eu$^{2+}$ (SALON)~\cite{hoerder2019sr} allowed to blue-shift the emission peak from 650~nm (SLA) to 614~nm, with similar performance than SLA. 
Numerous oxide-based phosphors were then developed with general formula M$_4$[Li$_3$SiO$_4$]$_4$ where M can be selected from Li$^+$, Na$^+$, K$^+$, Rb$^+$, Cs$^+$ and their combination~\cite{zhao2018next,liao2018learning,fang2020cuboid}. 
Most of these alkali lithosilicate phosphors, with O$_8$ cuboid environment provide green/cyan/blue emission. 

It is commonly accepted that the narrow-band emission of Eu-doped UCr$_4$C$_4$ structure is linked to the highly condensed host structure and the cuboid coordination environment. 
However, the microscopic origin of such narrow emission is not fully understood. 
In this respect, Huang-Rhys theory~\cite{huang1950theory} and its generalization~\cite{alkauskas2014,jin2021photoluminescence} fed by first-principles computations allows one to gain insights into the electron-phonon coupling accompanying electronic optical transitions. 
Mainly used in the context of defects for quantum information technologies~\cite{linderalv2021vibrational,gali2019ab,razinkovas2021vibrational,jin2021photoluminescence}, this theory was only exploited recently to obtain information on the phonon side bands of a few selected phosphors~\cite{bouquiaux2021importance,linderalv2020luminescence,wang2022role}.

Despite the pioneer status and technological relevance of SLA~\cite{wang2018down}, its phonon side bands with apparent vibronic signatures have not received any theoretical attention. 
Initial theoretical studies of SLA  focused on the bulk host material only~\cite{tolhurst2015investigations,tolhurst2016electronic}.
Recently, a time-dependent DFT approach for embedded clusters has clarified the electronic processes responsible for light absorption in SLA
but has not investigated emission and the PL spectrum~\cite{shafei2022electronic}. 

In SLA, the europium dopant can substitute the strontium atom in two inequivalent positions, which lead to two emission centers.
Thanks to their different decay time, time-resolved luminescence was used in 2016 to decompose the zero-phonon line (ZPLs), 15377 cm$^{-1}$ (1.906 eV) and 15780 cm$^{-1}$ (1.956 eV), and corresponding phonon side band of each center but their structural assignment was not done~\cite{tsai2016improvement}.  

In this work, we study from first-principles the vibronic processes occurring in the PL spectrum of SrLiAl$_3$N$_4$:Eu$^{2+}$ (SLA) phosphor.  
We simulate the PL spectra from the two luminescent sites and assign them to their specific microscopic environment
using predictive methods whose accuracy has previously been demonstrated~\cite{jia2016first,ponce2016understanding,jia2017first,jia2017assessment,jia2020design}.

We use large supercells computed with density functional theory (DFT) and the constrained-$\Delta$SCF method to obtain optimized Eu-4f ground state structures and Eu-5d excited state structures.
The atomic displacements induced by the 5d-4f transitions are projected onto the phonons modes of the system in order to obtain the Huang-Rhys (HR) spectral function which provides information on the nature of the phonons participating to the transition.
The HR spectral function is converged by increasing the supercell size using the embedding procedure proposed by Alkauskas \textit{et al.}~\cite{alkauskas2014}.
The PL lineshape is computed following the generating function approach and its temperature-dependent generalization~\cite{jin2021photoluminescence}. 
Finally, the effect of thermal expansion on PL properties is included via volumic quasi-harmonic approximation (QHA)~\cite{rignanese1996ab,carrier2007first} with a minimization of the Helmholtz free energy with respect to the global volume.
This allows one to estimate ZPL energies with increasing temperatures.

It is found that the emission lineshape with a ZPL energy around 1.906~eV and very narrow emission bandwidth~\cite{tsai2016improvement} is associated to the Sr site having a second coordination sphere composed of 3 Li and 5 Al while the one with higher ZPL energy around 1.956~eV and larger emission bandwidth is associated to the Sr site having second coordination sphere composed of 1 Li and 7 Al. 
The very distinct shapes of the two PL spectra originate from different promoted excited 5d orbitals (5d$_{z^2}$-like aligned along Sr channel or 5d$_{x^2-y^2}$-like pointing along Al/Li atoms), which lead to different 5d-4f atomic relaxation pattern, and hence different electron-phonon coupling.
By explicitly including the effect of thermal expansion on ZPL energies, inducing a blue-shift of around 30~meV for both sites between 10~K and 573~K, we find an excellent agreement between the simulated temperature dependent PL spectra and experiment.

The paper is structured as follows. 
We first present the theoretical background and the computational methodology. 
The results of the work are then presented. 
Both Sr sites are compared carefully for each property: the excited state geometry, the coupling with phonons and finally the temperature dependent PL spectra, including thermal expansion effect. 
Finally, we provide additional discussions and conclude this work.  

\section{Theory and computational methodology}
\label{sec:Theory}

\subsection{Huang-Rhys theory for the luminescence spectrum}

\begin{figure}[h!]
    \centering
	\includegraphics[width=0.99\linewidth]{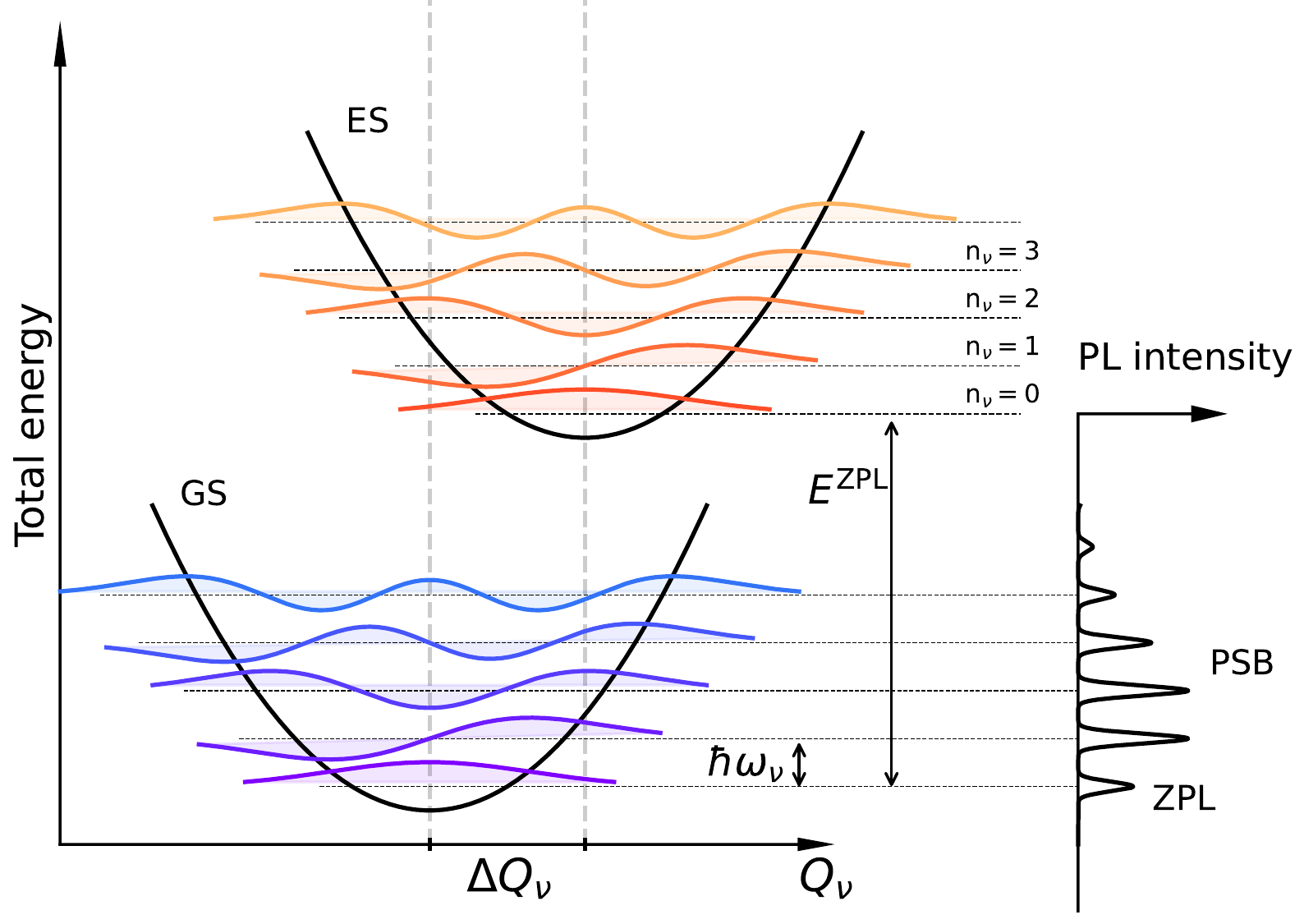}
	\caption{Schematic representation of the origin of photoluminescence (PL) spectra. On the left, ground state (GS) and  excited state (ES) energy curves are projected along phonon normal coordinate $Q_\nu$ (here only for one mode) and are approximated by harmonic functions with the same frequency $\omega_\nu$. Vibrational energy levels and corresponding eigenfunctions are shown as horizontal lines and colored areas. GS and ES are displaced by $\Delta Q_\nu$, the mass-weighted displacement between the  minimum of the GS and the ES curves. On the right, the PL spectrum is formed. 
	The zero-phonon line (ZPL) comes from the transition between the first vibrational level of the ES to the first vibrational level of the GS. Other transitions give the phonon sideband (PSB). The intensity of each peak is computed with the overlap between corresponding eigenfunctions.}
	\label{fig:displaced_HA}
\end{figure}

Within the Huang-Rhys theory and the Franck-Condon approximation~\cite{huang1950theory,lax1952franck}, ground state (GS) and excited state (ES) potential energy surfaces are projected along the harmonic phonon normal coordinates of the system $Q_{\nu}$. 
The GS and ES surfaces are assumed to be identical (same phonon frequencies $\omega_{\nu}$ and eigenmodes) except for a rigid offset $\Delta Q_{\nu}$ coming from linear electron-phonon interaction and an energy difference called zero-phonon line energy $E^{\rm ZPL}$. 
This reduces the problem to a displaced harmonic oscillator problem for each phonon mode $\nu$, as depicted on Fig. \ref{fig:displaced_HA}, for which an analytic expression to compute the associated phonon side band (PSB) exists. 
At 0~K, where only the  ES vibrational state $n = 0$ contributes, one has: $|\langle \chi_{n_{\nu}}^{\rm GS}|\chi_{0_\nu}^{\rm ES} \rangle|^2 = e^{-S_\nu}(S_{\nu})^{n_{\nu}} / (n_{\nu}!)$ with $S_{\nu}$ the Huang-Rhys factor of mode $\nu$. 
For a given photon energy $\hbar\omega$, the luminescence intensity is given by~\cite{jin2021photoluminescence,linderalv2020luminescence,alkauskas2014}:
 \begin{equation}
L(\hbar\omega,T) \propto \omega^3 A(\hbar\omega,T),
\label{eq:LA}
\end{equation}
where the lineshape function $A(\hbar\omega)$ is evaluated as the Fourier transform of the generating function $G(t,T)$~\cite{kubo1955application} :
\begin{align}\label{eq:L(hw)_generating}
A(\hbar\omega, T) &=  \int_{-\infty}^{+\infty} G(t,T)e^{i\omega t-\frac{\gamma}{\hbar}\abs{t}-i\frac{E^{\rm ZPL}}{\hbar}t}dt,\\
G(t,T)            &= e^{S(t)-S(0)+C(t,T)+C(-t,T)-2C(0,T)}, \label{eq:G_generating}
\end{align}
where $S(t)=\sum_{\nu}S_\nu e^{i\omega_{\nu}t}$ and $C(t,T)=\sum_{\nu} \overline{n}{_\nu}(T)S_\nu e^{i\omega_{\nu}t}$ are the Fourier transforms of the Huang-Rhys spectral functions and temperature weighted Huang-Rhys spectral functions, respectively.
$\gamma$ is the homogeneous Lorentzian broadening of each vibronic transition, and $\overline{n}{_\nu}(T)$ is the average occupation number of $\nu$-th phonon mode:
\begin{equation}
    \label{eq:average_phonon}
    \overline{n}{_\nu}(T)=\frac{1}{e^{\frac{\hbar\omega_{\nu}}{k_BT}}-1}.
\end{equation}

The ingredients for Eq.~\eqref{eq:L(hw)_generating} are the partial Huang-Rhys factor of each phonon mode $S_{\nu}$, the phonon frequencies $\omega_{\nu}$ and the zero-phonon line energy E$^{\rm ZPL}$. 
The $S_{\nu}$ is the mean number of phonon $\nu$ involved in the transition:
\begin{equation}\label{eq:S_nu}
S_{\nu}=\frac{\frac{1}{2}\omega_{\nu}^2\Delta Q_{\nu}^2}{\hbar\omega_{\nu}}=\frac{\omega_{\nu} \Delta Q_{\nu}^2}{2\hbar},
\end{equation}
where $\Delta Q_{\nu}$ is the mass-weighted atomic displacement projected along phonon mode $\nu$:
\begin{equation}
    \label{eq:Delta_Q_nu}
    \Delta Q_\nu=\sum_{\kappa\alpha}\sqrt{M_{\kappa}}\Delta R_{\kappa\alpha} e_{\nu,\kappa\alpha},
\end{equation}
where $\Delta \mathbf{R}_{\kappa} = \mathbf{R}_{\kappa}^{\rm GS}-\mathbf{R}_{\kappa}^{\rm ES}$ is the vector associated with the displacement of atom $\kappa$ between excited and ground state, $\mathbf{e}_{\nu,\kappa}$ are the phonon eigenvectors and $M_{\kappa}$ are the atomic masses.
Under the harmonic approximation, Eq.~\eqref{eq:Delta_Q_nu} becomes
\begin{equation}
    \Delta Q_\nu=\frac{1}{\omega_\nu^2}\sum_{\kappa\alpha}\frac{\Delta F_{\kappa\alpha} e_{\nu,\kappa\alpha}}{\sqrt{M_{\kappa}}},
    \label{eq:Delta_Q_nu_forces}
\end{equation}
where $\Delta \mathbf{F}_{\kappa}$ are the ground-state forces evaluated at the equilibrium excited-state structure.
This formulation is advantageous as the forces decay faster than the displacements, see Section 1.1 and 1.2 of the supplementary informations (SI)~\cite{supp}. 
We use DFT to obtain the ground-state optimized structure $\mathbf{R}_{\kappa}^{\rm GS}$, the forces $\Delta \mathbf{F}_{\kappa}$, the phonon eigenvectors $\mathbf{e}_{\nu,\kappa}$ and eigenfrequencies $\omega_{\nu}$. 
The $\Delta$SCF constrained-occupation method is used to optimize the excited-state structure $\mathbf{R}_{\kappa}^{\rm ES}$.

\subsection{Computational method}

Calculations are performed with density-functional theory (DFT) using ABINIT~\cite{Abinit2002,gonze2020abinit} with the PAW method~\cite{torrent2008implementation}.
The generalized gradient approximation (GGA-PBE) is used to treat exchange-correlation effects \cite{perdew1996generalized} and a Hubbard U=7~eV term is added on the $4f$ states of europium, consistently with our previous works~\cite{jia2017first}.
We find that the value of the U parameter has only a weak impact on the ZPL energy and 5d-4f  displacements, see Section 2 of the SI~\cite{supp}.
Calculations on europium-doped SLA are conducted with a 2$\times$2$\times$2 supercell containing 288 atoms. 
The two possible emission centers are treated as two independent systems. 
For all supercell calculations, the structures are relaxed below a maximal residual force of $10^{-4}$~Hartree/Bohr 
(about $0.005$~eV/\AA). 
The cut-off kinetic energy is 25~Ha (680~eV) with a single zone-centered \textbf{k}-point. 
For the treatment of the Eu $4f^65d^1$ excited state, we use the $\Delta$SCF method whereby the eigenfunctions associated to the highest predominantly Eu $4f$ band are forced to be unoccupied while the next predominantly Eu $5d$
energy band is constrained to be occupied. 
While the latter is more hybridized than the Eu $4f$ one, it is nevertheless refered to as a Eu $5d$ band.
The ZPL energy is computed as the difference of the total energy of the relaxed excited states and that of the ground state. 
Detailed information on the use of this approach can be found in Ref.~\cite{bouquiaux2021importance}.
This work focuses on the lowest excited state of the 4f$^6$5d$^1$ configuration and describes the vibronic features appearing in the emission spectrum of the two emission centers. The complete 4f$^6$5d$^1$ configuration, giving rise to a complex fine structure in the excitation band, cannot be captured with our methodology. Multiconfigurational methods for embedded-cluster \cite{joos2020insights,barandiaran2022luminescent} are more suitable to understand the 4f$^6$5d$^1$ manifold and the absorption spectra of Eu-doped phosphor materials.

\subsubsection{Phonons and embedding methodology}
Phonons modes of the undoped primitive cell of SLA containing 36 atoms are obtained by diagonalizing the dynamical matrices computed with DFPT with a 2$\times$2$\times$2  \textbf{k}-points and \textbf{q}-points grids and then Fourier interpolated on a fine 4$\times$4$\times$4 \textbf{q}-grid~\cite{gonze1997dynamical}. 
The phonon band structure Fourier interpolated along high-symmetry lines and the density of states can be found in Section 3 of the SI~\cite{supp}. 

Reaching the dilute limit requires to estimate the interatomic force constants (IFCs) on large defective supercells, which would be computationally too expensive with a direct approach. 
Hence a technique similar to the one proposed by Alkauskas \textit{et al.}~\cite{alkauskas2014} is followed. 
The IFCs of the pristine SLA, computed on a 4$\times$4$\times$4 \textbf{q}-grid, are mapped on the corresponding 4$\times$4$\times$4 supercell.
The IFCs of the Eu-doped system are computed on a smaller 2$\times$2$\times$2 supercell containing 288 atoms using a frozen-phonon approach as implemented in the \textsc{PHONOPY} package~\cite{togo2015first}. 
In order to construct the total IFCs, the following cutoff is applied.
If both atoms $\kappa$ and $\kappa'$ are separated from the dopant by a distance smaller than $R_c$=5.5\AA, then the IFCs computed with the 2$\times$2$\times$2 defect supercell are used.
For all other atomic pairs, the IFCs computed with the pristine system are used.
This procedure breaks the acoustic sum rule~\cite{jin2021photoluminescence} which we reimpose with $ C_{\kappa\alpha,\kappa\alpha}=-\sum_{\alpha\ne\beta}C_{\kappa\beta,\kappa\alpha}$, where $\alpha$ and $\beta$ refers to Cartesian coordinates. 
This embedding methodology allows one to obtain the phonon eigenfrequencies $\omega_{\nu}$ and eigenvectors $\mathbf{e}_{\nu,\kappa}$ of Eq.~\eqref{eq:Delta_Q_nu_forces} on the 4$\times$4$\times$4 supercell. 
The forces $\Delta \mathbf{F}_{\kappa}$ of Eq.~\eqref{eq:Delta_Q_nu_forces} are those of the 2$\times$2$\times$2 supercell and are set to zero elsewhere because of the short-range decay of the forces with respect to the Eu activator.
Convergence of the Huang-Rhys spectral function as a function of the size of the large supercell is reported in Section 1.3 of the SI~\cite{supp}. 

\subsubsection{Thermal expansion}
To compute the effect of thermal expansion, the volumic quasiharmonic approximation (QHA) is used, neglecting deviatoric thermal stresses and internal forces (v-ZSISA approximation)~\cite{carrier2008quasiharmonic,allan1996zero,masuki2023full}. 
The only contribution to the thermal expansion of the crystal is due to the coupling between phonons and the change in unit cell volume~\cite{ritz2019thermal}, with cell parameters and internal coordinates relaxed at fixed volumes. 
We obtain the temperature versus volume curve of the undoped SLA by minimizing the Helmholtz free energy (FE)~\cite{rignanese1996ab,carrier2007first,brousseau2022zero}:
\begin{equation}
    \label{F(V,T}
    F(V,T)=E(V)+F^{\rm vib}(V,T),
\end{equation}
where 
\begin{equation}
	\label{Fvib(V,T}
	F^{\rm vib}(V,T)=\sum_{\bm{q}\nu}\frac{\hbar\omega_{\bm{q}\nu}(V)}{2}+k_{\rm B}T \ln(1-e^{-\frac{\hbar\omega_{\bm{q}\nu}(V)}{k_{\rm B}T}}),
\end{equation}
where $E(V)$ is the static DFT energy versus volume curve, with all non-volumic degrees of freedom relaxed, and $F^{\rm vib}(V,T)$ is the vibration energy with the same cell and atomic geometry. 
The FE curve is constructed using seven configurations within a [-2\%,+4\%] volume change with respect to the static equilibrium volume. 
The minimum volume for a given temperature is obtained by fitting a Murnaghan equation of state~\cite{fu1983first} to the FE curve. 
Detailed information on the FE curves, temperature dependent thermal expansion coefficient and zero-point volume are reported in Section 4 of the SI~\cite{supp}.
To evaluate the effect of thermal expansion on the PL spectrum, we have scaled the lattice parameters obtained for the undoped SLA to the 288 atoms supercells, and re-computed ZPL energies at fixed volumes corresponding to the selected temperatures, see Fig.~S8 of the SI~\cite{supp}. 
This allowed us to compute both the shift of the ZPL energies with temperature as well as the effect of zero-point volume on the ZPL energy at 0~K. 

\section{Results and discussion}
\label{sec:Results}

\begin{figure}
\centering
\includegraphics[width=0.99\linewidth]{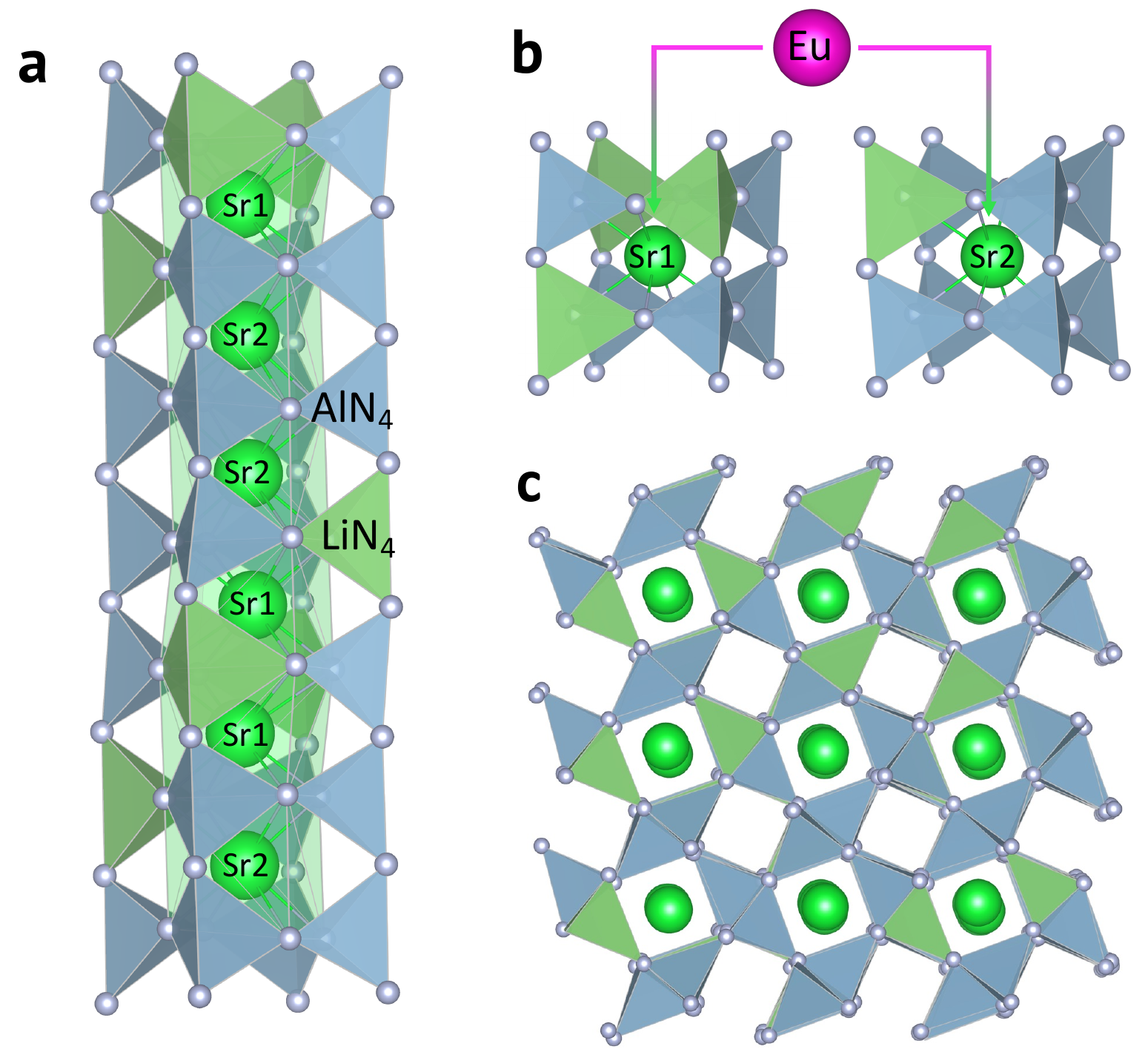}
\caption{\textbf{a} Crystal structure of SrLiAl$_3$N$_4$:Eu$^{2+}$ viewed along the Sr chain. \textbf{b} First and second shell environment of Sr1 and Sr2 sites that Eu atom can substitute. \textbf{c} Structure perpendicular to the Sr chains, with the characteristic UCr$_4$C$_4$ framework.}
\label{fig:Cristal_stru}
\end{figure}

\subsection{Ground state geometry}
\label{subsec:ground_state}

The SLA crystallizes in a triclinic crystal system ($P\overline{1}$ space group) with the typical channel structure observed in other UCr$_4$C$_4$ type phosphors, see Fig.~\ref{fig:Cristal_stru}.
Our computed lattice parameters are overestimated by about 0.3\%, which is common for PBE exchange and correlation functionals. 
The angles match within 0.05$^{\circ}$ difference.
Half of the channels are occupied by the Sr$^{2+}$ ions. 
These channels are built up with ordered tetrahedra alignment as seen in Fig.~\ref{fig:Cristal_stru}\textbf{a}, with one green [LiN$_4$]$^{11-}$ followed by three blue [AlN$_4$]$^{9-}$. 
Two Sr sites with very similar first-shell environment (Sr-N$_8$) but different second-shell can host the Eu activator and are shown in Fig.~\ref{fig:Cristal_stru}\textbf{b}, which leads to two emission centers.
The first site, denoted as Sr1 in this work, is surrounded by 3 [LiN$_4$]$^{11-}$ and 5 [AlN$_4$]$^{9-}$ tetrahedra, while the second, denoted as Sr2, is surrounded by one [LiN$_4$]$^{11-}$ and 7 [AlN$_4$]$^{9-}$ tetrahedra. 
We observe that within the Sr chain, the Sr1-Sr1 distance of 3.2~\AA\, is shorter than the Sr2-Sr2 distance of~3.42 \AA. 
The Sr1-Sr2 distance is in between with a value of 3.28~\AA. 
We compute that the Sr-N$_8$ cuboid has similar mean Sr-N distance in both sites (2.809~\AA{} and 2.804~\AA)
When the Sr atoms are replaced by Eu atoms, these distances change by less than 0.1\% due to similar atomic radii for the Eu$^{2+}$ and Sr$^{2+}$ atoms~\cite{shannon1976revised}. Additional informations can be found in Section 5 of the SI~\cite{supp}.
%

\subsection{Excited state geometry}
\label{subsec:Excited state}

\begin{figure*}
	\centering
	\includegraphics[width=0.99\linewidth]{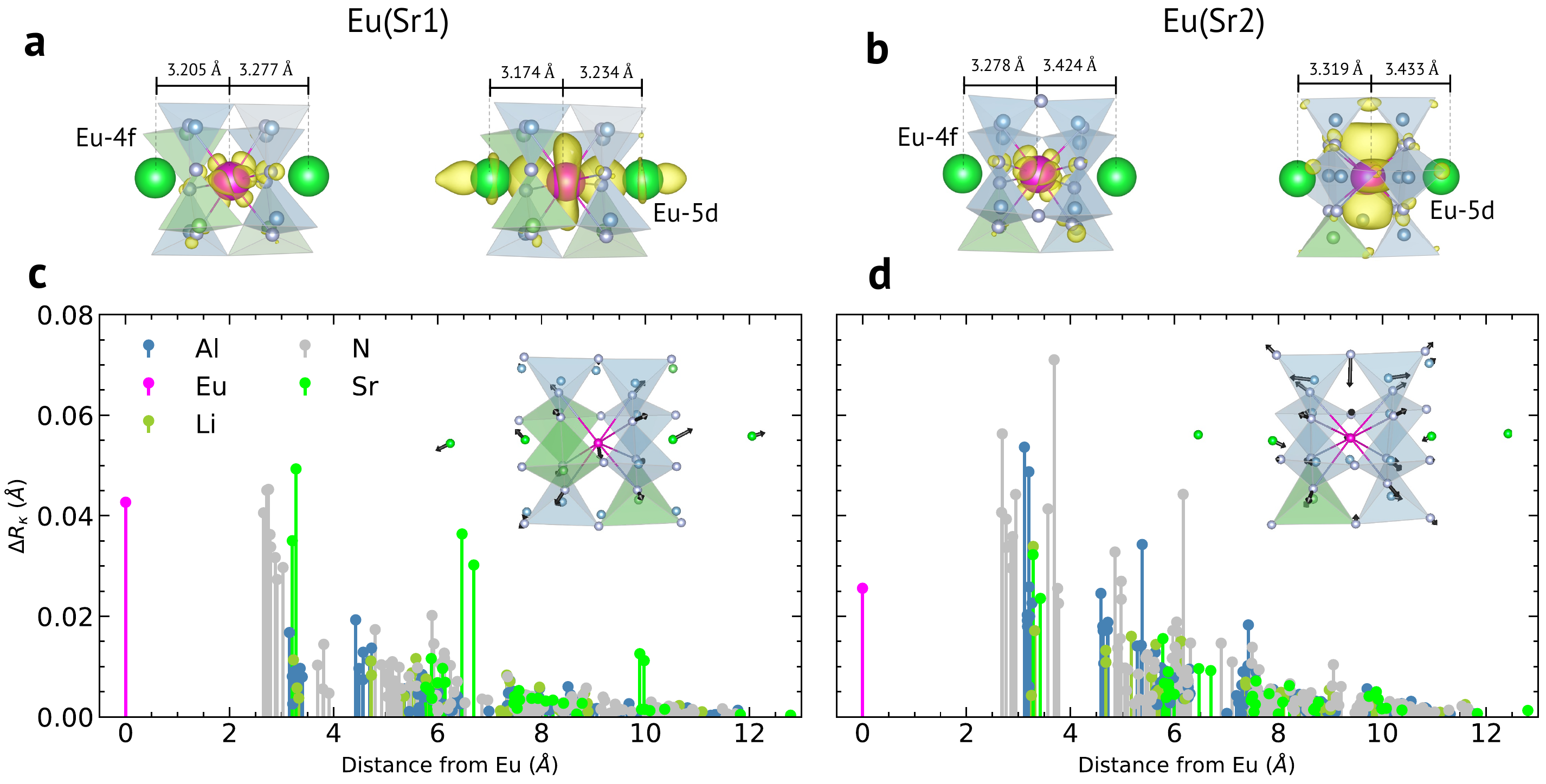}
	\caption{\label{fig:4f-5d_relaxation} The probability density of the highest occupied Kohn-Sham orbitals associated to the 4f$^7$($^8$S$_{7/2}$) ground and lowest 4f$^6$($^7$F$_J$)5d$^1$ excited states, for \textbf{a} Eu(Sr1) site and \textbf{b} Eu(Sr2) site. 
	Norm of the atomic displacements induced by the 5d to 4f transition (upon emission) for \textbf{c} Eu(Sr1) site and \textbf{d} Eu(Sr2) site as a function of the distance from Eu. 
	The insets show a three-dimensional view of these displacements near the Eu activator, scaled by 20 for clarity. }
\end{figure*}

The probability density of the highest occupied Kohn-Sham orbitals associated to the 4f$^7$($^8$S$_{7/2}$) ground and lowest 4f$^6$($^7$F$_J$)5d$^1$ excited states are presented in Fig.~\ref{fig:4f-5d_relaxation}\textbf{a-b} for both Sr sites. 
The computed energy levels are reported in Section 6 of the SI~\cite{supp}.
In a hypothetical pure cubic EuN$_8$ environment, 5d$_{z^{2}}$ and 5d$_{x^2-y^2}$ would be the lowest degenerate d-states. 
However, this degeneracy is lifted when considering a non-perfect EuN$_8$ cuboid with different second shell environment ([LiN$_4$]$^{11-}$ and [AlN$_4$]$^{9-}$ tetrahedra) and different Eu-Sr distances along the 1D Sr chain.

In the case of Eu(Sr1), we observe that a 5d$_{z^2}$-like orbital is stabilized along the Sr chain as in the case of Sr[Li$_2$Al$_2$O$_2$N$_2$]:Eu$^{2+}$~\cite{bouquiaux2021importance}. 
The 5d$_{x^2-y^2}$-like state is also located in the band gap, but 0.27~eV above the 5d$_{z^2}$-like state. 

In contrast, for Eu(Sr2), the opposite situation appears: a 5d$_{x^2-y^2}$-like orbital is stabilized, perpendicular to the Sr chain. 
The four lobes point toward the interstitial area of low electrostatic potential produced by nitrogen ligands.  
The inversion of the 5d states ordering appears to be related to the difference in second-shell environments, given the similarity of the Eu-N$_8$ cuboids. 
However, it is unclear whether the inversion is caused by the difference in Eu-Sr distances (steric effect) or the difference in tetrahedra types (coulombic effect). 
To explore this further, we have relaxed a fictitious system in which the Eu atom replaces Sr1, while keeping the Eu-Sr distances fixed at the relaxed Eu(Sr2) values.
Our findings indicate that the 5d$_{z^2}$ state remains stabilized, but the 5d$_{x^2-y^2}$ state is now 0.15 eV above. 
This suggests that the difference in tetrahedra types is the primary explanation for the inversion of the 5d states, with some contribution from the difference in Eu-Sr distances.

We stress that in Ref.~\cite{shafei2022electronic}, it was argued that the similar local geometry and electronic structure between the two sites leads to a similar emission spectrum.
However, our computations contradict this claim, as we have found that the two sites possess different electronic structures that result in distinct emission spectra, as it will be demonstrated later on.
We also find that for both sites, the computed energy separation between the empty 5d state and the conduction bottom (Sr-4d character) is 0.29~eV. 
Experimentally, the 5d-conduction band separation is estimated to be 0.28~eV~\cite{tolhurst2016electronic} based on a fitting of thermal quenching data with an exponential non-radiative decay rate from the 5d state to the conduction bottom through thermal excitation. 
Despite the good agreement between these values, it is important to remember that the Kohn-Sham energy levels calculated using GGA-PBE are only approximations of the actual energy levels and this agreement should be appreciated with caution.

The displacement field associated to the 5d-4f transition is presented in the insets of Fig.~\ref{fig:4f-5d_relaxation}\textbf{c-d}, scaled by 20 for clarity. 
The norm of these displacements is shown as a function of the distance from the Eu activator.
For both Sr sites, upon emission, going from the 5d to 4f state leads to an elongation of Eu-N bond lengths by 0.03-0.06~\AA. 
Indeed, upon 5d excitation, additional covalent interactions appear between the N ligands with the inner 4f hole that shortens the bond length~\cite{joos2020insights}.
Interestingly, the Eu atom moves in its cuboid cage as a result of its non-symmetrical environment. 
Looking now at the atomic rearrangements beyond the ﬁrst coordination shell, we see first that a significant fraction of the total displacements is outside this first shell, as observed in similar UCr$_4$C$_4$ phosphors~\cite{bouquiaux2021importance,wang2022role}. 
This fact might be linked to the rigid structure (we compute a Debye temperature of 878.5~K) that distributes the local strain caused by the localized 5d-4f excitation to atoms far away from the Eu activator.
Quantitatively, 61\% of the total mass-weighted displacements $\Delta Q=\sqrt{\sum_{\kappa}M_{\kappa}\Delta R_{\kappa}^2}$ is contained in the Eu-N$_8$ cuboid for Eu(Sr1) and  50\% for Eu(Sr2).
Given the high Eu mass, Eu atom contributes to this $\Delta Q$ up to 49\% in Eu(Sr1) and to 30\% in Eu(Sr2).

By inspecting closely the difference between Eu(Sr1) and Eu(Sr2), we observe that a main relaxation pattern in Eu(Sr1) is a long-range displacement of the Sr chain, as already observed in Sr[Li$_2$Al$_2$O$_2$N$_2$]:Eu$^{2+}$, which is a consequence of the 5d$_{z^{2}}$ orientation along this chain. 
Given the high Sr mass, we expect that the Sr displacements are predominant in shaping the electron-phonon spectral function of Eu(Sr1), as will be see later.
For Eu(Sr2), a major fraction of the displacements are distributed on the adjacent [LiN$_4$]$^{11-}$ and [AlN$_4$]$^{9-}$ tetrahedra, which is a consequence of the 5d$_{x^2-y^2}$ lobes pointing towards it. 
We provide the structural parameters for pristine SLA, doped SLA Eu(Sr1) and Eu(Sr2) in their 4f and 5d states in Tables~1, 2 and 3 of the SI~\cite{supp}. 
Finally, the energy between the computed relaxed excited and ground states provides an estimation of the ZPL energies, $E^{\rm {ZPL-Sr1}}$=1.916~eV and $E^{\rm {ZPL-Sr2}}$= 1.989~eV, close to the experimental ZPL energies of 1.906~eV and 1.956~eV, respectively.
In order to explain this difference in ZPL energies, one might invoke the difference in [EuN$_{8}$] cuboid 
volumes between the two sites, as it was done in Ref.~\cite{fang2020cuboid} to explain the variation of emission energies across different oxy-nitrides and alkali-lithosilicates. 
However, the very small difference between the two cuboids volumes ($\approx$ 1\%) is not sufficient to
explain the difference in ZPL energies. 
We believe that the different degree of ionicity experienced by the 5d orbital due to the different number of [LiN$_4$]$^{11-}$ and [AlN$_4$]$^{9-}$ tetrahedra is responsible for the difference in ZPL energies. 
As a rough estimate, Eu(Sr1) with 3[LiN$_4$]$^{11-}$ and 5[AlN$_4$]$^{9-}$ has a second shell formal charge of -78 e while Eu(Sr2) with 1[LiN$_4$]$^{11-}$ and 7[AlN$_4$]$^{9-}$ has a second shell formal charge of -74 e, where e is the absolute value of the electron charge.
In SALON, this second shell formal charge is -64 e. This is in line with the experimental ZPL energies of 1.906 eV, 1.956 eV and 2.03 eV~\cite{ruegenberg2023mixed}, respectively.
 
\subsection{Huang-Rhys spectral function}
\label{subsec:Spectral}

\begin{figure}[h!]
	\centering
	\includegraphics[width=0.99\linewidth]{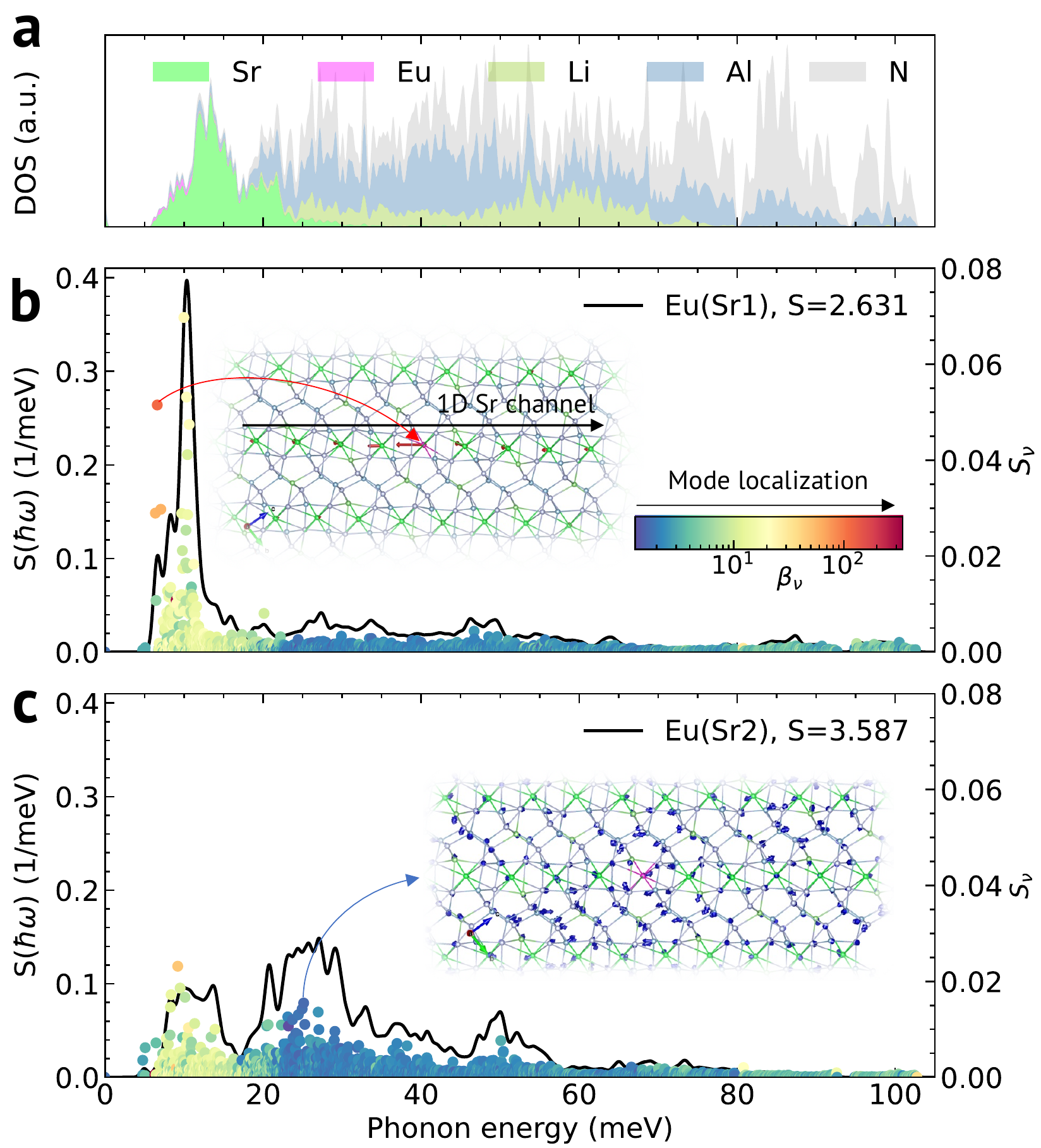}
	\caption{\label{fig:Spectral_function}Spectral decomposition of the Huang-Rhys function obtained from the $S_{\nu}$ (colored dots), $S(\hbar\omega)=\sum_{\nu} S_{\nu} \delta(\hbar\omega - \hbar\omega_{\nu})$ (black lines), using a 1~meV Gaussian broadening, for both sites. 
	In \textbf{a} we show the atom-projected phonon density-of-states. 
	The height of the dots in \textbf{b} -for Eu(Sr1)- and \textbf{c} -for Eu(Sr2)- provides the partial Huang-Rhys factor of each phonon mode, while the color indicates the degree of localization of each mode $\beta_{\nu}$, see text.  
	For each Sr site, two illustrative modes with high Huang-Rhys factor but different degree of localization are shown. }	
\end{figure}

The Huang-Rhys spectral decomposition is presented in Fig.~\ref{fig:Spectral_function}. 
Eq.~\eqref{eq:S_nu} was used with a supercell containing 2304 atoms (4$\times$4$\times$4 supercell) thanks to the embedding procedure, described in the theoretical section, which allows us to reach the dilute limit while keeping local and quasi-local modes brought by the Eu defect. 
The localization of the phonon modes can be characterized by computing the inverse participation ratio (IPR) defined as~\cite{alkauskas2014,jin2021photoluminescence,wang2022role}
\begin{equation}
	\label{IPR}
	\rm{IPR_{\nu}}=\frac{1}{\sum_{\kappa}|\langle \mathbf{e}_{\nu,\kappa}| \mathbf{e}_{\nu,\kappa} \rangle|^2},
\end{equation}
which indicates, roughly speaking, the number of atoms that participate to a phonon mode $\nu$. 
For example, $\rm{IPR_{\nu}}=1$ means that only one atom vibrates and therefore the phonon mode is maximally localized, while $\rm{IPR_{\nu}}=N$ means that N atoms vibrates in the supercell with the same amplitude. 
The localization ratio $\beta_{\nu}$ is defined by taking the ratio of the total number of atoms in the supercell $N$ and the $\rm{IPR}$,   
$\beta_{\nu} =N/\rm{IPR_{\nu}}$ where $\beta_{\nu} \approx 1$ represents a bulk-like delocalized mode while $\beta_{\nu} \gg 1$ corresponds to a quasi-local or local mode.
We show in Fig.~\ref{fig:Spectral_function}\textbf{b}-\textbf{c} the color-coded mode localization $\beta_{\nu}$, which combined with the atom-projected phonon density of states in Fig.~\ref{fig:Spectral_function}\textbf{a}, shows that the low-energy modes are more localized and dominate the electron-phonon coupling associated with the 5d-4f transition in the case of Eu(Sr1).

For Eu(Sr1), the spectral function is dominated by a large peak around 10~meV where low-frequency phonon modes associated to Sr and Eu displacements are involved. 
They mostly corresponds to long-wavelength collective displacements of the Sr atoms along the Sr chain that couple strongly with the above-described 5d-4f relaxation of the Sr channel containing Eu. 
We compute a total Huang-Rhys factor $S=\sum_{\nu}S_{\nu}$ of 2.631 in that case and find that the phonons with high partial HR factor $S_{\nu}$ can be either relatively delocalized or very localized. 
For instance, the highest coupling mode at 9.97~meV with $S_{\nu}$=0.07 has a $\beta_{\nu}$ of 23. 
This mode is associated to the collective displacements of all the Sr atoms in the chain and can be considered as a slightly perturbed bulk mode. 
In contrast, the third highest coupling mode at 6.6~meV is very localized, with a $S_{\nu}$=0.05 and a large $\beta_{\nu}$ of 115. 
As illustrated in the inset of Fig.~\ref{fig:Spectral_function}\textbf{b}, the atomic vibrations associated with this mode (red arrows) are localized around the Eu doping atom.

In the case of the second site Eu(Sr2), the spectral function indicates that phonon modes involving Eu and Sr atoms participate but to a lesser extent than in the Eu(Sr1) case. 
Additionally, a broad peak appears between 20~meV and 40~meV which is associated to modes involving Al and N atoms. 
The total Huang-Rhys factor of 3.587 indicates a stronger electron-phonon coupling. 
The highest coupling mode located at 9.2~meV with $S_{\nu}$=0.023 has a $\beta_{\nu}$ of 59. 
This mode is mostly associated with the Eu atom moving in its cage.  
Similarly to the Eu(Sr1) case, there is also a number of delocalized bulk modes that contribute.
Indicatively, the fifth highest coupling mode at 25.1~meV, with a $S_{\nu}$=0.015 has a $\beta_{\nu}$ of 1.7 and is illustrated in the inset of Fig.~\ref{fig:Spectral_function}\textbf{c}. The atomic vibrations associated to this mode (blue arrows) are distributed mainly on Al and N atoms in the whole system.
We analyze in detail the five phonon modes which contribute the most to the spectral function in Section~7 of the SI~\cite{supp}.

\subsection{Photoluminescence spectrum and its temperature dependence}
\label{subsec:Lum}
\begin{figure}[h!]
	\centering
	\includegraphics[width=0.99\linewidth]{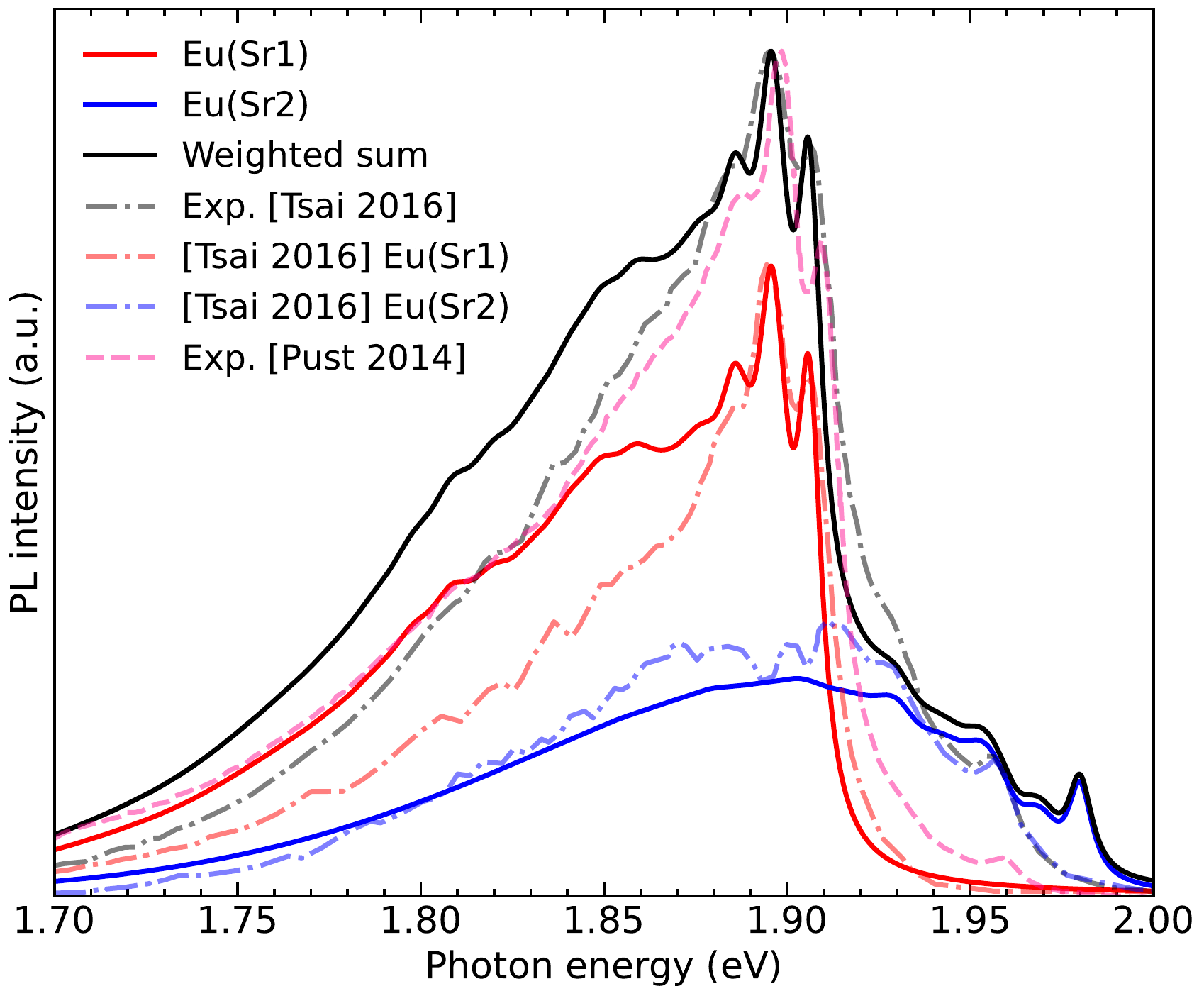}
	\caption{Computed photoluminescence intensities $L(\hbar\omega$) for both Sr sites at 0~K and their weighted sum compared with two low-temperature experiments~\cite{pust2014narrow,tsai2016improvement}. The red and blue dotted curves are experimental PL intensities of both sites from Ref.~\cite{tsai2016improvement}, where the total PL intensity (gray dotted curve) was deconvoluted with time-resolved spectroscopy. 
	The computed Eu(Sr1) zero-phonon line energy is aligned with the second highest peak of the experimental data from Ref.~\cite{tsai2016improvement} and the two weights for the black line are also taken from Ref.~\cite{tsai2016improvement}. Reproduced with permission from Ref.~\cite{pust2014narrow,tsai2016improvement}.  Copyright 2014 Springer Nature and Copyright 2016 John Wiley and Sons.}
	\label{fig:Low_T_PL}
\end{figure}

Using the generating function approach described in the theoretical section, we compute the photoluminescence (PL) intensity from the Huang-Rhys spectral function. 
Fig.~\ref{fig:Low_T_PL} compares the PL spectra from both sites at zero temperature with the experimental PL intensities at 10~K from Ref.~\cite{tsai2016improvement}, where the total PL intensity was deconvoluted with time-resolved spectroscopy (red and blue dotted curves). 
The total spectrum is also in good agreement with prior measurements at 6~K~\cite{pust2014narrow}.

\begin{figure*}
	\centering
	\includegraphics[width=0.99\linewidth]{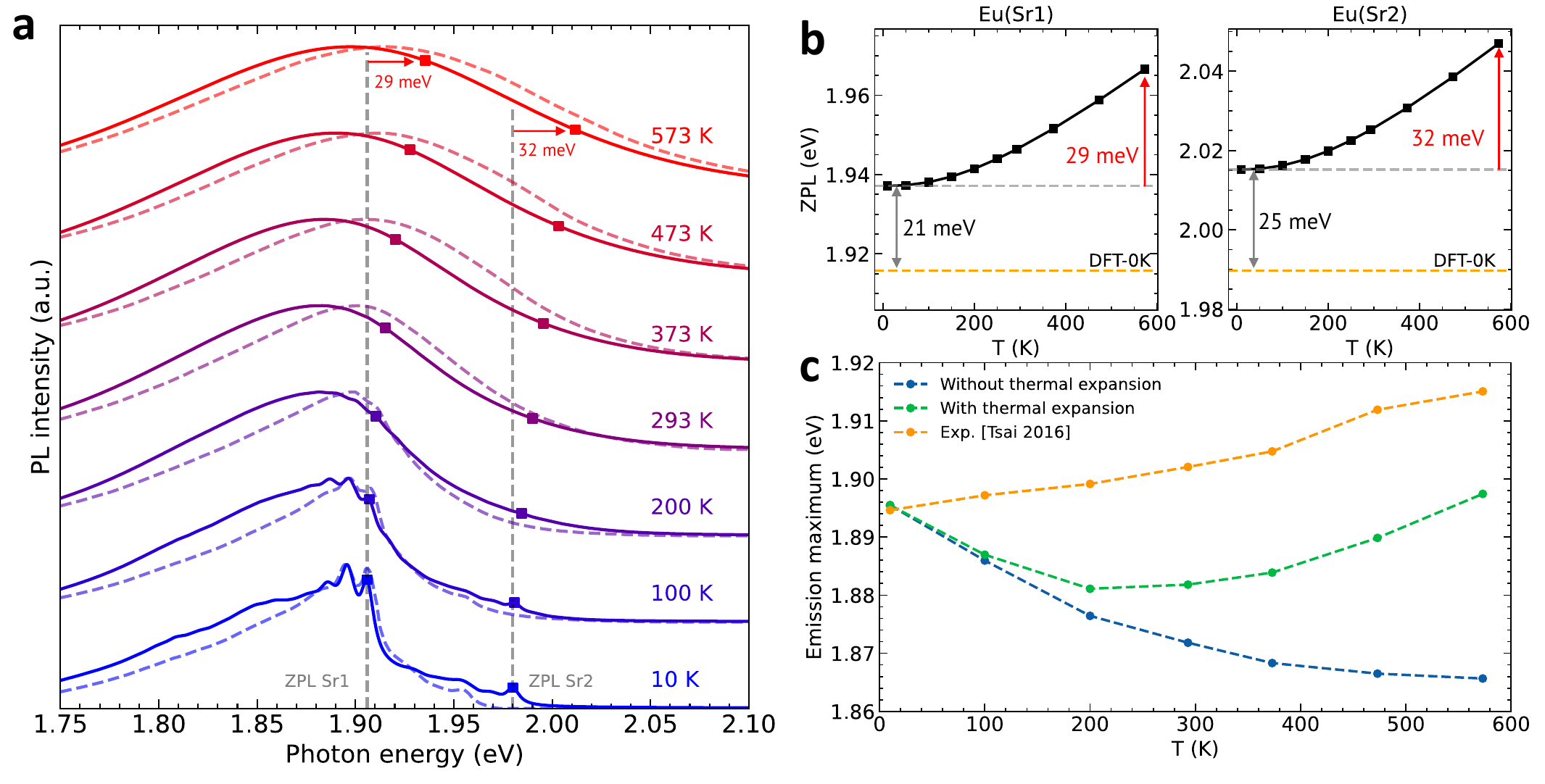}
	\caption{\textbf{a} Temperature-dependent photoluminescence (PL) spectrum computed with Eq.~\eqref{eq:L(hw)_generating} (plain curves) and compared with experimental data~\cite{tsai2016improvement} (dashed curves). 
	The zero-phonon line (ZPL) energies at 0~K are indicated by gray vertical dashed lines and the ZPL including thermal expansion effects are indicated with squares. 
The PL spectra for each temperature is shifted vertically for clarity.	
	\textbf{b} Effect of thermal expansion on the ZPL energies for both sites where the DFT values at 0~K are shown with horizontal orange dashed lines. 
The shift associated with the zero-point volume are 21 and 25~meV, respectively. 
Between 0~K and 573~K, thermal expansion leads to a blue shift of 29~meV and 32~meV, respectively.
\textbf{c} Photon energy of the emission maximum as a function of temperature. 
Ignoring thermal expansion leads to a red shift of the emission maximum. 
Adding the thermal expansion effect allows to obtain a better agreement with the experiment.}
	\label{fig:PL_T}
\end{figure*}

For definitiveness, we have (i) aligned our computed ZPL energy of Eu(Sr1) with the highest experimental peak~\cite{tsai2016improvement}, (ii) used a constant Lorentzian broadening of 25~meV, and (iii) used the experimental Eu(Sr1) to Eu(Sr2) total intensity ratio (computed as the ratio between the areas under the curves) from Ref.~\cite{tsai2016improvement}. We do not attempt in this work to compute 
these weights from first principles, which would require to 
compare correctly the energetics and the electric dipole matrix elements of both sites. 
Indeed, the weight of the second PL intensity (ZPL at 1.956~eV) is different between the two experiments, which could be explained by different synthesis conditions.
We note also that the energy difference between the experimental ZPL energies is 50~meV while the computed energy difference is 74~meV. 
This explains the misalignment observed in Fig.~\ref{fig:Low_T_PL} between the experimental Eu(Sr2) ZPL and the computed one.

The vibronic peaks appearing in the experimental spectra as well as the global shapes of both spectra match our computations which means that the sharper PL spectrum with ZPL at 1.906~eV can be assigned with confidence to the Sr1 site while the broader spectrum with ZPL at 1.956~eV is assigned to the Sr2 site.
When looking at the total experimental spectrum, the high-energy peak at 1.956~eV can be assigned to the ZPL of Eu(Sr2) center, the peak at 1.906~eV to the ZPL of Eu(Sr1) center, and the two next peaks to the one- and two-phonon contributions of strontium based phonon modes in Eu(Sr1). 
We still note that our computed total Huang-Rhys factor for Eu(Sr1)  S$_1$ seems slightly overestimated. In order to estimate the experimental S$_1$, the 5d-4f forces can be scaled in order to fit the experimental spectrum. Scaling the forces by 90\% (giving S$_1^{\rm{exp}}$ $\approx$ 2.13) allows one to obtain an excellent reproduction of the experimental phonon side-band (see figure S11).

Finally, all the elements can be placed together to compute the temperature-dependent PL spectrum of SLA using Eq.~\eqref{eq:L(hw)_generating} which we present in Fig.~\ref{fig:PL_T}\textbf{a}.
This achievement represents the main result of our work, with excellent experimental agreement.  
The effect of thermal expansion on the ZPL energy is quite similar for both sites and leads to a blue-shift of 29 and 32 meV from 0 to 573K , see Fig.~\ref{fig:PL_T}\textbf{b}. 
The ZPL shifts associated to the zero-point volume are estimated to be 21 and 25 meV.
With increasing temperature, the spectrum broadens, with a progressive loss of clear vibronic peaks. 

Fig.~\ref{fig:PL_T}\textbf{c} presents the position of the emission maximum as a function of temperature, both with and without accounting for thermal expansion. 
When thermal expansion is not considered, a red-shift between 10~K and 573~K is computed. This is attributed to the positioning of the vibronic transitions that are activated by temperature relative to the emission maximum.
Above 200K, where the position of the emission maximum is determined by the envelope of the vibronic transitions, accounting for thermal expansion yields the temperature dependence observed experimentally.
We notice that the $\omega^3$ dependence of $L(\hbar\omega) \propto \omega^3A(\hbar\omega)$ causes a blue shift of the emission maximum due to the broadening of the lineshape function $A(\hbar\omega)$ with temperature, even if $A(\hbar\omega)$ does not shift. This leads to a blue shift of +11 meV between 0K and 573K (see Section 8.2 of the SI).
We provide the site decomposition of this emission maximum shift and further details in the SI~\cite{supp}. 
Our analysis sheds new light on the puzzling phenomenon of temperature-induced shift in phosphor materials, which remains theoretically underexplored~\cite{yan2018origin,xu2022microscopic}. %
It highlights the importance of both the $\omega^3$ dependence of the PL intensity and the effect of thermal expansion.

\section{Conclusions}
\label{sec:Conclusion}

In this work, the vibronic processes occurring in the narrow-band emission SrLiAl$_3$N$_4$:Eu$^{2+}$ red phosphor is characterized from first principles. 
Two Sr sites can host the Eu activator, leading to two emission centers. 
The first, denoted as Sr1, is surrounded by 3 [LiN$_4$]$^{11-}$ and 5 [AlN$_4$]$^{9-}$ tetrahedra, while the second, denoted as Sr2, is surrounded by 1 [LiN$_4$]$^{11-}$ and 7 [AlN$_4$]$^{9-}$ tetrahedra.
We hence apply the constrained-$\Delta$SCF method to optimize Eu-5d excited state structures for both sites, independently. 
Different 5d orbital are stabilized: a 5d$_{z^2}$-like state aligned along the Sr channel for Eu(Sr1), and a 5d$_{x^2-y^2}$-like state pointing along Al/Li atoms for Eu(Sr2).

The 5d-4f atomic relaxation is dominated by a long-range displacement of the Sr chain for Eu(Sr1), as a consequence of the  5d$_{z^2}$ orientation. 
By projecting the atomic relaxation onto the phonon modes of a large doped supercell, we identify the phonon that couples the most with the 5d-4f transition. 
For Eu(Sr1),  low-frequency phonon modes associated to Sr and Eu displacements are involved. 
They correspond either to bulk-like collective displacements of the Sr chains or to very localized modes concentrated on the Sr chain around the Eu activator.  
We computed a total Huang-Rhys factor of $S = 2.631$ resulting in a small coupling with phonons of low frequency which yields a small bandwidth. 
The same situation was found in Sr[Li$_2$Al$_2$O$_2$N$_2$]:Eu$^{2+}$ suggesting that the small bandwidth of other UCr$_4$C$_4$-type phosphors with Ca/Sr/Ba channel can be explained in a similar way.  
For Eu(Sr2), as a consequence of a different 5d orbital orientation, modes associated to Sr and Eu displacements are less involved and delocalized bulk modes involving Al and N atoms contribute as well, leading to a higher Huang-Rhys factor $S = 3.587$. 
This larger coupling with phonons of higher frequencies yields a broader lineshape.
By comparing our computations with experimental low-temperature photo-luminescent spectra, we are able to assign the peak at 1.906~eV to the ZPL of Eu(Sr1) center, the two next peaks at 1.89 and 1.88~eV being the one and two-phonon contributions of strontium-based phonon modes, respectively. 
The peak at 1.956~eV is assigned to the ZPL of Eu(Sr2) center. 
Finally, we show the importance of thermal expansion effect on the ZPL energies which induces a blue-shift of around 30~meV for both sites between 10~K and 573~K. 
This results in a good agreement between our simulated temperature dependent PL spectra and experimental data, for both the broadening and the shift of the spectrum with temperature. 
Overall, this work offers a direct theoretical understanding of the observed spectrum of SrLiAl$_3$N$_4$:Eu$^{2+}$ which highlights the importance of the Sr channels in shaping the small bandwidth.
These findings are general and should apply to any UCr$_4$C$_4$-type phosphors.

\medskip
\textbf{Supporting informations} \par 
Details on the embedding approach, influence of the Hubbard U, phonon band structure of pristine SLA, additional details on thermal expansion, additional details on structural parameters, Kohn-Sham levels, dominant phonon modes, additional details on the emission shift with temperature.

\medskip
\textbf{Acknowledgments} \par 
Computational resources have been provided by the supercomputing facilities of the Université catholique de Louvain (CISM/UCL) and the Consortium des Équipements de Calcul Intensif en Fédération Wallonie Bruxelles (CÉCI) funded by the Fond de la Recherche Scientifique de Belgique (F.R.S.-FNRS) under convention 2.5020.11 and by the Walloon Region, as well as the Tier-1 supercomputer of the F\'ed\'eration Wallonie-Bruxelles, infrastructure funded by the Walloon Region under the grant agreement No. 1117545. 
We thank the Consortium des Équipements de Calcul Intensif, Belgium for awarding to this project an access to the LUMI supercomputer, owned by the EuroHPC Joint Undertaking, hosted by CSC (Finland) and the LUMI consortium through the project 465000061.
J.B. and S.P. acknowledge support from the F.R.S.-FNRS.
X.G. also acknowledges support from the F.R.S.-FNRS, under grant n°T.0103.19 (PDR - ALPS). 
This work was supported by the Communauté française de Belgique through the SURFASCOPE project (ARC 19/24-102. 
Y.J acknowledges the funding support from National Key R\&D Program (No.2022YFB3503800), Natural Science Foundation of Hebei Province (No.E2021203126) and Cultivation Project for Basic Research and Innovation of Yanshan University (No.2021LGQN033).
\medskip

\bibliography{main} 

\begin{thebibliography}{54}%
\makeatletter
\providecommand \@ifxundefined [1]{%
 \@ifx{#1\undefined}
}%
\providecommand \@ifnum [1]{%
 \ifnum #1\expandafter \@firstoftwo
 \else \expandafter \@secondoftwo
 \fi
}%
\providecommand \@ifx [1]{%
 \ifx #1\expandafter \@firstoftwo
 \else \expandafter \@secondoftwo
 \fi
}%
\providecommand \natexlab [1]{#1}%
\providecommand \enquote  [1]{``#1''}%
\providecommand \bibnamefont  [1]{#1}%
\providecommand \bibfnamefont [1]{#1}%
\providecommand \citenamefont [1]{#1}%
\providecommand \href@noop [0]{\@secondoftwo}%
\providecommand \href [0]{\begingroup \@sanitize@url \@href}%
\providecommand \@href[1]{\@@startlink{#1}\@@href}%
\providecommand \@@href[1]{\endgroup#1\@@endlink}%
\providecommand \@sanitize@url [0]{\catcode `\\12\catcode `\$12\catcode
  `\&12\catcode `\#12\catcode `\^12\catcode `\_12\catcode `\%12\relax}%
\providecommand \@@startlink[1]{}%
\providecommand \@@endlink[0]{}%
\providecommand \url  [0]{\begingroup\@sanitize@url \@url }%
\providecommand \@url [1]{\endgroup\@href {#1}{\urlprefix }}%
\providecommand \urlprefix  [0]{URL }%
\providecommand \Eprint [0]{\href }%
\providecommand \doibase [0]{https://doi.org/}%
\providecommand \selectlanguage [0]{\@gobble}%
\providecommand \bibinfo  [0]{\@secondoftwo}%
\providecommand \bibfield  [0]{\@secondoftwo}%
\providecommand \translation [1]{[#1]}%
\providecommand \BibitemOpen [0]{}%
\providecommand \bibitemStop [0]{}%
\providecommand \bibitemNoStop [0]{.\EOS\space}%
\providecommand \EOS [0]{\spacefactor3000\relax}%
\providecommand \BibitemShut  [1]{\csname bibitem#1\endcsname}%
\let\auto@bib@innerbib\@empty
\bibitem [{\citenamefont {Lin}\ \emph {et~al.}(2016)\citenamefont {Lin},
  \citenamefont {Meijerink},\ and\ \citenamefont {Liu}}]{lin2016critical}%
  \BibitemOpen
  \bibfield  {author} {\bibinfo {author} {\bibfnamefont {C.~C.}\ \bibnamefont
  {Lin}}, \bibinfo {author} {\bibfnamefont {A.}~\bibnamefont {Meijerink}},\
  and\ \bibinfo {author} {\bibfnamefont {R.-S.}\ \bibnamefont {Liu}},\
  }\href@noop {} {\bibfield  {journal} {\bibinfo  {journal} {The journal of
  physical chemistry letters}\ }\textbf {\bibinfo {volume} {7}},\ \bibinfo
  {pages} {495} (\bibinfo {year} {2016})}\BibitemShut {NoStop}%
\bibitem [{\citenamefont {Pust}\ \emph {et~al.}(2015)\citenamefont {Pust},
  \citenamefont {Schmidt},\ and\ \citenamefont {Schnick}}]{pust2015revolution}%
  \BibitemOpen
  \bibfield  {author} {\bibinfo {author} {\bibfnamefont {P.}~\bibnamefont
  {Pust}}, \bibinfo {author} {\bibfnamefont {P.~J.}\ \bibnamefont {Schmidt}},\
  and\ \bibinfo {author} {\bibfnamefont {W.}~\bibnamefont {Schnick}},\
  }\href@noop {} {\bibfield  {journal} {\bibinfo  {journal} {Nature materials}\
  }\textbf {\bibinfo {volume} {14}},\ \bibinfo {pages} {454} (\bibinfo {year}
  {2015})}\BibitemShut {NoStop}%
\bibitem [{\citenamefont {Lee}(2022)}]{lee20222022}%
  \BibitemOpen
  \bibfield  {author} {\bibinfo {author} {\bibfnamefont {K.}~\bibnamefont
  {Lee}},\ }\href@noop {} {\emph {\bibinfo {title} {2022 Solid-State Lighting
  R\&D Opportunities}}},\ \bibinfo {type} {Tech. Rep.}\ (\bibinfo
  {institution} {Guidehouse},\ \bibinfo {year} {2022})\BibitemShut {NoStop}%
\bibitem [{\citenamefont {Fang}\ \emph {et~al.}(2022)\citenamefont {Fang},
  \citenamefont {Bao}, \citenamefont {Huang},\ and\ \citenamefont
  {Liu}}]{fang2022evolutionary}%
  \BibitemOpen
  \bibfield  {author} {\bibinfo {author} {\bibfnamefont {M.-H.}\ \bibnamefont
  {Fang}}, \bibinfo {author} {\bibfnamefont {Z.}~\bibnamefont {Bao}}, \bibinfo
  {author} {\bibfnamefont {W.-T.}\ \bibnamefont {Huang}},\ and\ \bibinfo
  {author} {\bibfnamefont {R.-S.}\ \bibnamefont {Liu}},\ }\href@noop {}
  {\bibfield  {journal} {\bibinfo  {journal} {Chemical Reviews}\ }\textbf
  {\bibinfo {volume} {122}},\ \bibinfo {pages} {11474} (\bibinfo {year}
  {2022})}\BibitemShut {NoStop}%
\bibitem [{\citenamefont {Fang}\ \emph {et~al.}(2020)\citenamefont {Fang},
  \citenamefont {Mariano}, \citenamefont {Chen}, \citenamefont {Hu},\ and\
  \citenamefont {Liu}}]{fang2020cuboid}%
  \BibitemOpen
  \bibfield  {author} {\bibinfo {author} {\bibfnamefont {M.-H.}\ \bibnamefont
  {Fang}}, \bibinfo {author} {\bibfnamefont {C.~O.~M.}\ \bibnamefont
  {Mariano}}, \bibinfo {author} {\bibfnamefont {P.-Y.}\ \bibnamefont {Chen}},
  \bibinfo {author} {\bibfnamefont {S.-F.}\ \bibnamefont {Hu}},\ and\ \bibinfo
  {author} {\bibfnamefont {R.-S.}\ \bibnamefont {Liu}},\ }\href@noop {}
  {\bibfield  {journal} {\bibinfo  {journal} {Chemistry of Materials}\ }\textbf
  {\bibinfo {volume} {32}},\ \bibinfo {pages} {1748} (\bibinfo {year}
  {2020})}\BibitemShut {NoStop}%
\bibitem [{\citenamefont {Fang}\ \emph {et~al.}(2018)\citenamefont {Fang},
  \citenamefont {Lea{\~n}o~Jr},\ and\ \citenamefont {Liu}}]{fang2018control}%
  \BibitemOpen
  \bibfield  {author} {\bibinfo {author} {\bibfnamefont {M.-H.}\ \bibnamefont
  {Fang}}, \bibinfo {author} {\bibfnamefont {J.~L.}\ \bibnamefont
  {Lea{\~n}o~Jr}},\ and\ \bibinfo {author} {\bibfnamefont {R.-S.}\ \bibnamefont
  {Liu}},\ }\href@noop {} {\bibfield  {journal} {\bibinfo  {journal} {ACS
  Energy Letters}\ }\textbf {\bibinfo {volume} {3}},\ \bibinfo {pages} {2573}
  (\bibinfo {year} {2018})}\BibitemShut {NoStop}%
\bibitem [{\citenamefont {Pust}\ \emph
  {et~al.}(2014{\natexlab{a}})\citenamefont {Pust}, \citenamefont {Weiler},
  \citenamefont {Hecht}, \citenamefont {T{\"u}cks}, \citenamefont {Wochnik},
  \citenamefont {Hen{\ss}}, \citenamefont {Wiechert}, \citenamefont {Scheu},
  \citenamefont {Schmidt},\ and\ \citenamefont {Schnick}}]{pust2014narrow}%
  \BibitemOpen
  \bibfield  {author} {\bibinfo {author} {\bibfnamefont {P.}~\bibnamefont
  {Pust}}, \bibinfo {author} {\bibfnamefont {V.}~\bibnamefont {Weiler}},
  \bibinfo {author} {\bibfnamefont {C.}~\bibnamefont {Hecht}}, \bibinfo
  {author} {\bibfnamefont {A.}~\bibnamefont {T{\"u}cks}}, \bibinfo {author}
  {\bibfnamefont {A.~S.}\ \bibnamefont {Wochnik}}, \bibinfo {author}
  {\bibfnamefont {A.-K.}\ \bibnamefont {Hen{\ss}}}, \bibinfo {author}
  {\bibfnamefont {D.}~\bibnamefont {Wiechert}}, \bibinfo {author}
  {\bibfnamefont {C.}~\bibnamefont {Scheu}}, \bibinfo {author} {\bibfnamefont
  {P.~J.}\ \bibnamefont {Schmidt}},\ and\ \bibinfo {author} {\bibfnamefont
  {W.}~\bibnamefont {Schnick}},\ }\href@noop {} {\bibfield  {journal} {\bibinfo
   {journal} {Nature materials}\ }\textbf {\bibinfo {volume} {13}},\ \bibinfo
  {pages} {891} (\bibinfo {year} {2014}{\natexlab{a}})}\BibitemShut {NoStop}%
\bibitem [{\citenamefont {Pust}\ \emph
  {et~al.}(2014{\natexlab{b}})\citenamefont {Pust}, \citenamefont {Hintze},
  \citenamefont {Hecht}, \citenamefont {Weiler}, \citenamefont {Locher},
  \citenamefont {Zitnanska}, \citenamefont {Harm}, \citenamefont {Wiechert},
  \citenamefont {Schmidt},\ and\ \citenamefont {Schnick}}]{pust2014group}%
  \BibitemOpen
  \bibfield  {author} {\bibinfo {author} {\bibfnamefont {P.}~\bibnamefont
  {Pust}}, \bibinfo {author} {\bibfnamefont {F.}~\bibnamefont {Hintze}},
  \bibinfo {author} {\bibfnamefont {C.}~\bibnamefont {Hecht}}, \bibinfo
  {author} {\bibfnamefont {V.}~\bibnamefont {Weiler}}, \bibinfo {author}
  {\bibfnamefont {A.}~\bibnamefont {Locher}}, \bibinfo {author} {\bibfnamefont
  {D.}~\bibnamefont {Zitnanska}}, \bibinfo {author} {\bibfnamefont
  {S.}~\bibnamefont {Harm}}, \bibinfo {author} {\bibfnamefont {D.}~\bibnamefont
  {Wiechert}}, \bibinfo {author} {\bibfnamefont {P.~J.}\ \bibnamefont
  {Schmidt}},\ and\ \bibinfo {author} {\bibfnamefont {W.}~\bibnamefont
  {Schnick}},\ }\href@noop {} {\bibfield  {journal} {\bibinfo  {journal}
  {Chemistry of Materials}\ }\textbf {\bibinfo {volume} {26}},\ \bibinfo
  {pages} {6113} (\bibinfo {year} {2014}{\natexlab{b}})}\BibitemShut {NoStop}%
\bibitem [{\citenamefont {Schmiechen}\ \emph {et~al.}(2014)\citenamefont
  {Schmiechen}, \citenamefont {Schneider}, \citenamefont {Wagatha},
  \citenamefont {Hecht}, \citenamefont {Schmidt},\ and\ \citenamefont
  {Schnick}}]{schmiechen2014toward}%
  \BibitemOpen
  \bibfield  {author} {\bibinfo {author} {\bibfnamefont {S.}~\bibnamefont
  {Schmiechen}}, \bibinfo {author} {\bibfnamefont {H.}~\bibnamefont
  {Schneider}}, \bibinfo {author} {\bibfnamefont {P.}~\bibnamefont {Wagatha}},
  \bibinfo {author} {\bibfnamefont {C.}~\bibnamefont {Hecht}}, \bibinfo
  {author} {\bibfnamefont {P.~J.}\ \bibnamefont {Schmidt}},\ and\ \bibinfo
  {author} {\bibfnamefont {W.}~\bibnamefont {Schnick}},\ }\href@noop {}
  {\bibfield  {journal} {\bibinfo  {journal} {Chemistry of Materials}\ }\textbf
  {\bibinfo {volume} {26}},\ \bibinfo {pages} {2712} (\bibinfo {year}
  {2014})}\BibitemShut {NoStop}%
\bibitem [{\citenamefont {Hoerder}\ \emph {et~al.}(2019)\citenamefont
  {Hoerder}, \citenamefont {Seibald}, \citenamefont {Baumann}, \citenamefont
  {Schr{\"o}der}, \citenamefont {Peschke}, \citenamefont {Schmid},
  \citenamefont {Tyborski}, \citenamefont {Pust}, \citenamefont {Stoll},
  \citenamefont {Bergler} \emph {et~al.}}]{hoerder2019sr}%
  \BibitemOpen
  \bibfield  {author} {\bibinfo {author} {\bibfnamefont {G.~J.}\ \bibnamefont
  {Hoerder}}, \bibinfo {author} {\bibfnamefont {M.}~\bibnamefont {Seibald}},
  \bibinfo {author} {\bibfnamefont {D.}~\bibnamefont {Baumann}}, \bibinfo
  {author} {\bibfnamefont {T.}~\bibnamefont {Schr{\"o}der}}, \bibinfo {author}
  {\bibfnamefont {S.}~\bibnamefont {Peschke}}, \bibinfo {author} {\bibfnamefont
  {P.~C.}\ \bibnamefont {Schmid}}, \bibinfo {author} {\bibfnamefont
  {T.}~\bibnamefont {Tyborski}}, \bibinfo {author} {\bibfnamefont
  {P.}~\bibnamefont {Pust}}, \bibinfo {author} {\bibfnamefont {I.}~\bibnamefont
  {Stoll}}, \bibinfo {author} {\bibfnamefont {M.}~\bibnamefont {Bergler}},
  \emph {et~al.},\ }\href@noop {} {\bibfield  {journal} {\bibinfo  {journal}
  {Nature communications}\ }\textbf {\bibinfo {volume} {10}},\ \bibinfo {pages}
  {1} (\bibinfo {year} {2019})}\BibitemShut {NoStop}%
\bibitem [{\citenamefont {Zhao}\ \emph {et~al.}(2018)\citenamefont {Zhao},
  \citenamefont {Liao}, \citenamefont {Ning}, \citenamefont {Zhang},
  \citenamefont {Liu},\ and\ \citenamefont {Xia}}]{zhao2018next}%
  \BibitemOpen
  \bibfield  {author} {\bibinfo {author} {\bibfnamefont {M.}~\bibnamefont
  {Zhao}}, \bibinfo {author} {\bibfnamefont {H.}~\bibnamefont {Liao}}, \bibinfo
  {author} {\bibfnamefont {L.}~\bibnamefont {Ning}}, \bibinfo {author}
  {\bibfnamefont {Q.}~\bibnamefont {Zhang}}, \bibinfo {author} {\bibfnamefont
  {Q.}~\bibnamefont {Liu}},\ and\ \bibinfo {author} {\bibfnamefont
  {Z.}~\bibnamefont {Xia}},\ }\href@noop {} {\bibfield  {journal} {\bibinfo
  {journal} {Advanced Materials}\ }\textbf {\bibinfo {volume} {30}},\ \bibinfo
  {pages} {1802489} (\bibinfo {year} {2018})}\BibitemShut {NoStop}%
\bibitem [{\citenamefont {Liao}\ \emph {et~al.}(2018)\citenamefont {Liao},
  \citenamefont {Zhao}, \citenamefont {Molokeev}, \citenamefont {Liu},\ and\
  \citenamefont {Xia}}]{liao2018learning}%
  \BibitemOpen
  \bibfield  {author} {\bibinfo {author} {\bibfnamefont {H.}~\bibnamefont
  {Liao}}, \bibinfo {author} {\bibfnamefont {M.}~\bibnamefont {Zhao}}, \bibinfo
  {author} {\bibfnamefont {M.~S.}\ \bibnamefont {Molokeev}}, \bibinfo {author}
  {\bibfnamefont {Q.}~\bibnamefont {Liu}},\ and\ \bibinfo {author}
  {\bibfnamefont {Z.}~\bibnamefont {Xia}},\ }\href@noop {} {\bibfield
  {journal} {\bibinfo  {journal} {Angewandte Chemie}\ }\textbf {\bibinfo
  {volume} {130}},\ \bibinfo {pages} {11902} (\bibinfo {year}
  {2018})}\BibitemShut {NoStop}%
\bibitem [{\citenamefont {Huang}\ and\ \citenamefont
  {Rhys}(1950)}]{huang1950theory}%
  \BibitemOpen
  \bibfield  {author} {\bibinfo {author} {\bibfnamefont {K.}~\bibnamefont
  {Huang}}\ and\ \bibinfo {author} {\bibfnamefont {A.}~\bibnamefont {Rhys}},\
  }\href@noop {} {\bibfield  {journal} {\bibinfo  {journal} {Proceedings of the
  Royal Society of London. Series A. Mathematical and Physical Sciences}\
  }\textbf {\bibinfo {volume} {204}},\ \bibinfo {pages} {406} (\bibinfo {year}
  {1950})}\BibitemShut {NoStop}%
\bibitem [{\citenamefont {Alkauskas}\ \emph {et~al.}(2014)\citenamefont
  {Alkauskas}, \citenamefont {Buckley}, \citenamefont {Awschalom},\ and\
  \citenamefont {Van~de Walle}}]{alkauskas2014}%
  \BibitemOpen
  \bibfield  {author} {\bibinfo {author} {\bibfnamefont {A.}~\bibnamefont
  {Alkauskas}}, \bibinfo {author} {\bibfnamefont {B.~B.}\ \bibnamefont
  {Buckley}}, \bibinfo {author} {\bibfnamefont {D.~D.}\ \bibnamefont
  {Awschalom}},\ and\ \bibinfo {author} {\bibfnamefont {C.~G.}\ \bibnamefont
  {Van~de Walle}},\ }\href@noop {} {\bibfield  {journal} {\bibinfo  {journal}
  {New J. Phys.}\ }\textbf {\bibinfo {volume} {16}},\ \bibinfo {pages} {073026}
  (\bibinfo {year} {2014})}\BibitemShut {NoStop}%
\bibitem [{\citenamefont {Jin}\ \emph {et~al.}(2021)\citenamefont {Jin},
  \citenamefont {Govoni}, \citenamefont {Wolfowicz}, \citenamefont {Sullivan},
  \citenamefont {Heremans}, \citenamefont {Awschalom},\ and\ \citenamefont
  {Galli}}]{jin2021photoluminescence}%
  \BibitemOpen
  \bibfield  {author} {\bibinfo {author} {\bibfnamefont {Y.}~\bibnamefont
  {Jin}}, \bibinfo {author} {\bibfnamefont {M.}~\bibnamefont {Govoni}},
  \bibinfo {author} {\bibfnamefont {G.}~\bibnamefont {Wolfowicz}}, \bibinfo
  {author} {\bibfnamefont {S.~E.}\ \bibnamefont {Sullivan}}, \bibinfo {author}
  {\bibfnamefont {F.~J.}\ \bibnamefont {Heremans}}, \bibinfo {author}
  {\bibfnamefont {D.~D.}\ \bibnamefont {Awschalom}},\ and\ \bibinfo {author}
  {\bibfnamefont {G.}~\bibnamefont {Galli}},\ }\href@noop {} {\bibfield
  {journal} {\bibinfo  {journal} {Physical Review Materials}\ }\textbf
  {\bibinfo {volume} {5}},\ \bibinfo {pages} {084603} (\bibinfo {year}
  {2021})}\BibitemShut {NoStop}%
\bibitem [{\citenamefont {Linder{\"a}lv}\ \emph {et~al.}(2021)\citenamefont
  {Linder{\"a}lv}, \citenamefont {Wieczorek},\ and\ \citenamefont
  {Erhart}}]{linderalv2021vibrational}%
  \BibitemOpen
  \bibfield  {author} {\bibinfo {author} {\bibfnamefont {C.}~\bibnamefont
  {Linder{\"a}lv}}, \bibinfo {author} {\bibfnamefont {W.}~\bibnamefont
  {Wieczorek}},\ and\ \bibinfo {author} {\bibfnamefont {P.}~\bibnamefont
  {Erhart}},\ }\href@noop {} {\bibfield  {journal} {\bibinfo  {journal}
  {Physical Review B}\ }\textbf {\bibinfo {volume} {103}},\ \bibinfo {pages}
  {115421} (\bibinfo {year} {2021})}\BibitemShut {NoStop}%
\bibitem [{\citenamefont {Gali}(2019)}]{gali2019ab}%
  \BibitemOpen
  \bibfield  {author} {\bibinfo {author} {\bibfnamefont {{\'A}.}~\bibnamefont
  {Gali}},\ }\href@noop {} {\bibfield  {journal} {\bibinfo  {journal}
  {Nanophotonics}\ }\textbf {\bibinfo {volume} {8}},\ \bibinfo {pages} {1907}
  (\bibinfo {year} {2019})}\BibitemShut {NoStop}%
\bibitem [{\citenamefont {Razinkovas}\ \emph {et~al.}(2021)\citenamefont
  {Razinkovas}, \citenamefont {Doherty}, \citenamefont {Manson}, \citenamefont
  {Van~de Walle},\ and\ \citenamefont {Alkauskas}}]{razinkovas2021vibrational}%
  \BibitemOpen
  \bibfield  {author} {\bibinfo {author} {\bibfnamefont {L.}~\bibnamefont
  {Razinkovas}}, \bibinfo {author} {\bibfnamefont {M.~W.}\ \bibnamefont
  {Doherty}}, \bibinfo {author} {\bibfnamefont {N.~B.}\ \bibnamefont {Manson}},
  \bibinfo {author} {\bibfnamefont {C.~G.}\ \bibnamefont {Van~de Walle}},\ and\
  \bibinfo {author} {\bibfnamefont {A.}~\bibnamefont {Alkauskas}},\ }\href@noop
  {} {\bibfield  {journal} {\bibinfo  {journal} {Physical Review B}\ }\textbf
  {\bibinfo {volume} {104}},\ \bibinfo {pages} {045303} (\bibinfo {year}
  {2021})}\BibitemShut {NoStop}%
\bibitem [{\citenamefont {Bouquiaux}\ \emph {et~al.}(2021)\citenamefont
  {Bouquiaux}, \citenamefont {Ponc{\'e}}, \citenamefont {Jia}, \citenamefont
  {Miglio}, \citenamefont {Mikami},\ and\ \citenamefont
  {Gonze}}]{bouquiaux2021importance}%
  \BibitemOpen
  \bibfield  {author} {\bibinfo {author} {\bibfnamefont {J.}~\bibnamefont
  {Bouquiaux}}, \bibinfo {author} {\bibfnamefont {S.}~\bibnamefont
  {Ponc{\'e}}}, \bibinfo {author} {\bibfnamefont {Y.}~\bibnamefont {Jia}},
  \bibinfo {author} {\bibfnamefont {A.}~\bibnamefont {Miglio}}, \bibinfo
  {author} {\bibfnamefont {M.}~\bibnamefont {Mikami}},\ and\ \bibinfo {author}
  {\bibfnamefont {X.}~\bibnamefont {Gonze}},\ }\href@noop {} {\bibfield
  {journal} {\bibinfo  {journal} {Advanced Optical Materials}\ }\textbf
  {\bibinfo {volume} {9}},\ \bibinfo {pages} {2100649} (\bibinfo {year}
  {2021})}\BibitemShut {NoStop}%
\bibitem [{\citenamefont {Linder{\"a}lv}\ \emph {et~al.}(2020)\citenamefont
  {Linder{\"a}lv}, \citenamefont {{\AA}berg},\ and\ \citenamefont
  {Erhart}}]{linderalv2020luminescence}%
  \BibitemOpen
  \bibfield  {author} {\bibinfo {author} {\bibfnamefont {C.}~\bibnamefont
  {Linder{\"a}lv}}, \bibinfo {author} {\bibfnamefont {D.}~\bibnamefont
  {{\AA}berg}},\ and\ \bibinfo {author} {\bibfnamefont {P.}~\bibnamefont
  {Erhart}},\ }\href@noop {} {\bibfield  {journal} {\bibinfo  {journal}
  {Chemistry of Materials}\ }\textbf {\bibinfo {volume} {33}},\ \bibinfo
  {pages} {73} (\bibinfo {year} {2020})}\BibitemShut {NoStop}%
\bibitem [{\citenamefont {Wang}\ \emph {et~al.}(2022)\citenamefont {Wang},
  \citenamefont {Huang}, \citenamefont {Zhao}, \citenamefont {Tanner},
  \citenamefont {Zhou},\ and\ \citenamefont {Ning}}]{wang2022role}%
  \BibitemOpen
  \bibfield  {author} {\bibinfo {author} {\bibfnamefont {X.}~\bibnamefont
  {Wang}}, \bibinfo {author} {\bibfnamefont {X.}~\bibnamefont {Huang}},
  \bibinfo {author} {\bibfnamefont {M.}~\bibnamefont {Zhao}}, \bibinfo {author}
  {\bibfnamefont {P.~A.}\ \bibnamefont {Tanner}}, \bibinfo {author}
  {\bibfnamefont {X.}~\bibnamefont {Zhou}},\ and\ \bibinfo {author}
  {\bibfnamefont {L.}~\bibnamefont {Ning}},\ }\href@noop {} {\bibfield
  {journal} {\bibinfo  {journal} {Inorganic Chemistry}\ }\textbf {\bibinfo
  {volume} {61}},\ \bibinfo {pages} {7617} (\bibinfo {year}
  {2022})}\BibitemShut {NoStop}%
\bibitem [{\citenamefont {Wang}\ \emph {et~al.}(2018)\citenamefont {Wang},
  \citenamefont {Xie}, \citenamefont {Suehiro}, \citenamefont {Takeda},\ and\
  \citenamefont {Hirosaki}}]{wang2018down}%
  \BibitemOpen
  \bibfield  {author} {\bibinfo {author} {\bibfnamefont {L.}~\bibnamefont
  {Wang}}, \bibinfo {author} {\bibfnamefont {R.-J.}\ \bibnamefont {Xie}},
  \bibinfo {author} {\bibfnamefont {T.}~\bibnamefont {Suehiro}}, \bibinfo
  {author} {\bibfnamefont {T.}~\bibnamefont {Takeda}},\ and\ \bibinfo {author}
  {\bibfnamefont {N.}~\bibnamefont {Hirosaki}},\ }\href@noop {} {\bibfield
  {journal} {\bibinfo  {journal} {Chemical reviews}\ }\textbf {\bibinfo
  {volume} {118}},\ \bibinfo {pages} {1951} (\bibinfo {year}
  {2018})}\BibitemShut {NoStop}%
\bibitem [{\citenamefont {Tolhurst}\ \emph {et~al.}(2015)\citenamefont
  {Tolhurst}, \citenamefont {Boyko}, \citenamefont {Pust}, \citenamefont
  {Johnson}, \citenamefont {Schnick},\ and\ \citenamefont
  {Moewes}}]{tolhurst2015investigations}%
  \BibitemOpen
  \bibfield  {author} {\bibinfo {author} {\bibfnamefont {T.~M.}\ \bibnamefont
  {Tolhurst}}, \bibinfo {author} {\bibfnamefont {T.~D.}\ \bibnamefont {Boyko}},
  \bibinfo {author} {\bibfnamefont {P.}~\bibnamefont {Pust}}, \bibinfo {author}
  {\bibfnamefont {N.~W.}\ \bibnamefont {Johnson}}, \bibinfo {author}
  {\bibfnamefont {W.}~\bibnamefont {Schnick}},\ and\ \bibinfo {author}
  {\bibfnamefont {A.}~\bibnamefont {Moewes}},\ }\href@noop {} {\bibfield
  {journal} {\bibinfo  {journal} {Advanced Optical Materials}\ }\textbf
  {\bibinfo {volume} {3}},\ \bibinfo {pages} {546} (\bibinfo {year}
  {2015})}\BibitemShut {NoStop}%
\bibitem [{\citenamefont {Tolhurst}\ \emph {et~al.}(2016)\citenamefont
  {Tolhurst}, \citenamefont {Schmiechen}, \citenamefont {Pust}, \citenamefont
  {Schmidt}, \citenamefont {Schnick},\ and\ \citenamefont
  {Moewes}}]{tolhurst2016electronic}%
  \BibitemOpen
  \bibfield  {author} {\bibinfo {author} {\bibfnamefont {T.~M.}\ \bibnamefont
  {Tolhurst}}, \bibinfo {author} {\bibfnamefont {S.}~\bibnamefont
  {Schmiechen}}, \bibinfo {author} {\bibfnamefont {P.}~\bibnamefont {Pust}},
  \bibinfo {author} {\bibfnamefont {P.~J.}\ \bibnamefont {Schmidt}}, \bibinfo
  {author} {\bibfnamefont {W.}~\bibnamefont {Schnick}},\ and\ \bibinfo {author}
  {\bibfnamefont {A.}~\bibnamefont {Moewes}},\ }\href@noop {} {\bibfield
  {journal} {\bibinfo  {journal} {Advanced Optical Materials}\ }\textbf
  {\bibinfo {volume} {4}},\ \bibinfo {pages} {584} (\bibinfo {year}
  {2016})}\BibitemShut {NoStop}%
\bibitem [{\citenamefont {Shafei}\ \emph {et~al.}(2022)\citenamefont {Shafei},
  \citenamefont {Maganas}, \citenamefont {Strobel}, \citenamefont {Schmidt},
  \citenamefont {Schnick},\ and\ \citenamefont {Neese}}]{shafei2022electronic}%
  \BibitemOpen
  \bibfield  {author} {\bibinfo {author} {\bibfnamefont {R.}~\bibnamefont
  {Shafei}}, \bibinfo {author} {\bibfnamefont {D.}~\bibnamefont {Maganas}},
  \bibinfo {author} {\bibfnamefont {P.~J.}\ \bibnamefont {Strobel}}, \bibinfo
  {author} {\bibfnamefont {P.~J.}\ \bibnamefont {Schmidt}}, \bibinfo {author}
  {\bibfnamefont {W.}~\bibnamefont {Schnick}},\ and\ \bibinfo {author}
  {\bibfnamefont {F.}~\bibnamefont {Neese}},\ }\href@noop {} {\bibfield
  {journal} {\bibinfo  {journal} {Journal of the American Chemical Society}\
  }\textbf {\bibinfo {volume} {144}},\ \bibinfo {pages} {8038} (\bibinfo {year}
  {2022})}\BibitemShut {NoStop}%
\bibitem [{\citenamefont {Tsai}\ \emph {et~al.}(2016)\citenamefont {Tsai},
  \citenamefont {Nguyen}, \citenamefont {Lazarowska}, \citenamefont {Mahlik},
  \citenamefont {Grinberg},\ and\ \citenamefont {Liu}}]{tsai2016improvement}%
  \BibitemOpen
  \bibfield  {author} {\bibinfo {author} {\bibfnamefont {Y.-T.}\ \bibnamefont
  {Tsai}}, \bibinfo {author} {\bibfnamefont {H.-D.}\ \bibnamefont {Nguyen}},
  \bibinfo {author} {\bibfnamefont {A.}~\bibnamefont {Lazarowska}}, \bibinfo
  {author} {\bibfnamefont {S.}~\bibnamefont {Mahlik}}, \bibinfo {author}
  {\bibfnamefont {M.}~\bibnamefont {Grinberg}},\ and\ \bibinfo {author}
  {\bibfnamefont {R.-S.}\ \bibnamefont {Liu}},\ }\href@noop {} {\bibfield
  {journal} {\bibinfo  {journal} {Angewandte Chemie International Edition}\
  }\textbf {\bibinfo {volume} {55}},\ \bibinfo {pages} {9652} (\bibinfo {year}
  {2016})}\BibitemShut {NoStop}%
\bibitem [{\citenamefont {Jia}\ \emph {et~al.}(2016)\citenamefont {Jia},
  \citenamefont {Miglio}, \citenamefont {Ponc{\'e}}, \citenamefont {Gonze},\
  and\ \citenamefont {Mikami}}]{jia2016first}%
  \BibitemOpen
  \bibfield  {author} {\bibinfo {author} {\bibfnamefont {Y.}~\bibnamefont
  {Jia}}, \bibinfo {author} {\bibfnamefont {A.}~\bibnamefont {Miglio}},
  \bibinfo {author} {\bibfnamefont {S.}~\bibnamefont {Ponc{\'e}}}, \bibinfo
  {author} {\bibfnamefont {X.}~\bibnamefont {Gonze}},\ and\ \bibinfo {author}
  {\bibfnamefont {M.}~\bibnamefont {Mikami}},\ }\href@noop {} {\bibfield
  {journal} {\bibinfo  {journal} {Physical Review B}\ }\textbf {\bibinfo
  {volume} {93}},\ \bibinfo {pages} {155111} (\bibinfo {year}
  {2016})}\BibitemShut {NoStop}%
\bibitem [{\citenamefont {Ponc{\'e}}\ \emph {et~al.}(2016)\citenamefont
  {Ponc{\'e}}, \citenamefont {Jia}, \citenamefont {Giantomassi}, \citenamefont
  {Mikami},\ and\ \citenamefont {Gonze}}]{ponce2016understanding}%
  \BibitemOpen
  \bibfield  {author} {\bibinfo {author} {\bibfnamefont {S.}~\bibnamefont
  {Ponc{\'e}}}, \bibinfo {author} {\bibfnamefont {Y.}~\bibnamefont {Jia}},
  \bibinfo {author} {\bibfnamefont {M.}~\bibnamefont {Giantomassi}}, \bibinfo
  {author} {\bibfnamefont {M.}~\bibnamefont {Mikami}},\ and\ \bibinfo {author}
  {\bibfnamefont {X.}~\bibnamefont {Gonze}},\ }\href@noop {} {\bibfield
  {journal} {\bibinfo  {journal} {The Journal of Physical Chemistry C}\
  }\textbf {\bibinfo {volume} {120}},\ \bibinfo {pages} {4040} (\bibinfo {year}
  {2016})}\BibitemShut {NoStop}%
\bibitem [{\citenamefont {Jia}\ \emph {et~al.}(2017{\natexlab{a}})\citenamefont
  {Jia}, \citenamefont {Miglio}, \citenamefont {Ponc{\'e}}, \citenamefont
  {Mikami},\ and\ \citenamefont {Gonze}}]{jia2017first}%
  \BibitemOpen
  \bibfield  {author} {\bibinfo {author} {\bibfnamefont {Y.}~\bibnamefont
  {Jia}}, \bibinfo {author} {\bibfnamefont {A.}~\bibnamefont {Miglio}},
  \bibinfo {author} {\bibfnamefont {S.}~\bibnamefont {Ponc{\'e}}}, \bibinfo
  {author} {\bibfnamefont {M.}~\bibnamefont {Mikami}},\ and\ \bibinfo {author}
  {\bibfnamefont {X.}~\bibnamefont {Gonze}},\ }\href@noop {} {\bibfield
  {journal} {\bibinfo  {journal} {Phys. Rev. B}\ }\textbf {\bibinfo {volume}
  {96}},\ \bibinfo {pages} {125132} (\bibinfo {year}
  {2017}{\natexlab{a}})}\BibitemShut {NoStop}%
\bibitem [{\citenamefont {Jia}\ \emph {et~al.}(2017{\natexlab{b}})\citenamefont
  {Jia}, \citenamefont {Ponc{\'e}}, \citenamefont {Miglio}, \citenamefont
  {Mikami},\ and\ \citenamefont {Gonze}}]{jia2017assessment}%
  \BibitemOpen
  \bibfield  {author} {\bibinfo {author} {\bibfnamefont {Y.}~\bibnamefont
  {Jia}}, \bibinfo {author} {\bibfnamefont {S.}~\bibnamefont {Ponc{\'e}}},
  \bibinfo {author} {\bibfnamefont {A.}~\bibnamefont {Miglio}}, \bibinfo
  {author} {\bibfnamefont {M.}~\bibnamefont {Mikami}},\ and\ \bibinfo {author}
  {\bibfnamefont {X.}~\bibnamefont {Gonze}},\ }\href@noop {} {\bibfield
  {journal} {\bibinfo  {journal} {Advanced Optical Materials}\ }\textbf
  {\bibinfo {volume} {5}},\ \bibinfo {pages} {1600997} (\bibinfo {year}
  {2017}{\natexlab{b}})}\BibitemShut {NoStop}%
\bibitem [{\citenamefont {Jia}\ \emph {et~al.}(2020)\citenamefont {Jia},
  \citenamefont {Ponc{\'e}}, \citenamefont {Miglio}, \citenamefont {Mikami},\
  and\ \citenamefont {Gonze}}]{jia2020design}%
  \BibitemOpen
  \bibfield  {author} {\bibinfo {author} {\bibfnamefont {Y.}~\bibnamefont
  {Jia}}, \bibinfo {author} {\bibfnamefont {S.}~\bibnamefont {Ponc{\'e}}},
  \bibinfo {author} {\bibfnamefont {A.}~\bibnamefont {Miglio}}, \bibinfo
  {author} {\bibfnamefont {M.}~\bibnamefont {Mikami}},\ and\ \bibinfo {author}
  {\bibfnamefont {X.}~\bibnamefont {Gonze}},\ }\href@noop {} {\bibfield
  {journal} {\bibinfo  {journal} {Journal of Luminescence}\ }\textbf {\bibinfo
  {volume} {224}},\ \bibinfo {pages} {117258} (\bibinfo {year}
  {2020})}\BibitemShut {NoStop}%
\bibitem [{\citenamefont {Rignanese}\ \emph {et~al.}(1996)\citenamefont
  {Rignanese}, \citenamefont {Michenaud},\ and\ \citenamefont
  {Gonze}}]{rignanese1996ab}%
  \BibitemOpen
  \bibfield  {author} {\bibinfo {author} {\bibfnamefont {G.-M.}\ \bibnamefont
  {Rignanese}}, \bibinfo {author} {\bibfnamefont {J.-P.}\ \bibnamefont
  {Michenaud}},\ and\ \bibinfo {author} {\bibfnamefont {X.}~\bibnamefont
  {Gonze}},\ }\href@noop {} {\bibfield  {journal} {\bibinfo  {journal}
  {Physical Review B}\ }\textbf {\bibinfo {volume} {53}},\ \bibinfo {pages}
  {4488} (\bibinfo {year} {1996})}\BibitemShut {NoStop}%
\bibitem [{\citenamefont {Carrier}\ \emph {et~al.}(2007)\citenamefont
  {Carrier}, \citenamefont {Wentzcovitch},\ and\ \citenamefont
  {Tsuchiya}}]{carrier2007first}%
  \BibitemOpen
  \bibfield  {author} {\bibinfo {author} {\bibfnamefont {P.}~\bibnamefont
  {Carrier}}, \bibinfo {author} {\bibfnamefont {R.}~\bibnamefont
  {Wentzcovitch}},\ and\ \bibinfo {author} {\bibfnamefont {J.}~\bibnamefont
  {Tsuchiya}},\ }\href@noop {} {\bibfield  {journal} {\bibinfo  {journal}
  {Physical Review B}\ }\textbf {\bibinfo {volume} {76}},\ \bibinfo {pages}
  {064116} (\bibinfo {year} {2007})}\BibitemShut {NoStop}%
\bibitem [{\citenamefont {Lax}(1952)}]{lax1952franck}%
  \BibitemOpen
  \bibfield  {author} {\bibinfo {author} {\bibfnamefont {M.}~\bibnamefont
  {Lax}},\ }\href@noop {} {\bibfield  {journal} {\bibinfo  {journal} {The
  Journal of chemical physics}\ }\textbf {\bibinfo {volume} {20}},\ \bibinfo
  {pages} {1752} (\bibinfo {year} {1952})}\BibitemShut {NoStop}%
\bibitem [{\citenamefont {Kubo}\ and\ \citenamefont
  {Toyozawa}(1955)}]{kubo1955application}%
  \BibitemOpen
  \bibfield  {author} {\bibinfo {author} {\bibfnamefont {R.}~\bibnamefont
  {Kubo}}\ and\ \bibinfo {author} {\bibfnamefont {Y.}~\bibnamefont
  {Toyozawa}},\ }\href@noop {} {\bibfield  {journal} {\bibinfo  {journal}
  {Progress of Theoretical Physics}\ }\textbf {\bibinfo {volume} {13}},\
  \bibinfo {pages} {160} (\bibinfo {year} {1955})}\BibitemShut {NoStop}%
\bibitem [{sup()}]{supp}%
  \BibitemOpen
  \href@noop {} {\bibinfo {title} {See supplementary material at [add
  link]}}\BibitemShut {NoStop}%
\bibitem [{\citenamefont {Gonze}\ \emph {et~al.}(2002)\citenamefont {Gonze},
  \citenamefont {Beuken}, \citenamefont {Caracas}, \citenamefont {Detraux},
  \citenamefont {Fuchs}, \citenamefont {Rignanese}, \citenamefont {Sindic},
  \citenamefont {Verstraete}, \citenamefont {Zerah}, \citenamefont {Jollet},
  \citenamefont {Torrent}, \citenamefont {Roy}, \citenamefont {Mikami},
  \citenamefont {Ghosez}, \citenamefont {Raty},\ and\ \citenamefont
  {Allan}}]{Abinit2002}%
  \BibitemOpen
  \bibfield  {author} {\bibinfo {author} {\bibfnamefont {X.}~\bibnamefont
  {Gonze}}, \bibinfo {author} {\bibfnamefont {J.~M.}\ \bibnamefont {Beuken}},
  \bibinfo {author} {\bibfnamefont {R.}~\bibnamefont {Caracas}}, \bibinfo
  {author} {\bibfnamefont {F.}~\bibnamefont {Detraux}}, \bibinfo {author}
  {\bibfnamefont {M.}~\bibnamefont {Fuchs}}, \bibinfo {author} {\bibfnamefont
  {G.~M.}\ \bibnamefont {Rignanese}}, \bibinfo {author} {\bibfnamefont
  {L.}~\bibnamefont {Sindic}}, \bibinfo {author} {\bibfnamefont
  {M.}~\bibnamefont {Verstraete}}, \bibinfo {author} {\bibfnamefont
  {G.}~\bibnamefont {Zerah}}, \bibinfo {author} {\bibfnamefont
  {F.}~\bibnamefont {Jollet}}, \bibinfo {author} {\bibfnamefont
  {M.}~\bibnamefont {Torrent}}, \bibinfo {author} {\bibfnamefont
  {A.}~\bibnamefont {Roy}}, \bibinfo {author} {\bibfnamefont {M.}~\bibnamefont
  {Mikami}}, \bibinfo {author} {\bibfnamefont {P.}~\bibnamefont {Ghosez}},
  \bibinfo {author} {\bibfnamefont {J.~Y.}\ \bibnamefont {Raty}},\ and\
  \bibinfo {author} {\bibfnamefont {D.~C.}\ \bibnamefont {Allan}},\ }\href@noop
  {} {\bibfield  {journal} {\bibinfo  {journal} {Comput. Mater. Sci.}\ }\textbf
  {\bibinfo {volume} {25}},\ \bibinfo {pages} {478 } (\bibinfo {year}
  {2002})}\BibitemShut {NoStop}%
\bibitem [{\citenamefont {Gonze}\ \emph {et~al.}(2020)\citenamefont {Gonze},
  \citenamefont {Amadon}, \citenamefont {Antonius}, \citenamefont {Arnardi},
  \citenamefont {Baguet}, \citenamefont {Beuken}, \citenamefont {Bieder},
  \citenamefont {Bottin}, \citenamefont {Bouchet}, \citenamefont {Bousquet},
  \citenamefont {Brouwer}, \citenamefont {Bruneval}, \citenamefont {Brunin},
  \citenamefont {Cavignac}, \citenamefont {Charraud}, \citenamefont {Chen},
  \citenamefont {Côté}, \citenamefont {Cottenier}, \citenamefont {Denier},
  \citenamefont {Geneste}, \citenamefont {Ghosez}, \citenamefont {Giantomassi},
  \citenamefont {Gillet}, \citenamefont {Gingras}, \citenamefont {Hamann},
  \citenamefont {Hautier}, \citenamefont {He}, \citenamefont {Helbig},
  \citenamefont {Holzwarth}, \citenamefont {Jia}, \citenamefont {Jollet},
  \citenamefont {Lafargue-Dit-Hauret}, \citenamefont {Lejaeghere},
  \citenamefont {Marques}, \citenamefont {Martin}, \citenamefont {Martins},
  \citenamefont {Miranda}, \citenamefont {Naccarato}, \citenamefont {Persson},
  \citenamefont {Petretto}, \citenamefont {Planes}, \citenamefont {Pouillon},
  \citenamefont {Prokhorenko}, \citenamefont {Ricci}, \citenamefont
  {Rignanese}, \citenamefont {Romero}, \citenamefont {Schmitt}, \citenamefont
  {Torrent}, \citenamefont {{van Setten}}, \citenamefont {{Van Troeye}},
  \citenamefont {Verstraete}, \citenamefont {Zérah},\ and\ \citenamefont
  {Zwanziger}}]{gonze2020abinit}%
  \BibitemOpen
  \bibfield  {author} {\bibinfo {author} {\bibfnamefont {X.}~\bibnamefont
  {Gonze}}, \bibinfo {author} {\bibfnamefont {B.}~\bibnamefont {Amadon}},
  \bibinfo {author} {\bibfnamefont {G.}~\bibnamefont {Antonius}}, \bibinfo
  {author} {\bibfnamefont {F.}~\bibnamefont {Arnardi}}, \bibinfo {author}
  {\bibfnamefont {L.}~\bibnamefont {Baguet}}, \bibinfo {author} {\bibfnamefont
  {J.-M.}\ \bibnamefont {Beuken}}, \bibinfo {author} {\bibfnamefont
  {J.}~\bibnamefont {Bieder}}, \bibinfo {author} {\bibfnamefont
  {F.}~\bibnamefont {Bottin}}, \bibinfo {author} {\bibfnamefont
  {J.}~\bibnamefont {Bouchet}}, \bibinfo {author} {\bibfnamefont
  {E.}~\bibnamefont {Bousquet}}, \bibinfo {author} {\bibfnamefont
  {N.}~\bibnamefont {Brouwer}}, \bibinfo {author} {\bibfnamefont
  {F.}~\bibnamefont {Bruneval}}, \bibinfo {author} {\bibfnamefont
  {G.}~\bibnamefont {Brunin}}, \bibinfo {author} {\bibfnamefont
  {T.}~\bibnamefont {Cavignac}}, \bibinfo {author} {\bibfnamefont {J.-B.}\
  \bibnamefont {Charraud}}, \bibinfo {author} {\bibfnamefont {W.}~\bibnamefont
  {Chen}}, \bibinfo {author} {\bibfnamefont {M.}~\bibnamefont {Côté}},
  \bibinfo {author} {\bibfnamefont {S.}~\bibnamefont {Cottenier}}, \bibinfo
  {author} {\bibfnamefont {J.}~\bibnamefont {Denier}}, \bibinfo {author}
  {\bibfnamefont {G.}~\bibnamefont {Geneste}}, \bibinfo {author} {\bibfnamefont
  {P.}~\bibnamefont {Ghosez}}, \bibinfo {author} {\bibfnamefont
  {M.}~\bibnamefont {Giantomassi}}, \bibinfo {author} {\bibfnamefont
  {Y.}~\bibnamefont {Gillet}}, \bibinfo {author} {\bibfnamefont
  {O.}~\bibnamefont {Gingras}}, \bibinfo {author} {\bibfnamefont {D.~R.}\
  \bibnamefont {Hamann}}, \bibinfo {author} {\bibfnamefont {G.}~\bibnamefont
  {Hautier}}, \bibinfo {author} {\bibfnamefont {X.}~\bibnamefont {He}},
  \bibinfo {author} {\bibfnamefont {N.}~\bibnamefont {Helbig}}, \bibinfo
  {author} {\bibfnamefont {N.}~\bibnamefont {Holzwarth}}, \bibinfo {author}
  {\bibfnamefont {Y.}~\bibnamefont {Jia}}, \bibinfo {author} {\bibfnamefont
  {F.}~\bibnamefont {Jollet}}, \bibinfo {author} {\bibfnamefont
  {W.}~\bibnamefont {Lafargue-Dit-Hauret}}, \bibinfo {author} {\bibfnamefont
  {K.}~\bibnamefont {Lejaeghere}}, \bibinfo {author} {\bibfnamefont {M.~A.}\
  \bibnamefont {Marques}}, \bibinfo {author} {\bibfnamefont {A.}~\bibnamefont
  {Martin}}, \bibinfo {author} {\bibfnamefont {C.}~\bibnamefont {Martins}},
  \bibinfo {author} {\bibfnamefont {H.~P.}\ \bibnamefont {Miranda}}, \bibinfo
  {author} {\bibfnamefont {F.}~\bibnamefont {Naccarato}}, \bibinfo {author}
  {\bibfnamefont {K.}~\bibnamefont {Persson}}, \bibinfo {author} {\bibfnamefont
  {G.}~\bibnamefont {Petretto}}, \bibinfo {author} {\bibfnamefont
  {V.}~\bibnamefont {Planes}}, \bibinfo {author} {\bibfnamefont
  {Y.}~\bibnamefont {Pouillon}}, \bibinfo {author} {\bibfnamefont
  {S.}~\bibnamefont {Prokhorenko}}, \bibinfo {author} {\bibfnamefont
  {F.}~\bibnamefont {Ricci}}, \bibinfo {author} {\bibfnamefont {G.-M.}\
  \bibnamefont {Rignanese}}, \bibinfo {author} {\bibfnamefont {A.~H.}\
  \bibnamefont {Romero}}, \bibinfo {author} {\bibfnamefont {M.~M.}\
  \bibnamefont {Schmitt}}, \bibinfo {author} {\bibfnamefont {M.}~\bibnamefont
  {Torrent}}, \bibinfo {author} {\bibfnamefont {M.~J.}\ \bibnamefont {{van
  Setten}}}, \bibinfo {author} {\bibfnamefont {B.}~\bibnamefont {{Van
  Troeye}}}, \bibinfo {author} {\bibfnamefont {M.~J.}\ \bibnamefont
  {Verstraete}}, \bibinfo {author} {\bibfnamefont {G.}~\bibnamefont {Zérah}},\
  and\ \bibinfo {author} {\bibfnamefont {J.~W.}\ \bibnamefont {Zwanziger}},\
  }\href@noop {} {\bibfield  {journal} {\bibinfo  {journal} {Comput. Phys.
  Commun.}\ }\textbf {\bibinfo {volume} {248}},\ \bibinfo {pages} {107042}
  (\bibinfo {year} {2020})}\BibitemShut {NoStop}%
\bibitem [{\citenamefont {Torrent}\ \emph {et~al.}(2008)\citenamefont
  {Torrent}, \citenamefont {Jollet}, \citenamefont {Bottin}, \citenamefont
  {Z{\'e}rah},\ and\ \citenamefont {Gonze}}]{torrent2008implementation}%
  \BibitemOpen
  \bibfield  {author} {\bibinfo {author} {\bibfnamefont {M.}~\bibnamefont
  {Torrent}}, \bibinfo {author} {\bibfnamefont {F.}~\bibnamefont {Jollet}},
  \bibinfo {author} {\bibfnamefont {F.}~\bibnamefont {Bottin}}, \bibinfo
  {author} {\bibfnamefont {G.}~\bibnamefont {Z{\'e}rah}},\ and\ \bibinfo
  {author} {\bibfnamefont {X.}~\bibnamefont {Gonze}},\ }\href@noop {}
  {\bibfield  {journal} {\bibinfo  {journal} {Comput. Mater. Sci.}\ }\textbf
  {\bibinfo {volume} {42}},\ \bibinfo {pages} {337} (\bibinfo {year}
  {2008})}\BibitemShut {NoStop}%
\bibitem [{\citenamefont {Perdew}\ \emph {et~al.}(1996)\citenamefont {Perdew},
  \citenamefont {Burke},\ and\ \citenamefont
  {Ernzerhof}}]{perdew1996generalized}%
  \BibitemOpen
  \bibfield  {author} {\bibinfo {author} {\bibfnamefont {J.~P.}\ \bibnamefont
  {Perdew}}, \bibinfo {author} {\bibfnamefont {K.}~\bibnamefont {Burke}},\ and\
  \bibinfo {author} {\bibfnamefont {M.}~\bibnamefont {Ernzerhof}},\ }\href@noop
  {} {\bibfield  {journal} {\bibinfo  {journal} {Phys. Rev. Lett.}\ }\textbf
  {\bibinfo {volume} {77}},\ \bibinfo {pages} {3865} (\bibinfo {year}
  {1996})}\BibitemShut {NoStop}%
\bibitem [{\citenamefont {Joos}\ \emph {et~al.}(2020)\citenamefont {Joos},
  \citenamefont {Smet}, \citenamefont {Seijo},\ and\ \citenamefont
  {Barandiar{\'a}n}}]{joos2020insights}%
  \BibitemOpen
  \bibfield  {author} {\bibinfo {author} {\bibfnamefont {J.~J.}\ \bibnamefont
  {Joos}}, \bibinfo {author} {\bibfnamefont {P.~F.}\ \bibnamefont {Smet}},
  \bibinfo {author} {\bibfnamefont {L.}~\bibnamefont {Seijo}},\ and\ \bibinfo
  {author} {\bibfnamefont {Z.}~\bibnamefont {Barandiar{\'a}n}},\ }\href@noop {}
  {\bibfield  {journal} {\bibinfo  {journal} {Inorganic Chemistry Frontiers}\
  }\textbf {\bibinfo {volume} {7}},\ \bibinfo {pages} {871} (\bibinfo {year}
  {2020})}\BibitemShut {NoStop}%
\bibitem [{\citenamefont {Barandiar{\'a}n}\ \emph {et~al.}(2022)\citenamefont
  {Barandiar{\'a}n}, \citenamefont {Joos},\ and\ \citenamefont
  {Seijo}}]{barandiaran2022luminescent}%
  \BibitemOpen
  \bibfield  {author} {\bibinfo {author} {\bibfnamefont {Z.}~\bibnamefont
  {Barandiar{\'a}n}}, \bibinfo {author} {\bibfnamefont {J.}~\bibnamefont
  {Joos}},\ and\ \bibinfo {author} {\bibfnamefont {L.}~\bibnamefont {Seijo}},\
  }\href@noop {} {\emph {\bibinfo {title} {Luminescent Materials: A Quantum
  Chemical Approach for Computer-Aided Discovery and Design}}},\ Vol.\ \bibinfo
  {volume} {322}\ (\bibinfo  {publisher} {Springer Nature},\ \bibinfo {year}
  {2022})\BibitemShut {NoStop}%
\bibitem [{\citenamefont {Gonze}\ and\ \citenamefont
  {Lee}(1997)}]{gonze1997dynamical}%
  \BibitemOpen
  \bibfield  {author} {\bibinfo {author} {\bibfnamefont {X.}~\bibnamefont
  {Gonze}}\ and\ \bibinfo {author} {\bibfnamefont {C.}~\bibnamefont {Lee}},\
  }\href@noop {} {\bibfield  {journal} {\bibinfo  {journal} {Physical Review
  B}\ }\textbf {\bibinfo {volume} {55}},\ \bibinfo {pages} {10355} (\bibinfo
  {year} {1997})}\BibitemShut {NoStop}%
\bibitem [{\citenamefont {Togo}\ and\ \citenamefont
  {Tanaka}(2015)}]{togo2015first}%
  \BibitemOpen
  \bibfield  {author} {\bibinfo {author} {\bibfnamefont {A.}~\bibnamefont
  {Togo}}\ and\ \bibinfo {author} {\bibfnamefont {I.}~\bibnamefont {Tanaka}},\
  }\href@noop {} {\bibfield  {journal} {\bibinfo  {journal} {Scripta
  Materialia}\ }\textbf {\bibinfo {volume} {108}},\ \bibinfo {pages} {1}
  (\bibinfo {year} {2015})}\BibitemShut {NoStop}%
\bibitem [{\citenamefont {Carrier}\ \emph {et~al.}(2008)\citenamefont
  {Carrier}, \citenamefont {Justo},\ and\ \citenamefont
  {Wentzcovitch}}]{carrier2008quasiharmonic}%
  \BibitemOpen
  \bibfield  {author} {\bibinfo {author} {\bibfnamefont {P.}~\bibnamefont
  {Carrier}}, \bibinfo {author} {\bibfnamefont {J.~F.}\ \bibnamefont {Justo}},\
  and\ \bibinfo {author} {\bibfnamefont {R.~M.}\ \bibnamefont {Wentzcovitch}},\
  }\href@noop {} {\bibfield  {journal} {\bibinfo  {journal} {Physical Review
  B}\ }\textbf {\bibinfo {volume} {78}},\ \bibinfo {pages} {144302} (\bibinfo
  {year} {2008})}\BibitemShut {NoStop}%
\bibitem [{\citenamefont {Allan}\ \emph {et~al.}(1996)\citenamefont {Allan},
  \citenamefont {Barron},\ and\ \citenamefont {Bruno}}]{allan1996zero}%
  \BibitemOpen
  \bibfield  {author} {\bibinfo {author} {\bibfnamefont {N.}~\bibnamefont
  {Allan}}, \bibinfo {author} {\bibfnamefont {T.}~\bibnamefont {Barron}},\ and\
  \bibinfo {author} {\bibfnamefont {J.}~\bibnamefont {Bruno}},\ }\href@noop {}
  {\bibfield  {journal} {\bibinfo  {journal} {The Journal of chemical physics}\
  }\textbf {\bibinfo {volume} {105}},\ \bibinfo {pages} {8300} (\bibinfo {year}
  {1996})}\BibitemShut {NoStop}%
\bibitem [{\citenamefont {Masuki}\ \emph {et~al.}(2023)\citenamefont {Masuki},
  \citenamefont {Nomoto}, \citenamefont {Arita},\ and\ \citenamefont
  {Tadano}}]{masuki2023full}%
  \BibitemOpen
  \bibfield  {author} {\bibinfo {author} {\bibfnamefont {R.}~\bibnamefont
  {Masuki}}, \bibinfo {author} {\bibfnamefont {T.}~\bibnamefont {Nomoto}},
  \bibinfo {author} {\bibfnamefont {R.}~\bibnamefont {Arita}},\ and\ \bibinfo
  {author} {\bibfnamefont {T.}~\bibnamefont {Tadano}},\ }\href@noop {}
  {\bibfield  {journal} {\bibinfo  {journal} {arXiv preprint arXiv:2302.04537}\
  } (\bibinfo {year} {2023})}\BibitemShut {NoStop}%
\bibitem [{\citenamefont {Ritz}\ \emph {et~al.}(2019)\citenamefont {Ritz},
  \citenamefont {Li},\ and\ \citenamefont {Benedek}}]{ritz2019thermal}%
  \BibitemOpen
  \bibfield  {author} {\bibinfo {author} {\bibfnamefont {E.~T.}\ \bibnamefont
  {Ritz}}, \bibinfo {author} {\bibfnamefont {S.~J.}\ \bibnamefont {Li}},\ and\
  \bibinfo {author} {\bibfnamefont {N.~A.}\ \bibnamefont {Benedek}},\
  }\href@noop {} {\bibfield  {journal} {\bibinfo  {journal} {Journal of Applied
  Physics}\ }\textbf {\bibinfo {volume} {126}},\ \bibinfo {pages} {171102}
  (\bibinfo {year} {2019})}\BibitemShut {NoStop}%
\bibitem [{\citenamefont {Brousseau-Couture}\ \emph {et~al.}(2022)\citenamefont
  {Brousseau-Couture}, \citenamefont {Godbout}, \citenamefont {C{\^o}t{\'e}},\
  and\ \citenamefont {Gonze}}]{brousseau2022zero}%
  \BibitemOpen
  \bibfield  {author} {\bibinfo {author} {\bibfnamefont {V.}~\bibnamefont
  {Brousseau-Couture}}, \bibinfo {author} {\bibfnamefont {{\'E}.}~\bibnamefont
  {Godbout}}, \bibinfo {author} {\bibfnamefont {M.}~\bibnamefont
  {C{\^o}t{\'e}}},\ and\ \bibinfo {author} {\bibfnamefont {X.}~\bibnamefont
  {Gonze}},\ }\href@noop {} {\bibfield  {journal} {\bibinfo  {journal}
  {Physical Review B}\ }\textbf {\bibinfo {volume} {106}},\ \bibinfo {pages}
  {085137} (\bibinfo {year} {2022})}\BibitemShut {NoStop}%
\bibitem [{\citenamefont {Fu}\ and\ \citenamefont {Ho}(1983)}]{fu1983first}%
  \BibitemOpen
  \bibfield  {author} {\bibinfo {author} {\bibfnamefont {C.-L.}\ \bibnamefont
  {Fu}}\ and\ \bibinfo {author} {\bibfnamefont {K.-M.}\ \bibnamefont {Ho}},\
  }\href@noop {} {\bibfield  {journal} {\bibinfo  {journal} {Physical Review
  B}\ }\textbf {\bibinfo {volume} {28}},\ \bibinfo {pages} {5480} (\bibinfo
  {year} {1983})}\BibitemShut {NoStop}%
\bibitem [{\citenamefont {Shannon}(1976)}]{shannon1976revised}%
  \BibitemOpen
  \bibfield  {author} {\bibinfo {author} {\bibfnamefont {R.~D.}\ \bibnamefont
  {Shannon}},\ }\href@noop {} {\bibfield  {journal} {\bibinfo  {journal} {Acta
  crystallographica section A: crystal physics, diffraction, theoretical and
  general crystallography}\ }\textbf {\bibinfo {volume} {32}},\ \bibinfo
  {pages} {751} (\bibinfo {year} {1976})}\BibitemShut {NoStop}%
\bibitem [{\citenamefont {Ruegenberg}\ \emph {et~al.}(2023)\citenamefont
  {Ruegenberg}, \citenamefont {Garc{\'\i}a-Fuente}, \citenamefont {Seibald},
  \citenamefont {Baumann}, \citenamefont {Hoerder}, \citenamefont {Fiedler},
  \citenamefont {Urland}, \citenamefont {Huppertz}, \citenamefont {Meijerink},\
  and\ \citenamefont {Suta}}]{ruegenberg2023mixed}%
  \BibitemOpen
  \bibfield  {author} {\bibinfo {author} {\bibfnamefont {F.}~\bibnamefont
  {Ruegenberg}}, \bibinfo {author} {\bibfnamefont {A.}~\bibnamefont
  {Garc{\'\i}a-Fuente}}, \bibinfo {author} {\bibfnamefont {M.}~\bibnamefont
  {Seibald}}, \bibinfo {author} {\bibfnamefont {D.}~\bibnamefont {Baumann}},
  \bibinfo {author} {\bibfnamefont {G.}~\bibnamefont {Hoerder}}, \bibinfo
  {author} {\bibfnamefont {T.}~\bibnamefont {Fiedler}}, \bibinfo {author}
  {\bibfnamefont {W.}~\bibnamefont {Urland}}, \bibinfo {author} {\bibfnamefont
  {H.}~\bibnamefont {Huppertz}}, \bibinfo {author} {\bibfnamefont
  {A.}~\bibnamefont {Meijerink}},\ and\ \bibinfo {author} {\bibfnamefont
  {M.}~\bibnamefont {Suta}},\ }\href@noop {} {\bibfield  {journal} {\bibinfo
  {journal} {Advanced Optical Materials}\ ,\ \bibinfo {pages} {2202732}}
  (\bibinfo {year} {2023})}\BibitemShut {NoStop}%
\bibitem [{\citenamefont {Yan}(2018)}]{yan2018origin}%
  \BibitemOpen
  \bibfield  {author} {\bibinfo {author} {\bibfnamefont {S.}~\bibnamefont
  {Yan}},\ }\href@noop {} {\bibfield  {journal} {\bibinfo  {journal} {Optical
  Materials}\ }\textbf {\bibinfo {volume} {79}},\ \bibinfo {pages} {172}
  (\bibinfo {year} {2018})}\BibitemShut {NoStop}%
\bibitem [{\citenamefont {Xu}\ \emph {et~al.}(2022)\citenamefont {Xu},
  \citenamefont {Huang}, \citenamefont {Cheng}, \citenamefont {Whangbo},\ and\
  \citenamefont {Deng}}]{xu2022microscopic}%
  \BibitemOpen
  \bibfield  {author} {\bibinfo {author} {\bibfnamefont {J.}~\bibnamefont
  {Xu}}, \bibinfo {author} {\bibfnamefont {X.}~\bibnamefont {Huang}}, \bibinfo
  {author} {\bibfnamefont {X.}~\bibnamefont {Cheng}}, \bibinfo {author}
  {\bibfnamefont {M.-H.}\ \bibnamefont {Whangbo}},\ and\ \bibinfo {author}
  {\bibfnamefont {S.}~\bibnamefont {Deng}},\ }\href@noop {} {\bibfield
  {journal} {\bibinfo  {journal} {Angewandte Chemie International Edition}\
  }\textbf {\bibinfo {volume} {61}},\ \bibinfo {pages} {e202116404} (\bibinfo
  {year} {2022})}\BibitemShut {NoStop}%
\end{thebibliography}%


\begin{thebibliography}{1}

\bibitem{alkauskas2014}
A.~Alkauskas, B.~B. Buckley, D.~D. Awschalom, and C.~G. Van~de Walle,
  ``{First-principles theory of the luminescence lineshape for the triplet
  transition in diamond NV centres},'' {\em New J. Phys.}, vol.~16, no.~7,
  p.~073026, 2014.

\bibitem{pust2014narrow}
P.~Pust, V.~Weiler, C.~Hecht, A.~T{\"u}cks, A.~S. Wochnik, A.-K. Hen{\ss},
  D.~Wiechert, C.~Scheu, P.~J. Schmidt, and W.~Schnick, ``{Narrow-band
  red-emitting Sr[LiAl$_3$N$_4$]: Eu$^{2+}$ as a next-generation LED-phosphor
  material},'' {\em Nature materials}, vol.~13, no.~9, pp.~891--896, 2014.

\bibitem{shannon1976revised}
R.~D. Shannon, ``Revised effective ionic radii and systematic studies of
  interatomic distances in halides and chalcogenides,'' {\em Acta
  crystallographica section A: crystal physics, diffraction, theoretical and
  general crystallography}, vol.~32, no.~5, pp.~751--767, 1976.

\bibitem{tsai2016improvement}
Y.-T. Tsai, H.-D. Nguyen, A.~Lazarowska, S.~Mahlik, M.~Grinberg, and R.-S. Liu,
  ``{Improvement of the Water Resistance of a Narrow-Band Red-Emitting
  SrLiAl$_3$N$_4$:Eu$^{2+}$ Phosphor Synthesized under High Isostatic Pressure
  through Coating with an Organosilica Layer},'' {\em Angewandte Chemie
  International Edition}, vol.~55, no.~33, pp.~9652--9656, 2016.

\bibitem{reshchikov2005luminescence}
M.~A. Reshchikov and H.~Morko{\c{c}}, ``{Luminescence properties of defects in
  GaN},'' {\em Journal of applied physics}, vol.~97, no.~6, pp.~5--19, 2005.

\bibitem{jia2017first}
Y.~Jia, A.~Miglio, S.~Ponc{\'e}, M.~Mikami, and X.~Gonze, ``First-principles
  study of the luminescence of {Eu$^{2+}$}-doped phosphors,'' {\em Phys. Rev.
  B}, vol.~96, no.~12, p.~125132, 2017.

\end{thebibliography}

\medskip

\end{document}


{\Large Supporting informations}
\vspace{2cm}

\textbf{\Large A First-Principles Explanation of the Luminescent Line Shape of SrLiAl$_3$N$_4$:Eu$^{2+}$  Phosphor for Light-Emitting Diode Applications}
\vspace{0.5cm}

\textit{Julien Bouquiaux *, Samuel Ponc\'{e}, Yongchao Jia, Anna Miglio, Masayoshi Mikami, Xavier Gonze}

\tableofcontents

\section{Embedding approach}

\subsection{Forces vs Displacements}

For a given phonon mode $\nu$, the corresponding partial Huang-Rhys factor $S_{\nu}=\frac{\omega_{\nu}\Delta Q_{\nu}^2}{2\hbar}$ is computed thanks to the mass-weighted displacement between ground and excited states projected on this phonon mode:
\begin{equation}
	\label{eq:Delta_Q_nu}
	\Delta Q_\nu=\sum_{\kappa\alpha}\sqrt{M_{\kappa}}\Delta R_{\kappa\alpha}{e_{\nu,\kappa\alpha}},
\end{equation}
where $\kappa$ labels the atoms and $\alpha$ the cartesian direction. Under the harmonic approximation,
\begin{equation}
	\label{eq:Harmonic}
	M_{\kappa}\omega_{\nu}^2e_{\nu,\kappa\alpha}=\sum_{\kappa'}C_{\kappa\alpha,\kappa'\alpha'}e_{\nu,\kappa\alpha},
\end{equation}
with the interatomic force constants (IFCs) $C_{\kappa\alpha,\kappa'\alpha'}=\frac{\Delta F_{\kappa'\alpha'}}{\Delta R_{\kappa\alpha}}$, Eq.~\eqref{eq:Delta_Q_nu} rewrites:
\begin{equation}
	\Delta Q_\nu=\frac{1}{\omega_\nu^2}\sum_{\kappa\alpha}\frac{\Delta F_{\kappa\alpha} e_{\nu,\kappa\alpha}}{\sqrt{M_{\kappa}}}.
	\label{eq:Delta_Q_nu_forces}
\end{equation}

\begin{figure}[h!]
	\centering
	\includegraphics[width=0.7\linewidth]{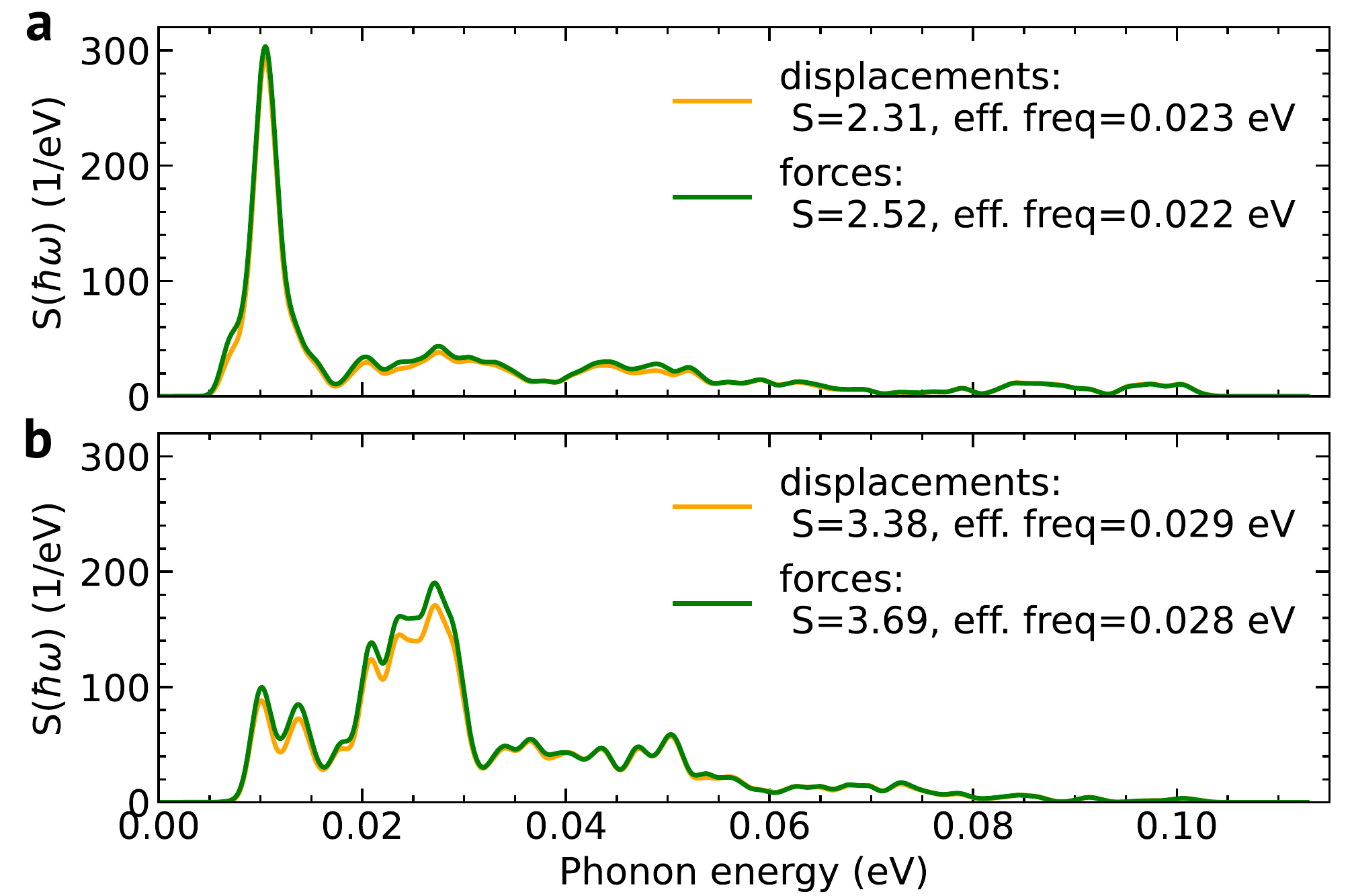}
	\caption{\label{fig:forces_vs_dis} Huang-Rhys spectral function $S(\hbar\omega)=\sum_{\nu}S_{\nu}\delta(\omega-\omega_{\nu})$ computed with the displacements using Eq.~\eqref{eq:Delta_Q_nu} (orange line) and the forces using Eq.~\eqref{eq:Delta_Q_nu_forces} (green line), on a 288-atoms supercell, for site Eu(Sr1) (a) and Eu(Sr2) (b). The effective frequency is computed as $\Omega_{\rm{eff}}^2=\sum_{\nu}p_{\nu}\omega_\nu^2$ with  $p_{\nu}=(\Delta Q_{\nu}/\Delta Q)^2$ the weight by which each phonon mode contributes to the atomic position changes when a electronic transition occurs.}
\end{figure}

The validity of this approximation was first tested by computing the Huang-Rhys spectral function with Eqs.~\eqref{eq:Delta_Q_nu} and \eqref{eq:Delta_Q_nu_forces} with displacements computed on a 2$\times$2$\times$2 supercell (288 atoms) and forces computed in the ground state at excited state equilibrium positions as depicted on Fig.~ \ref{fig:forces_vs_dis}(b). 
Phonons are computed with finite difference in the Eu-4f ground state in both approach. 
The result is shown on Fig.~\ref{fig:forces_vs_dis}(a). 
It can be observed that for both sites, the shape of the spectral function, and thus the effective frequency, is almost identical between two approaches. 
On the other hand, the use of the forces leads to an overestimation of the total Huang-Rhys factor $S=\sum_{\nu}S_{\nu}$ of 7\% for site 1 and of  11.5\% for site 2.

\subsection{Benefits of using the forces}

\begin{figure}[h!]
	\centering
	\includegraphics[width=0.9\linewidth]{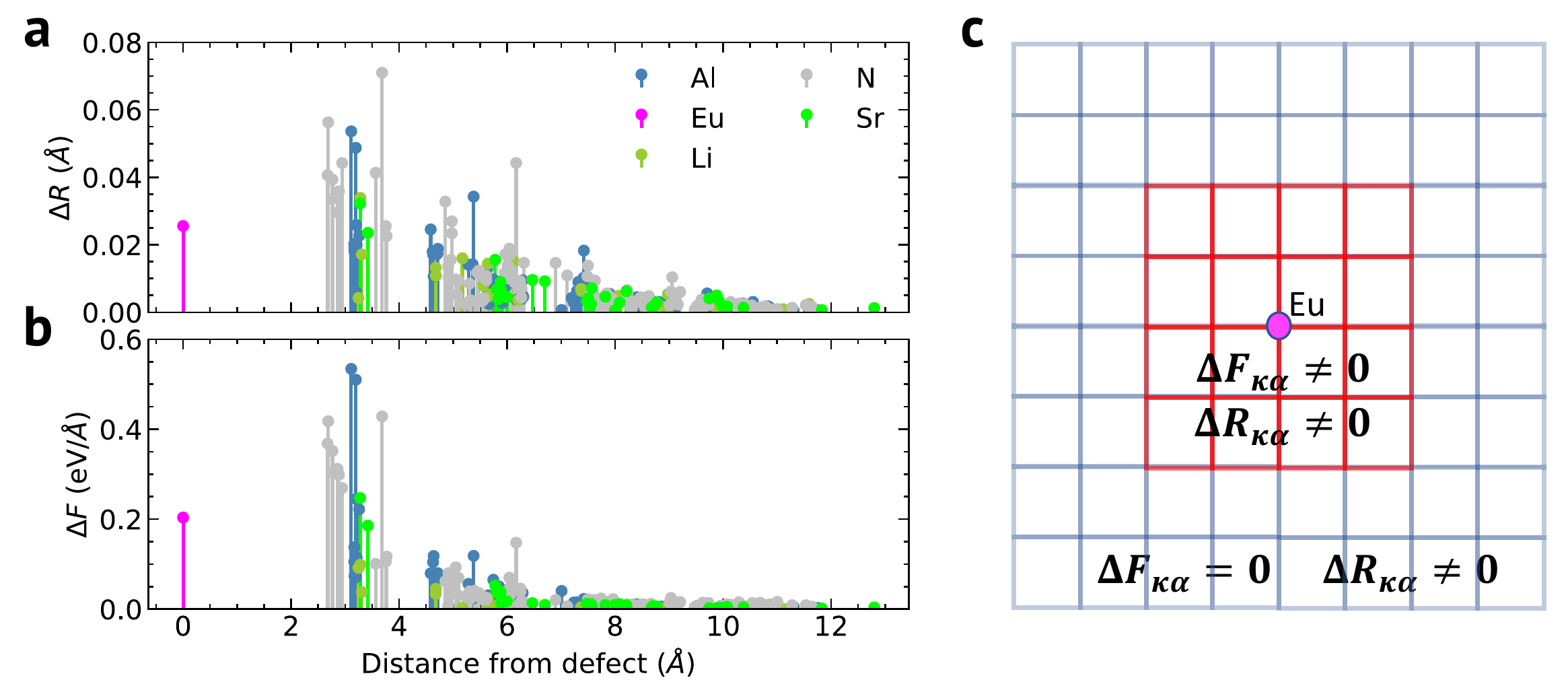}
	\caption{\label{fig:Force_vs_dis_site2}(a) Norm of the displacement induced by the 5d-4f transition for Eu(Sr2) as a function of the distance from the Eu atom and (b) norm of the forces in the 4f ground state at the 5d equilibrium atomic positions as a function of the distance from the Eu atom. The decay of forces (b) is much faster than the decay of displacements (a).
	(c) Cartoon of a small red supercell containing the Eu(Sr) substitutional defect where the forces are the ones computed with DFT. 
	Outside this red supercell, the forces are set to zero because of their short-range decay while the displacements are non-zero and are computed with the IFCs computed in this large blue supercell.}
\end{figure}

The use of Eq.~\eqref{eq:Delta_Q_nu_forces} allows increasing the supercell size and hence minimize finite-size effects. 
Indeed, the forces decay faster to zero compared to the displacements with respect to the distance from the defect as illustrated in Fig.~\ref{fig:Force_vs_dis_site2}\textbf{a-b} for site 2. 
We assume that the forces computed within the small red supercell (here 288 atoms), see Fig.~\ref{fig:Force_vs_dis_site2}\textbf{c}, are already converged and without finite-size effect because of this rapid decay, and that the forces outside this supercell are essentially zero. 
This means that we can use the forces computed in this red supercell, and deduce the displacements at much larger distances in the blue supercell via the IFCs computed within this large supercell. 
This subtle point is actually directly included when using Eq.~\eqref{eq:Delta_Q_nu_forces} in the blue supercell. 

\subsection{Computation of the IFCs in large supercells with defects}
\begin{figure}[h!]
	\centering	\includegraphics[width=\linewidth]{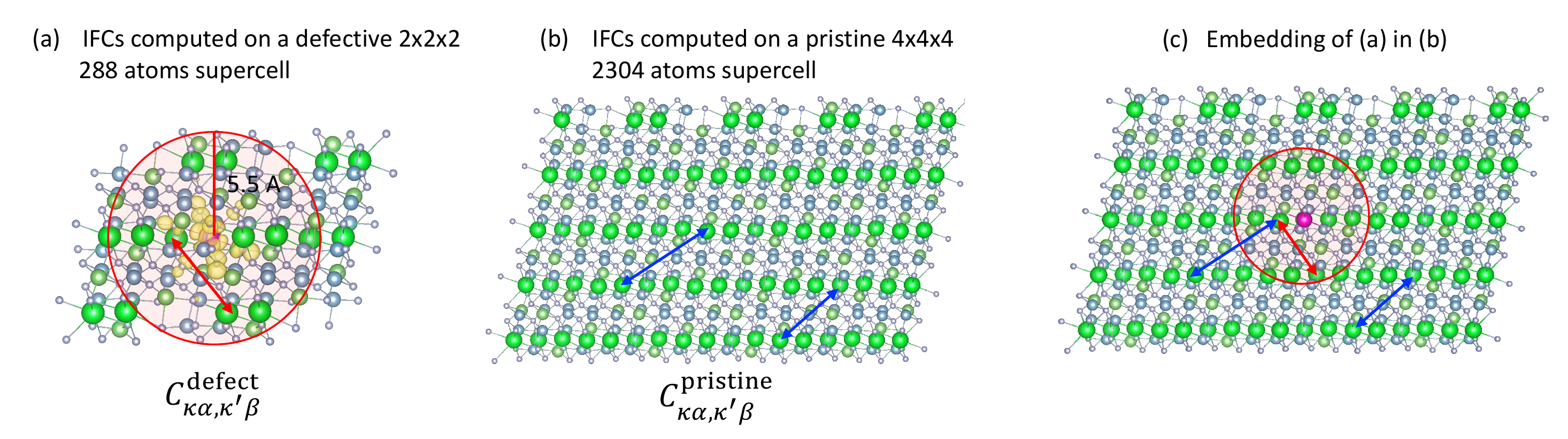}
	\caption{Procedure to compute phonons modes in defective large supercell, see text for details.}
	\label{fig:Phonons_emb}
\end{figure}

Reaching the dilute limit requires to estimate the IFCs in large supercells ($>$1000 atoms), which is computationally too demanding with a direct approach.
Hence we follow an approach similar to the one proposed by Alkauskas \textit{et al.}~\cite{alkauskas2014}.
We need to compute the phonons modes with displacement vectors $e_{\nu,\kappa\alpha}$ and frequencies $\omega_\nu$ by both approaching the dilute limit (including long-wavelength phonons), and keeping the localized modes brought by the defect. For this, we diagonalize modified IFCs following the scheme shown on figure \ref{fig:Phonons_emb} : 
\\
First, we compute the IFCs of the defect system with finite difference in a 2$\times$2$\times$2 supercell (which also corresponds to the 288 atoms, red supercell of Figure S2 (c), but different size could be possible).
\\
We then compute the IFCs of the pristine system with DFPT in a 36 atoms unit cell on a coarse 2$\times$2$\times$2 \textbf{k}-point and \textbf{q}-point grids that we Fourier interpolate on a 4$\times$4$\times$4 fine \textbf{q}-grid. 
These IFCs are then mapped on the corresponding 4$\times$4$\times$4 supercell.
In order to construct the IFCs of a large defective  4$\times$4$\times$4 supercell, we use the following rule. 
If both atoms $\kappa$ and $\kappa'$ are separated from the defect by a distance smaller than an cut-off radius $R_c$, then we use the IFCs computed with the small defect supercell. 
For all other atomic pairs, we use the IFCs computed with the pristine system. 
\\
We have used a cut-off of 5.5~\AA, which corresponds to the radius of the largest sphere contained in the 2x2x2 supercell.
This procedure breaks the acoustic sum rule. Hence we reinforce it by setting :
\begin{equation}
	\label{eq:ASR}
	C_{\kappa\alpha,\kappa\alpha}=-\sum_{\alpha\ne\beta}C_{\kappa\beta,\kappa\alpha}.
\end{equation}
\begin{figure}[h!]
	\centering
	\includegraphics[width=0.99\linewidth]{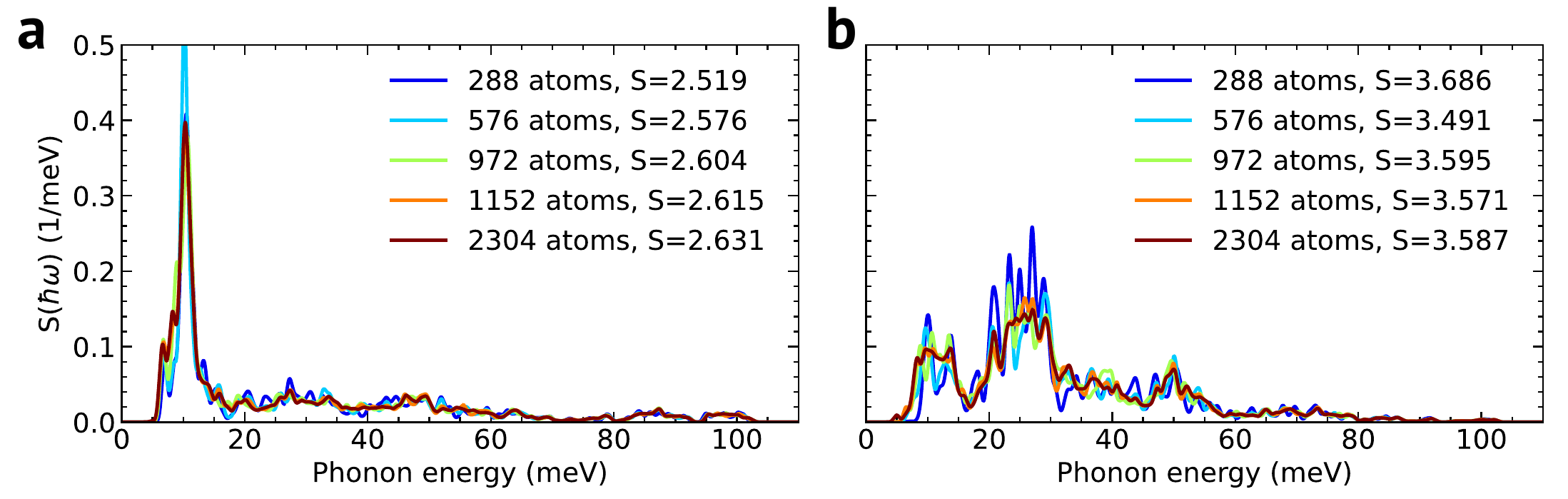}
	\caption{\label{fig:Conv_HR} Convergence of the Huang-Rhys spectral function as a function of the size of the large supercell 2$\times$2$\times$2, 4$\times$2$\times$2, 3$\times$3$\times$3, 4$\times$4$\times$2 and 4$\times$4$\times$4, for \textbf{a} site 1 and \textbf{b} site 2.}
\end{figure}

We have performed in Fig.~\ref{fig:Conv_HR} a convergence study of the Huang-Rhys spectral function as a function of the size of the large supercell. 
The result is shown in  \ref{fig:Conv_HR} with [2,2,2],[4,2,2],[3,3,3],[4,4,2] and [4,4,4] supercells containing respectively 288, 576, 972, 1152 and 2304 atoms. We have also tried a cut-off off 5~\AA. The total Huang-Rhys factors of both sites change by less than 1\% with no noticeable change on the shape of the spectral function. 

\pagebreak
\section{Influence of the Hubbard U value}

\begin{figure}[h!]
	\centering
	\includegraphics[width=0.7\linewidth]{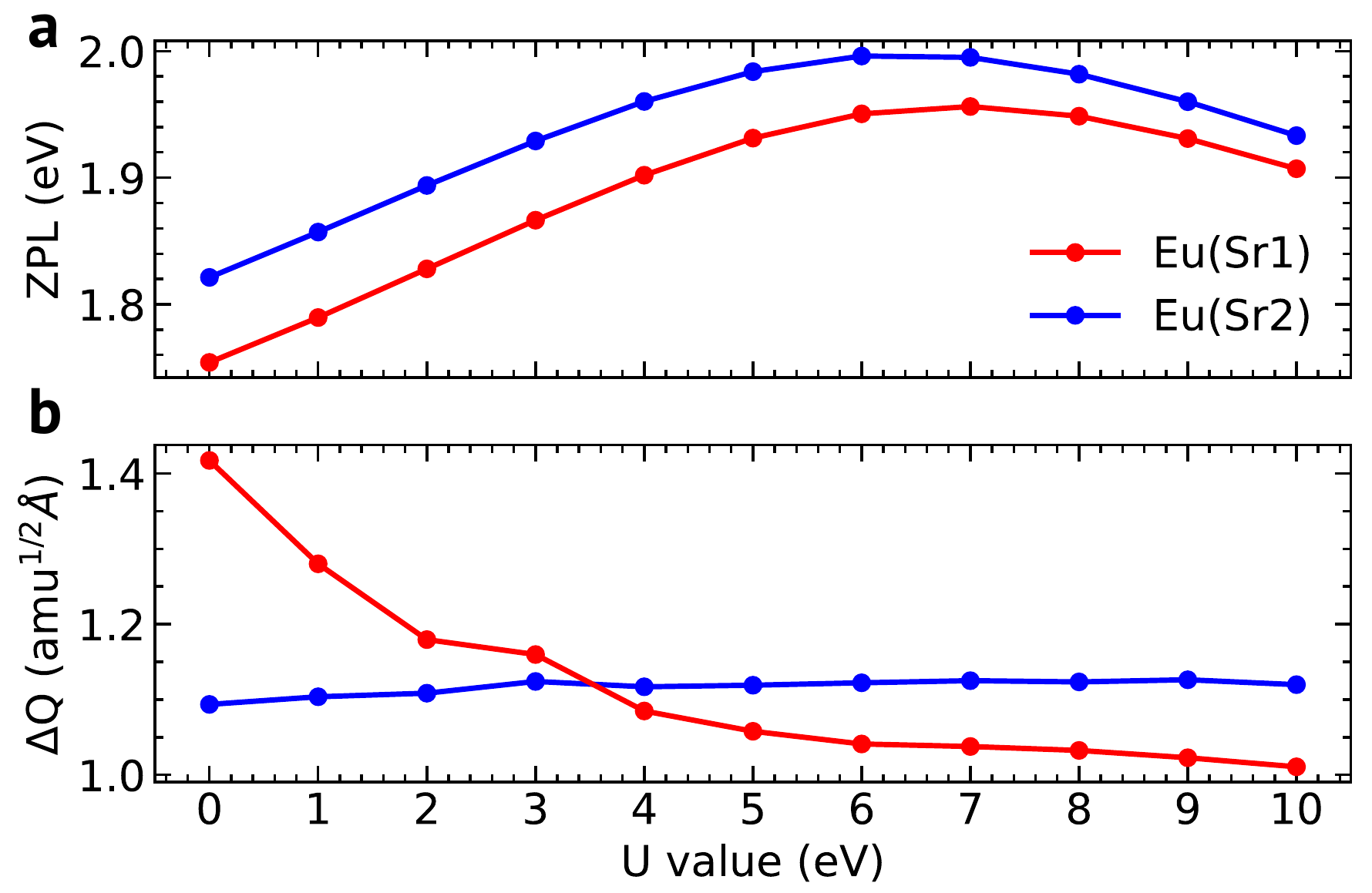}
	\caption{\label{fig:U_study}(a) ZPL energies and (b) total mass weighted displacement $\Delta Q$ as a function of the U value applied on the Eu-4f states. Calculations were done on a small 72-atoms supercell.}
\end{figure}

The Fig.~\ref{fig:U_study} presents the impact of the value of the U parameter applied on the $4f$ state of europium on the zero phonon line (ZPL) and the total mass weighted displacement $\Delta$Q.
The calculations are done on a 2$\times$1$\times$1 supercell containing 72 atoms where one strontium atom is replaced by an europium one at site 1 (red) or site 2 (blue).  
We find that the results are insensitive to the value of U in the range 5 to 9~eV with a maximal deviation of  3\% on $\Delta$Q of Sr1. 

\pagebreak
\section{Phonon band structure of pristine SLA}

\begin{figure}[h!]
	\centering
	\includegraphics[width=0.7\linewidth]{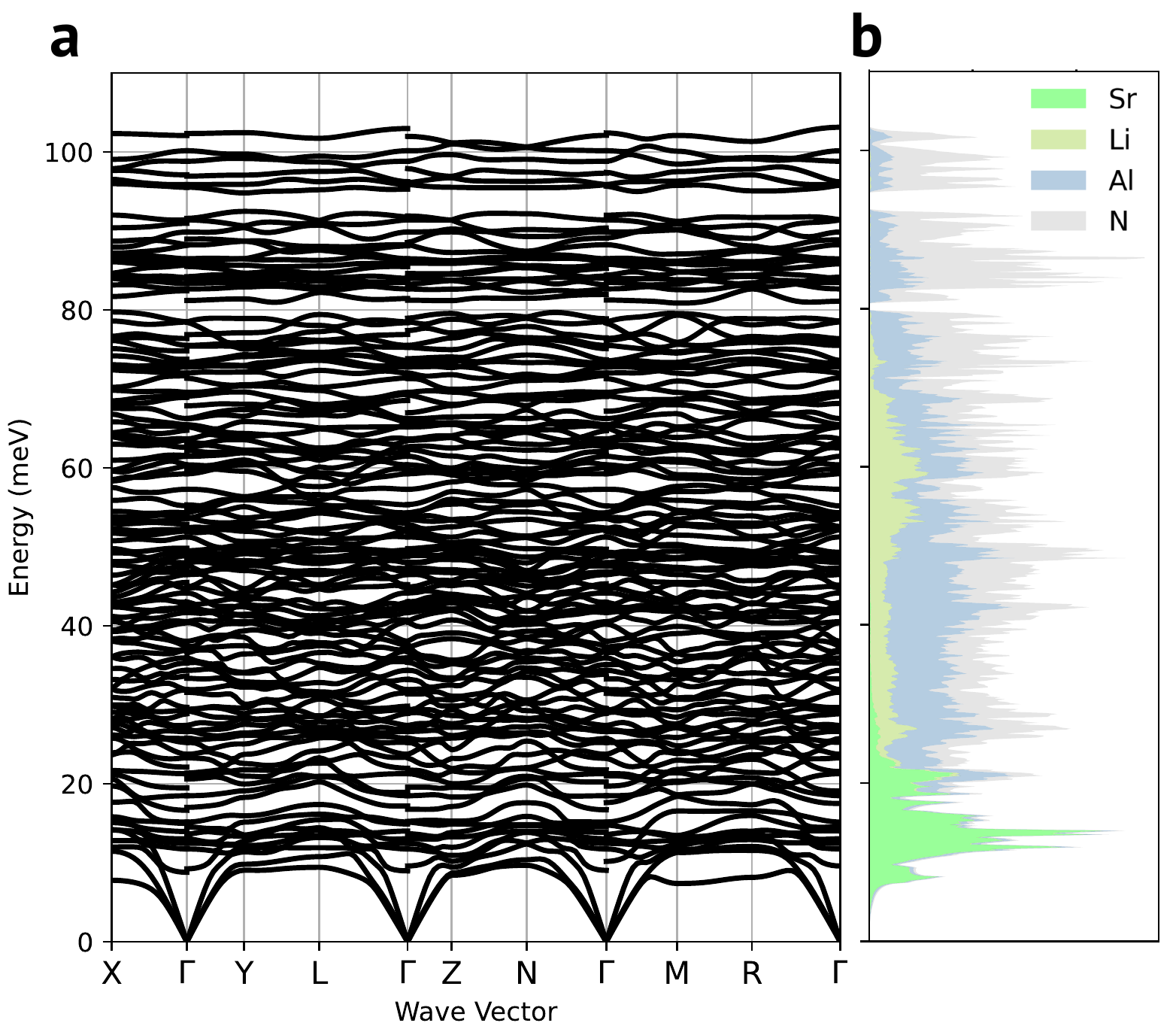}
	\caption{\label{fig:phbands_phdos} \textbf{a} Phonon band structure and \textbf{b} atom-projected density of states of pristine SrLiAl$_3$N$_4$ obtained with DFPT.}
\end{figure}

The phonon bandstructure obtained with density functional perturbation theory for the undoped SrLiAl$_3$N$_4$ primitive cell is given in Fig.~\ref{fig:phbands_phdos} using a 2$\times$2$\times$2 \textbf{k}-point and \textbf{q}-point grids.

\section{Thermal expansion, additional details}

\begin{figure}[h!]
	\centering
	\includegraphics[width=0.9\linewidth]{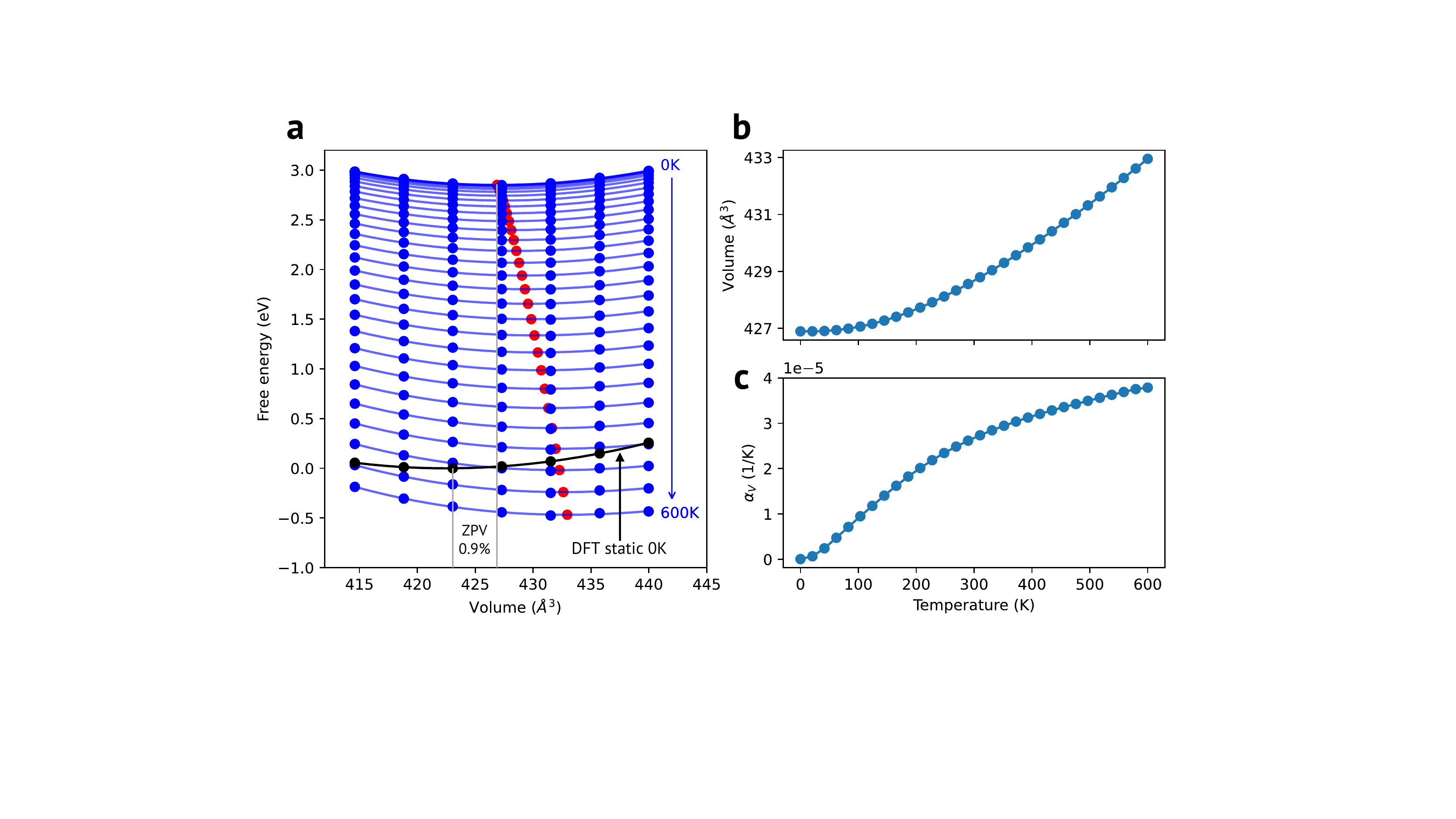}
	\caption{\textbf{a} Helmholtz free energy as a function of volume from 0~K to 600~K (see Eq.~(8) of the main text), using seven configurations within a [-2\%,+4\%] volume change with respect to the static undoped unit-cell equilibrium volume. 
	The minimum volume for a given T (red dots) is obtained using the Murnaghan equation of state. 
	The DFT static curve is displayed in black. 
	A zero-point volume (ZPV) of 0.9\% is computed. 
	\textbf{b} Change of volume with temperature and \textbf{c} change of volumic thermal expansion coefficient ($\alpha_V=\frac{1}{V}\frac{\partial V}{\partial T}$) with temperature.	}
	\label{fig:QHA_procedure}
\end{figure}

In Fig.~\ref{fig:QHA_procedure}, the Helmholtz free energy as a function of volume for different temperatures, see Eq.~(9) of the main text, is displayed with blue dots, from 0~K to 600~K,  using seven configurations within a [-2\%,+4\%] volume change with respect to the static unit-cell equilibrium volume. 
For each volume, we relaxed the structure with constrained volume and we computed the phonon density-of-states with DFPT using a  2$\times$2$\times$2 \textbf{k}-point and \textbf{q}-point grid and Fourier-interpolated it to a 15$\times$15$\times$15 fine \textbf{q}-grid.
The minimum volume for a given T (red dots) is obtained using Murnaghan equation of state for the fitting (blue curves). 
The DFT static curve is displayed in black. 
This allows to extract the zero-point volume (ZPV) which is computed to be an increase of 0.9\% between the minimum volume of the DFT static curve and the minimum volume of the Helmholtz free energy at 0~K. 
The volume dilation with temperature is shown in Fig.~\ref{fig:QHA_procedure}\textbf{b}, from which we extract the temperature-dependent volumic thermal expansion coefficient ($\alpha_V=\frac{1}{V}\frac{\partial V}{\partial T}$) shown in Fig.~\ref{fig:QHA_procedure}\textbf{c}.

\begin{figure}[h!]
	\centering
	\includegraphics[width=0.7\linewidth]{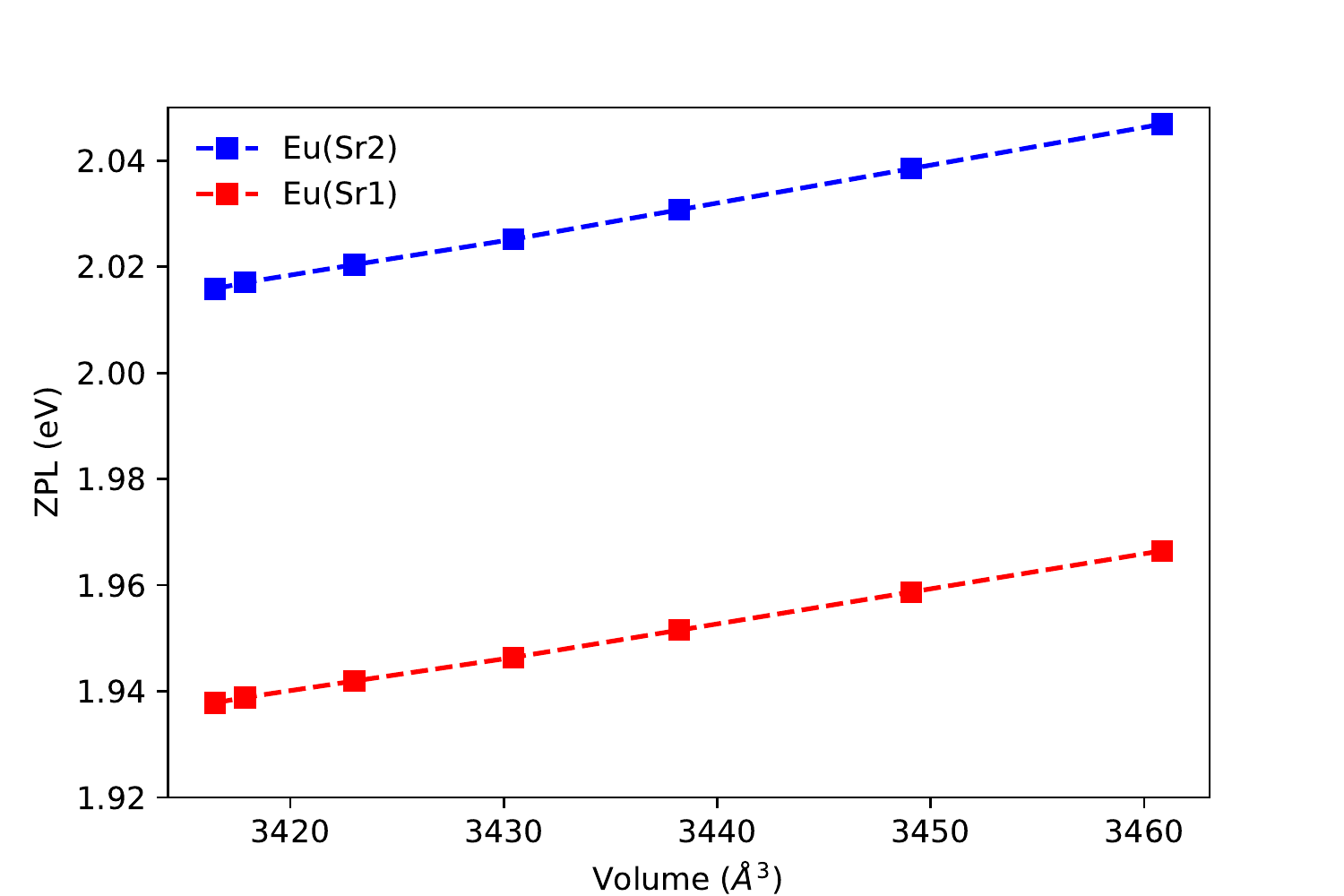}
	\caption{Change of zero-phonon line (ZPL) with volume where the different volumes are those corresponding to the selected temperatures. 
	A linear relationship is found for both sites with a slope of 0.66~meV/$\AA^3$ for site 1 and 0.72~meV/$\AA^3$ for site 2.}
	\label{fig:ZPL_V}
\end{figure}

We assessed the impact of thermal expansion on ZPL energies by scaling the lattice parameters of the bulk unit-cell SLA to the 288 atom supercells. 
We then performed the complete $\Delta$SCF procedure at fixed volumes corresponding to selected temperatures (10~K, 100~K, 200~K, 293~K, 373~K, 473~K, 573~K). 
This method enabled us to evaluate the temperature-dependent shift of the ZPL energies and the zero-point volume effect on the ZPL energy at 0~K.
The ZPL vs volume curve is shown on figure \ref{fig:ZPL_V} while the corresponding ZPL vs T curve is shown in the main inset of figure (6) of the main text. 
A linear relationship is found for both sites with a slope of 0.66 meV/$\AA^3$ for site 1 and 0.72 meV/$\AA^3$ for site 2

\clearpage
\section{Structural parameters}

The table~\ref{tab:Pristine_SLA} compares the relaxed cell parameters as computed in this work for bulk SLA with the experimental results~\cite{pust2014narrow}. 
We find an overestimation of the structural parameters by about 0.3\%, which is common for PBE exchange and correlation functional. 
The angles match within 0.05$^{\circ}$.
Tables~\ref{tab:doped_SLA_Sr1} and \ref{tab:doped_SLA_Sr2} show the local structural parameters around Sr1 and Sr2 in the pristine structure, as well as when replaced by Eu in the 4f ground or 5d excited state.
In the pristine SLA, we first observe that within the Sr chain, the Sr1-Sr1 distance of 3.2~\AA\, is shorter than the Sr2-Sr2 distance of 3.42~\AA. 
The Sr1-Sr2 distance is in between, with a value of 3.28~\AA. 
The mean distance between the Sr atom and its eight N neighbors is very similar in both sites (2.809~\AA{} and 2.804~\AA). 
When the Sr atoms are replaced by Eu atom, these distances change by less than 0.1\% due to similar atomic radii of the Eu$^{2+}$ and Sr$^{2+}$ atoms~\cite{shannon1976revised}.
When going from the 4f ground state to the 5d excited state, the mean Eu-N$_8$ shrinks in both sites by 1.32\% and 1.35\%, respectively. 
Regarding the Sr chain, their atomic distances Eu(Sr1)-Sr1 and Eu(Sr1)-Sr2 decrease by 1\% and 1.3\%, respectively. 
In the case of Eu(Sr2) instead, the distances elongate by 1.2\% and 0.3\%, respectively.

\begin{table}[h!]
	\centering
	\begin{tabular}{lrr}
		\toprule
		{} &  Exp.~\cite{pust2014narrow} &  This work \\
		\midrule
		a   [\AA]        &             5.866 &      5.883 \\
		b   [\AA]         &             7.511 &      7.532 \\
		c   [\AA]          &             9.965 &      9.995 \\
		 $\alpha$ [$^{\circ}$]    &            83.603 &     83.656 \\
		 $\beta$  [$^{\circ}$]   &            76.772 &     76.837 \\
		 $\gamma$ [$^{\circ}$]     &            79.565 &     79.596 \\
		Cell volume [\AA$^3$] &           419.250 &    423.047 \\
		\bottomrule
	\end{tabular}
    \caption{\label{tab:Pristine_SLA} Cell parameters of the undoped SLA as computed in this work compared to experimental results~\cite{pust2014narrow}.}
\end{table}

\begin{table}[h!]
	\centering
	\begin{tabular}{lrrr}
		\toprule
		{} &  pristine (X=Sr1) &  X=Eu(Sr1)-4f &  X=Eu(Sr1)-5d \\
		\midrule
		X-Sr1 distance  [\AA]     &             3.202 &         3.205 &         3.174 \\
		X-Sr2 distance   [\AA]   &             3.278 &         3.277 &         3.234 \\
		Mean X-N distance  [\AA]  &             2.809 &         2.813 &         2.776 \\
		\bottomrule
	\end{tabular}
	\caption{Local distances around the Sr1 atom in the pristine structure, or when replaced by an Eu atom in the 4f or 5d electronic state.}
	 \label{tab:doped_SLA_Sr1}
\end{table}

\begin{table}[h!]
	\centering
	\begin{tabular}{lrrr}
		\toprule
		{} &  pristine (X=Sr2) &  X=Eu(Sr2)-4f &  X=Eu(Sr2)-5d \\
		\midrule
		X-Sr1 distance   [\AA]   &             3.278 &         3.280 &         3.319 \\
		X-Sr2 distance  [\AA]    &             3.424 &         3.423 &         3.433 \\
		Mean X-N distance  [\AA] &             2.804 &         2.806 &         2.768 \\
		\bottomrule
	\end{tabular}
    \caption{ \label{tab:doped_SLA_Sr2}Local distances around Sr2 atom in the pristine structure, or when replaced by Eu atom in the 4f or 5d electronic state.}
   
\end{table}

\clearpage

\section{Kohn Sham levels}

\begin{figure}[h!]
	\centering
	\includegraphics[width=0.9\linewidth]{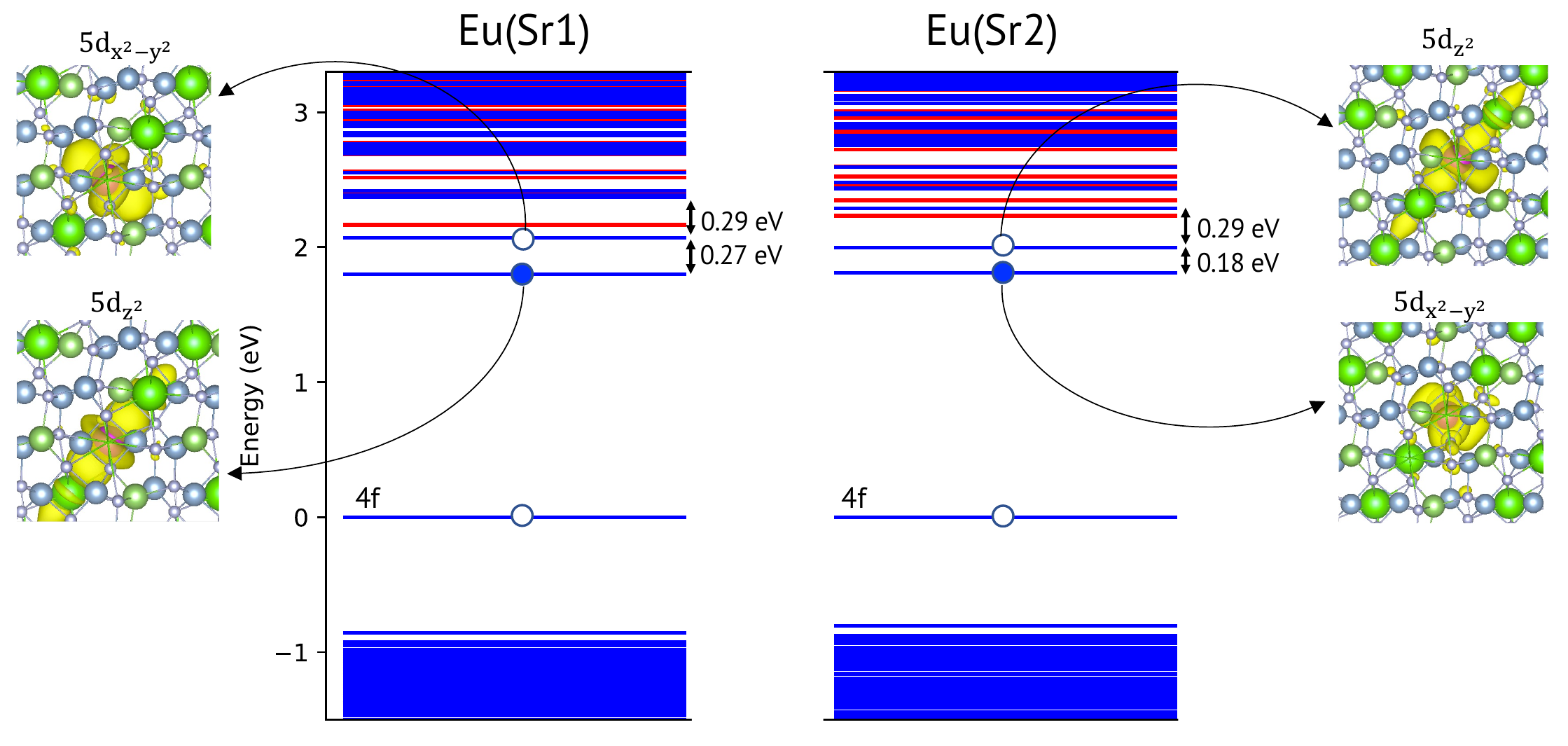}
	\caption{\label{fig:defect_levels} Kohn-Sham energies computed in the relaxed 288-atoms supercell at the PBE+U level for both Eu(Sr1) and Eu(Sr2) systems, following the constrained occupation method. 
	Blue (red) lines refers to spin up (down) states. 
	For Eu(Sr1), a 5d$_{z^2}$-like orbital is stabilized along the Sr chain while an empty 5d$_{x^2-y^2}$-like state is located in the band gap 0.27~eV above.  
	For Eu(Sr2), a 5d$_{x^2-y^2}$-like orbital is stabilized perpendicular to this Sr chain.  
	The empty 5d$_{z^2}$-like state is located 0.18~eV above. 
	The computed energy separation between the empty 5d states and the conduction bottom (Sr-4d character) is 0.29~eV.
	}
\end{figure}

In Fig.~\ref{fig:defect_levels}, we compare the localized DFT eigenstates at the zone center for the 2$\times$2$\times$2 supercell where one atom of Sr has been replaced by an Eu in the two inequivalent positions.  
We observe a character inversion of the lowest two 5d conduction states between the Eu(Sr1) and Eu(Sr2) cases.
For comparison, the energy have been shifted to be 0 at the position of the highest unoccupied 4f state.  

\clearpage

\section{Dominant phonon modes in the Huang-Rhys spectral function}

\begin{figure}[h!]
	\centering
	\includegraphics[width=0.88\linewidth]{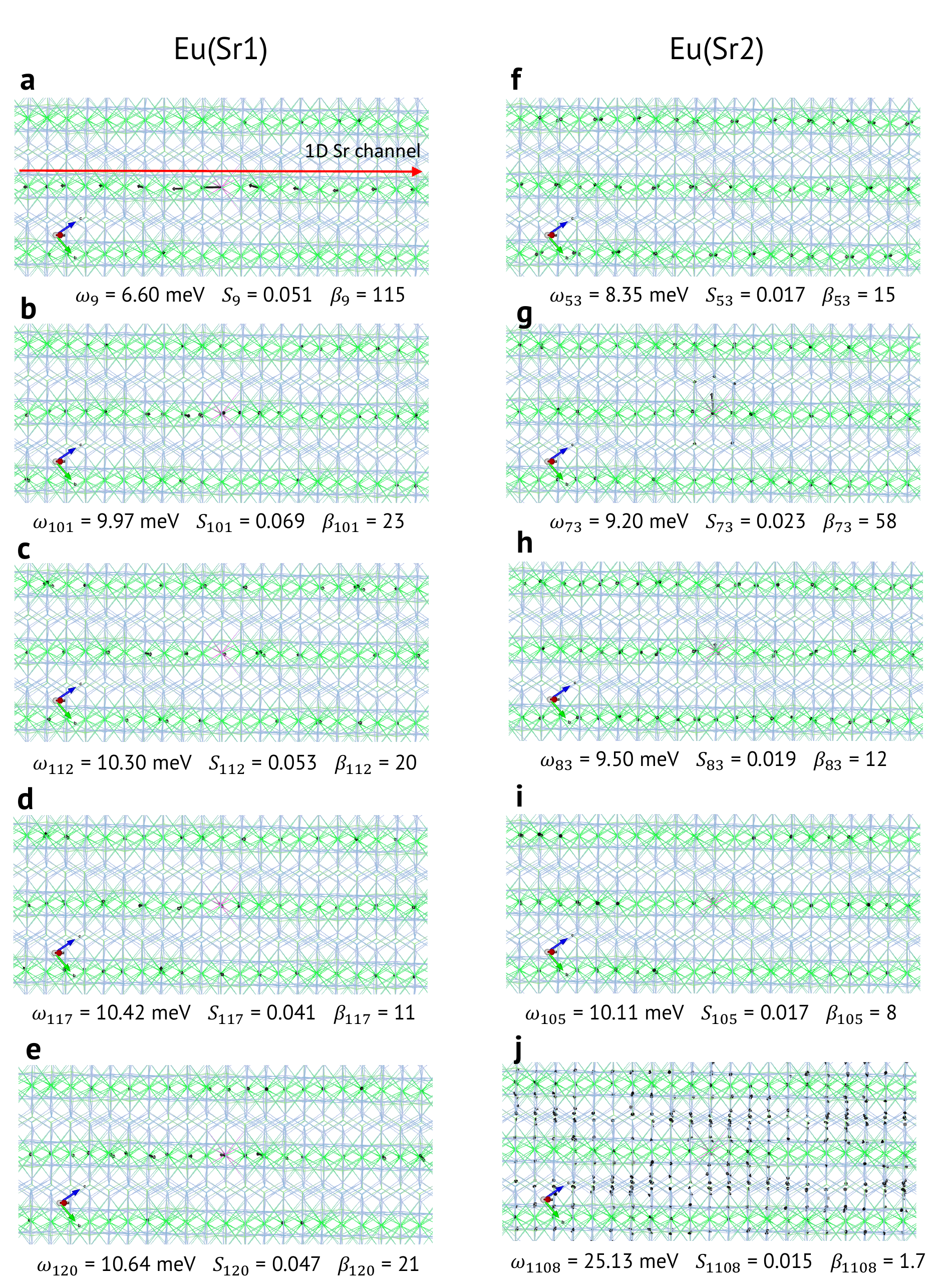}
	\caption{\label{fig:phonons_modes}
Atomic displacements associated with the five phonon modes with the highest partial Huang-Rhys factor $S_{\nu}$ 
for \textbf{a}-\textbf{e} the Eu(Sr1) and \textbf{f}-\textbf{j} the Eu(Sr2) sites, where $\omega_{\nu}$ is the phonon frequency and $\beta_{\nu}$ indicates the degree of localization. 
A $\beta_{\nu} \approx 1 $ describes delocalized modes while a $\beta_{\nu} \gg 1 $ describes (quasi)-local modes.
The displacements arrows have been scaled by 10 for \textbf{a}-\textbf{i} and by 20 for \textbf{j} for clarity.}
\end{figure}

We present in Fig.~\ref{fig:phonons_modes} the atomic displacements of the five dominant phonon modes, computed in a 4$\times$4$\times$4 supercell with 2304 atoms and a total of 6912 possible vibrational modes. 

\clearpage
\section{Eu(Sr1) PL spectrum with scaled forces}
\begin{figure}[h!]
	\centering
	\includegraphics[width=0.7\linewidth]{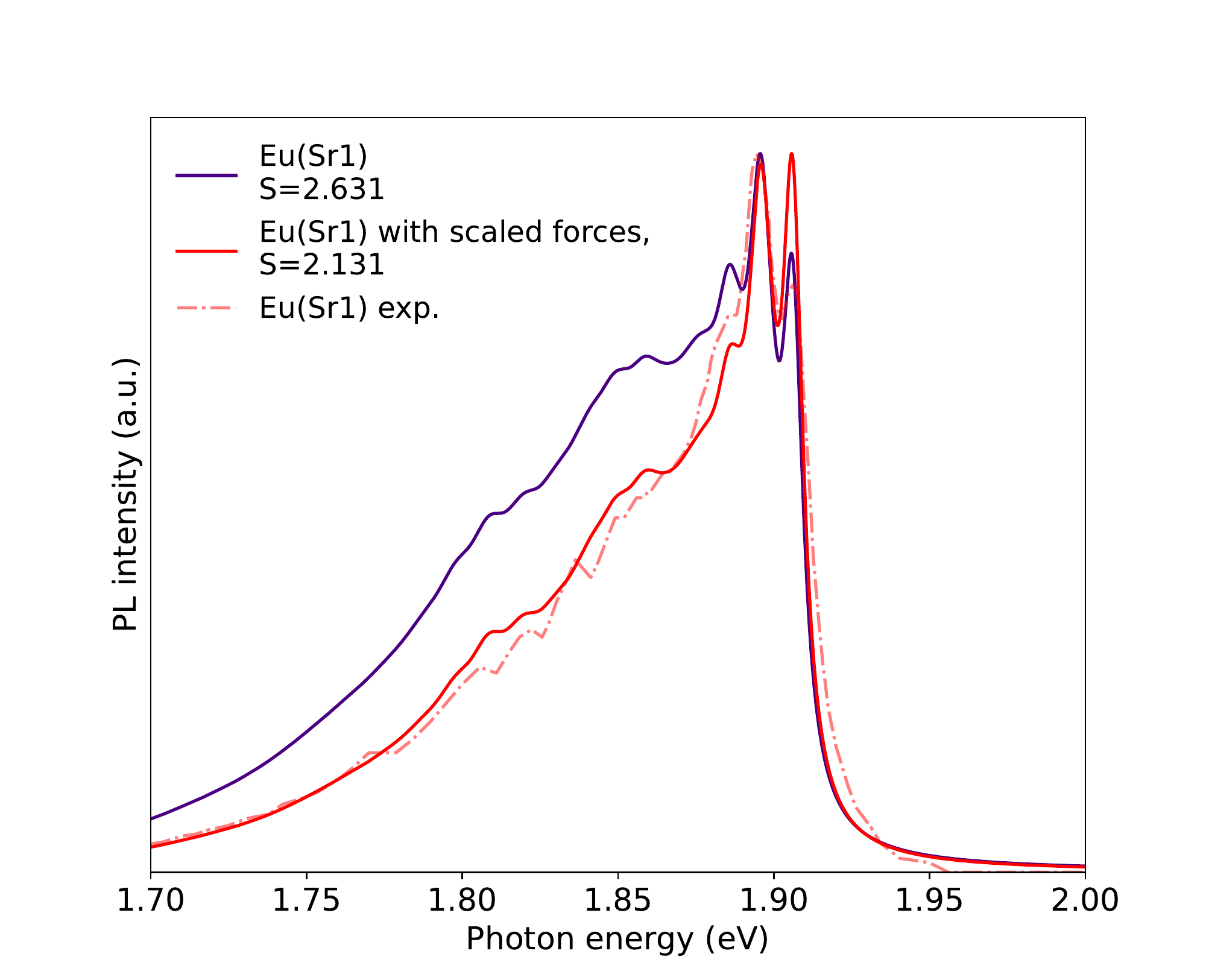}
	\caption{0K PL spectrum of Eu(Sr1) using 5d-4f forces from first principles, or scaled by 90\% or from experiment. The 90\% scaling yields an excellent reproduction of the experimental phonon side-band \cite{tsai2016improvement}. The reason for the overestimation of the forces remains unknown.}
        \label{fig:scaled_forces}
\end{figure}

In order to estimate the Huang-Rhys factor S1 of Eu(Sr1), we have scaled the 5d-4f forces in order to fit the experimental spectrum from Ref.~\cite{tsai2016improvement}. We find that scaling the forces by  90\% allows to obtain an excellent reproduction of the experimental phonon side-band, as shown figure \ref{fig:scaled_forces}. It shows that the true S1 should be around 2.13. This also suggests that the computed forces are slightly overestimated. The origin of this overestimmation remains unknown. We note that the Huang-Rhys spectral function is sensitive to the square of the forces (2.63*0.9$^2$$\approx$2.13), highlighting the difficulty in obtaining accurate Huang-Rhys factor and correct ZPL/phonon-side band weights.
\clearpage
\section{Emission spectrum shift with temperature}
We study in this section the temperature evolution of the photoluminescence spectra (PL) emission peak for the two sites and isolate the effect of thermal expansion and $\omega^3$ dependence of the PL intensity $L(\hbar\omega)\propto\omega^3A(\hbar\omega)$. 
\subsection{Thermal expansion effect}
\begin{figure}[h!]
	\centering
	\includegraphics[width=0.8\linewidth]{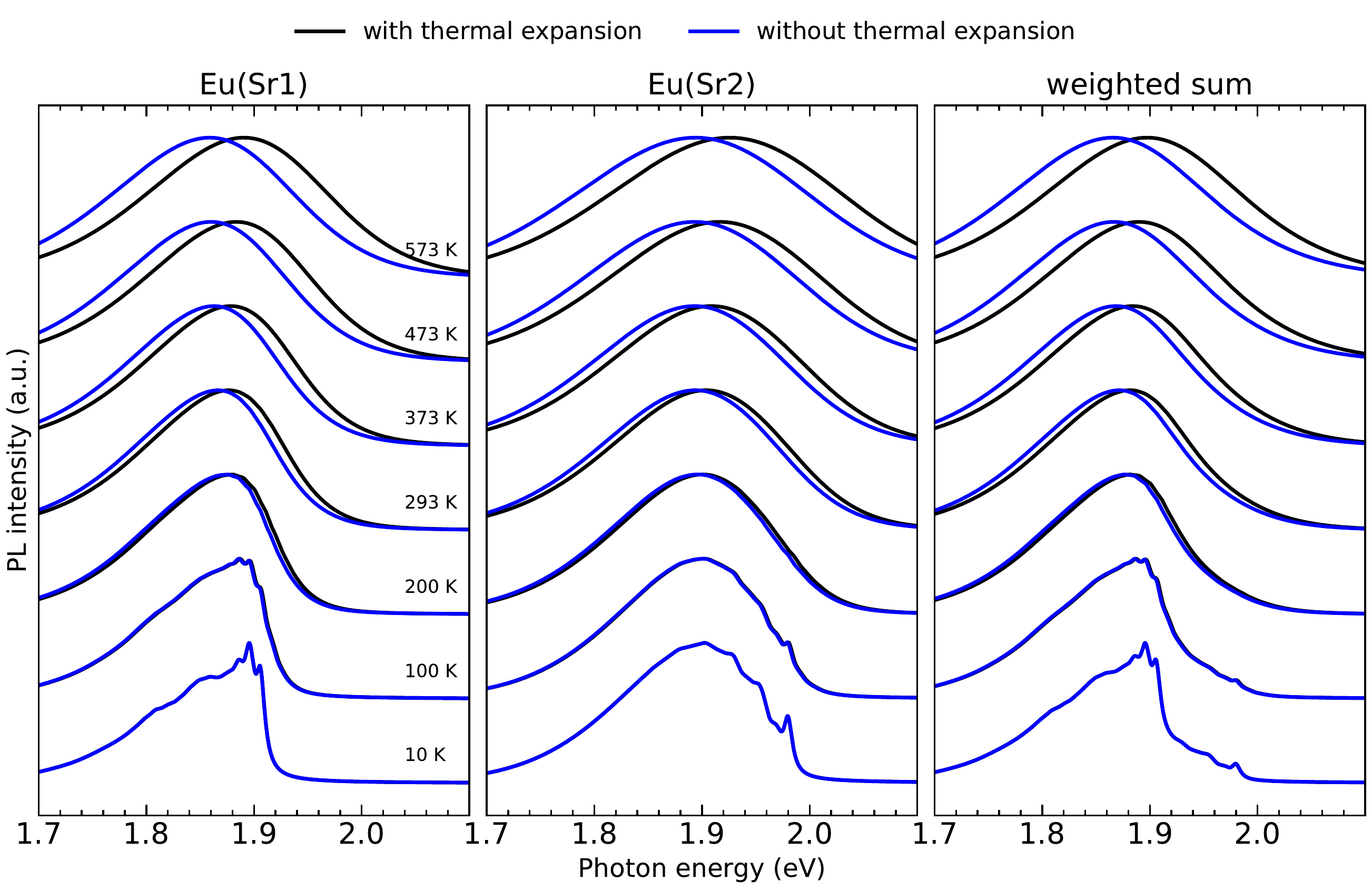}
	\caption{\label{fig:T_shift_spectra} Photoluminescence intensities of Eu(Sr1), Eu(Sr2) and their weighted sum. 
	Two cases are compared:  without thermal expansion and with thermal expansion.}
\end{figure}

\begin{figure}[h!]
	\centering
	\includegraphics[width=0.75\linewidth]{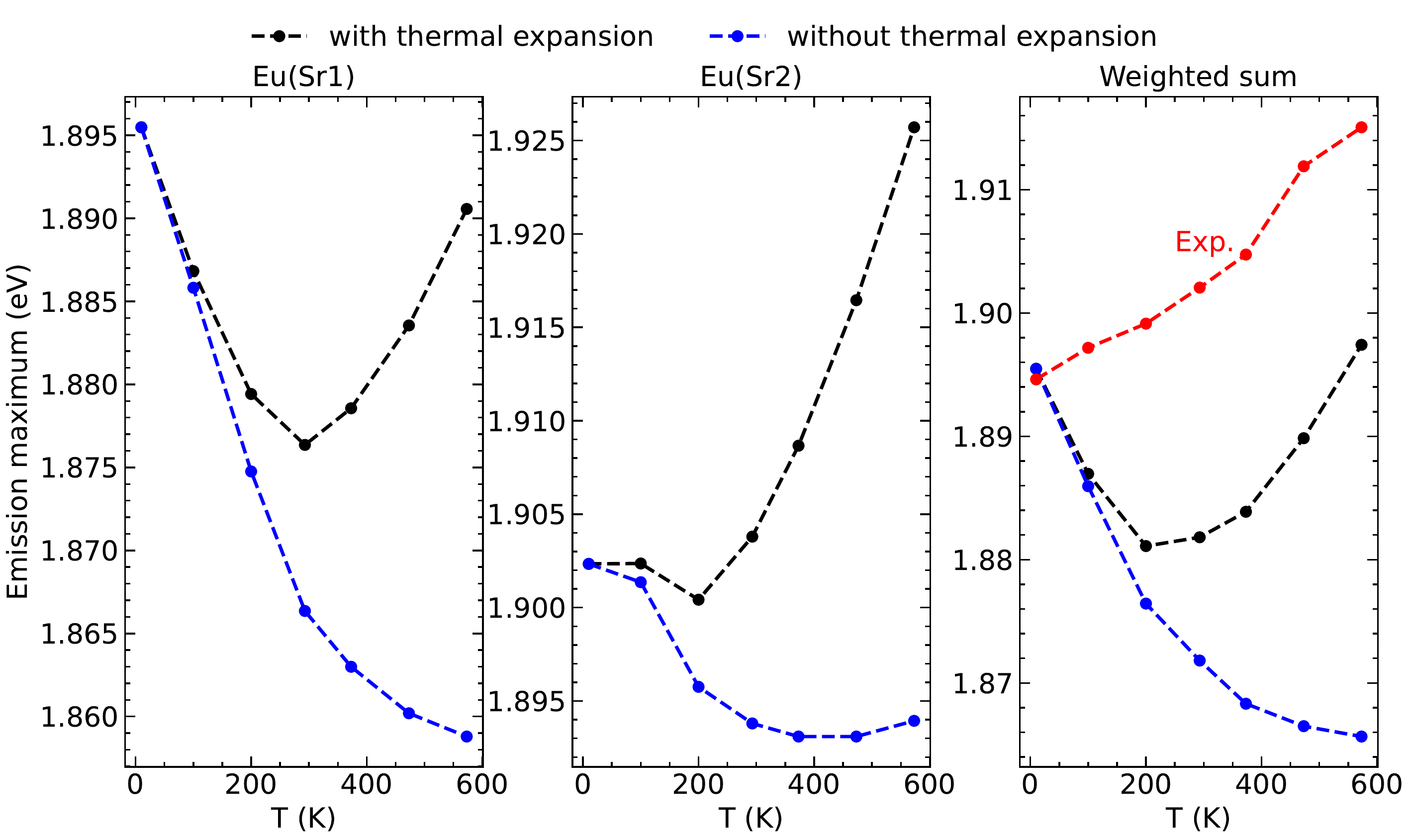}
	\caption{\label{fig:E_max_shift_T} Photon energy of the emission maximum as a function of temperature for both sites and their weighted sum. 
		Two cases are compared:  without thermal expansion and with thermal expansion. 
		In the case of the weighted sum, the experimental data from Ref.~\cite{tsai2016improvement} is presented in red.}
\end{figure}

In Fig.~\ref{fig:T_shift_spectra} we present the temperature evolution of the PL $L(\hbar\omega)$ for the two sites with, and without thermal expansion effect included. 
In Fig.~\ref{fig:E_max_shift_T} we show the emission peak position as a function of temperature.
%

\subsection{$\omega^3$ dependence effect}

\begin{figure}[h!]
	\centering
	\includegraphics[width=1\linewidth]{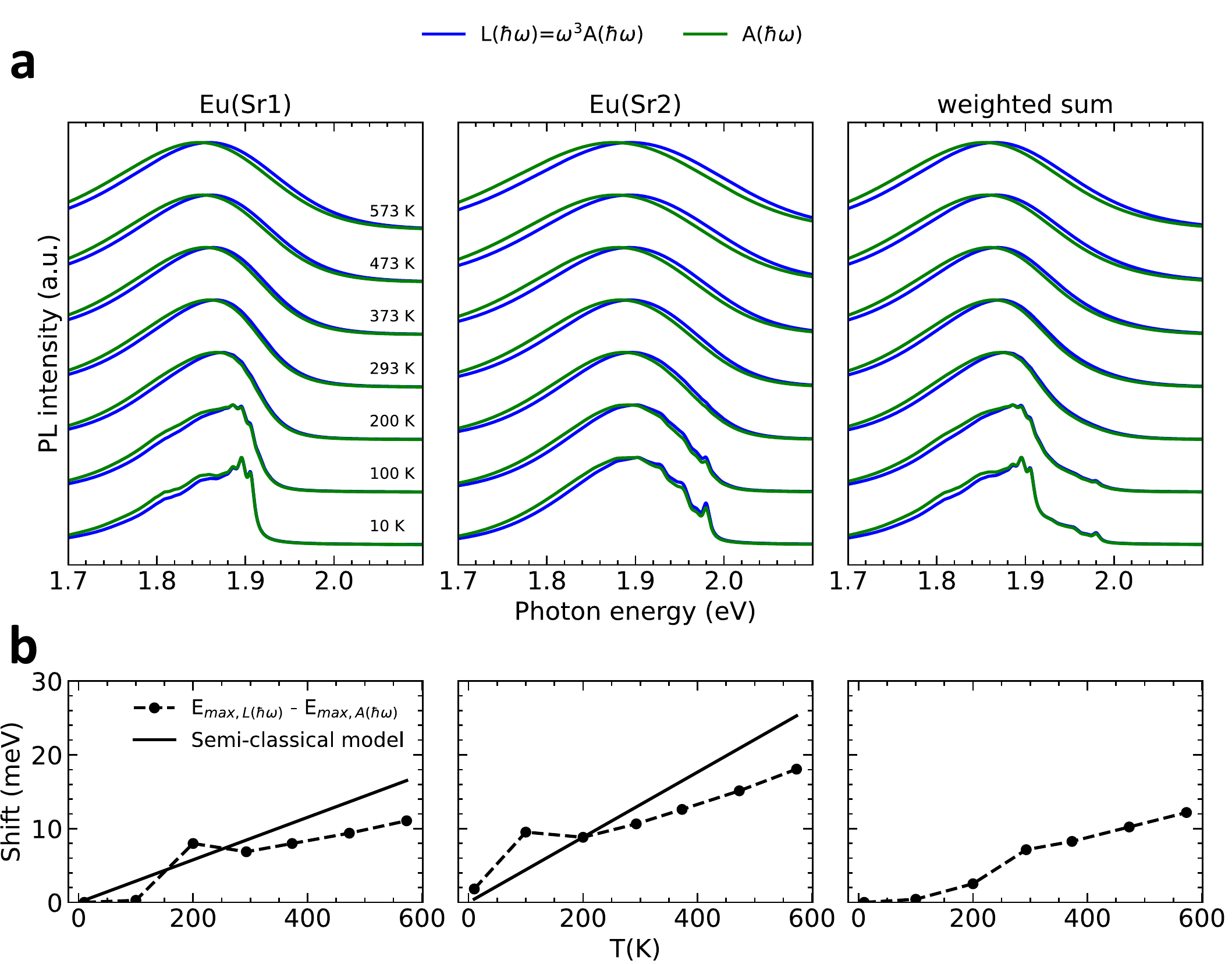}
	\caption{\label{fig:E_max_shift_T_omega_3}
 \textbf{a} Temperature dependent PL intensities $L(\hbar\omega)\propto \omega^3A(\hbar\omega)$ (blue curves) of Eu(Sr1), Eu(Sr2) and their weighted sum compared with the lineshape function $A(\hbar\omega)$ (green curves). \textbf{b} Shift  as computed with Eq.~\eqref{eq:shit} following a semi-classical model (plain lines) compared with the difference between the emission maximum of $L(\hbar\omega) \propto \omega^3A(\hbar\omega)$ with the emission maximum of $A(\hbar\omega)$ (dotted lines).}
\end{figure}

The $\omega^3$ dependence of $L(\hbar\omega) \propto \omega^3A(\hbar\omega)$ causes a blue shift of the emission maximum due to the broadening of the lineshape function $A(\hbar\omega)$ with temperature, even if $A(\hbar\omega)$ does not shift. This fact is illustrated on Fig~\ref{fig:E_max_shift_T_omega_3}\textbf{a}.

We note that within a semi-classical treatment of the PL spectrum, one can obtain from a one-dimensional configuration coordinate model the emisison shift with temperature~\cite{reshchikov2005luminescence,jia2017first}: 
\begin{equation}
	\label{eq:shit}
	E_{em}(T)-E_{em}(0)=\left(\frac{\Omega_g^2-\Omega_e^2}{\Omega_e^2}+\frac{8\Omega_g^4\Delta S(0)}{\Omega_e^2(\Omega_g^2+\Omega_e^2)E_{em}(0)}\right)k_{B}T,
\end{equation}
where the $\Omega_g$ and $\Omega_e$ are the effective frequencies of the ground and excited state.
$\Delta S(0)$ is the Stokes shift and $E_{em}(0)$ is the emission maximum at 0K. 
The first term of Eq.~\eqref{eq:shit} accounts for a potential difference in the Born-Oppenheimer curvatures of the ground and excited state which we find to be negligible for SLA.
The second term is obtained when considering the $\omega^3$ dependence. 
Within this formulation, the vibronic peaks are not treated explicitely and the PL spectrum possesses a gaussian-like shape. 

We compare on Fig~\ref{fig:E_max_shift_T_omega_3}\textbf{b} the shift of Eq.~\eqref{eq:shit} with the difference between the emission maximum of $L(\hbar\omega) \propto \omega^3A(\hbar\omega)$ (blue curve) with the emission maximum of $A(\hbar\omega)$ (green curve). 
For Eu(Sr1), Eq.~\eqref{eq:shit} yields $E_{em}(T=573K)-E_{em}(0)$=+16 meV whereas the more accurate computation yields +11 meV. 
For Eu(Sr2), Eq.~\eqref{eq:shit} yields $E_{em}(T=573K)-E_{em}(0)$=+25 meV whereas the more accurate computation yields +18 meV.
 It is worth noting that these discrepancies are reasonable since Eq.~\eqref{eq:shit} is only applicable under conditions of strong electron-phonon coupling.
\clearpage
\medskip

\bibliography{main}